\documentclass[11pt]{article}
\pdfoutput=1

\usepackage{color}
\usepackage{mhchem}
\usepackage{xcolor}
\usepackage{colortbl}
\usepackage{hhline}
\usepackage{cite}
\usepackage{hyperref}
\hypersetup{colorlinks=true,linkcolor=red,anchorcolor=black,citecolor=green}
\usepackage[toc,page]{appendix}
\usepackage{mathtools}
\usepackage{amsfonts}
\usepackage{bbold}
\usepackage{textcomp}
\usepackage[DIV13]{typearea}
\usepackage{amsmath, amsthm, amssymb, mathtools,empheq,latexsym,dsfont}
\usepackage{bbm}
\usepackage{amsmath,amssymb,latexsym,dsfont}
\usepackage{slashed, simplewick}
\usepackage[utf8]{inputenc}
\usepackage{graphicx}
\usepackage{graphicx,placeins}
\usepackage{makeidx}
\usepackage[font=small,labelfont=bf]{caption}
\usepackage{nicefrac}
\usepackage{subfigure}
\usepackage{array, bigdelim,multirow,multicol}
\usepackage{mathrsfs}
\usepackage{amsmath}
\usepackage{adjustbox}
\usepackage{booktabs}
\usepackage{indentfirst}
\usepackage{booktabs}
\usepackage{tabulary}
\usepackage{longtable}
\usepackage{fancybox}
\usepackage{graphicx}
\usepackage{bm}
\usepackage{bbm}
\usepackage{tikz}
\usepackage{diagbox}
\usepackage{enumitem}
\usepackage[integrals]{wasysym}
\allowdisplaybreaks
\graphicspath{{immagini/}}
\definecolor{Gray}{gray}{0.92}

\newcommand{\ignore}[1]{}

\newcommand{\be}{\begin{equation}}
\newcommand{\ee}{\end{equation}}
\newcommand{\bea}{\begin{eqnarray}}
\newcommand{\eea}{\end{eqnarray}}

\definecolor{lightred}{rgb}{1,0.4,0.4}
\definecolor{lightgreen}{rgb}{0.4,1,0.4}
\definecolor{LightCyan}{rgb}{0.88,1,1}

\newcounter{Thm}[section]
\renewcommand{\theThm}{\arabic{section}.\arabic{Thm}}

\newcounter{nodecount}
\tikzstyle{every picture}+=[remember picture,baseline]
\tikzstyle{every node}+=[inner sep=0pt,anchor=base,
text depth=.25ex,outer sep=1.5pt]
\tikzstyle{every path}+=[thick, rounded corners]
\tikzset{
        plabel/.style={inner sep=2pt}
}

\makeatletter
\@addtoreset{equation}{section}
\makeatother

\begin{document}
\title{\begin{center}
		{\Large\bf Lepton models from non-holomorphic $A^{\prime}_{5}$ modular flavor symmetry}
\end{center}}
\date{}
\author{Cai-Chang Li$^{a,b,c}$\footnote{E-mail: {\tt
			ccli@nwu.edu.cn}},  \
	Gui-Jun Ding$^{d,e}$\footnote{E-mail: {\tt dinggj@ustc.edu.cn}} \ \\*[20pt]
	\centerline{\begin{minipage}{\linewidth}
			\begin{center}
				$^a${\it\small School of Physics, Northwest University, Xi'an 710127, China}\\[2mm]
				$^b${\it\small Shaanxi Key Laboratory for Theoretical Physics Frontiers, Xi'an 710127, China}\\[2mm]
				$^c${\it\small NSFC-SPTP Peng Huanwu Center for Fundamental Theory, Xi'an 710127, China}\\[2mm]
				$^d${\it \small Department of Modern Physics, University of Science and Technology of China,\\
					Hefei, Anhui 230026, China}\\[2mm]
                $^e${\it \small College of Physics, Guizhou University, Guiyang 550025, China}\\[2mm]
			\end{center}
	\end{minipage}} \\[10mm]}
\maketitle

\thispagestyle{empty}

\centerline{\large\bf Abstract}
\begin{quote}
\indent
In the framework of non-holomorphic modular invariance approach, we have systematically constructed all minimal lepton models based on the non-holomorphic $A^{\prime}_{5}$ modular symmetry from a bottom-up approach.  In these models, the Yukawa couplings are described by polyharmonic Maa{\ss} forms of integer weights at level $N=5$. Under the assumption of Majorana neutrinos, both the Weinberg operator and the type-I seesaw mechanism are considered for neutrino mass generation. All minimal models are found to be based on generalized CP (gCP) symmetry, and each of them depends on five real dimensionless parameters and two overall scales. Through comprehensive numerical scanning, we obtain 6 (4) phenomenologically viable Weinberg operator models and 94 (76) phenomenologically viable seesaw models for normal (inverted) ordering neutrino masses. For each viable model, we present predictions for key neutrino properties, such as lepton masses, CP violation phases, mixing angles, effective Majorana mass for neutrinoless double beta decay and the kinematical mass in beta decay. Furthermore, we provide detailed numerical analysis for two representative models  to illustrate our results.

\end{quote}
\newpage
%%%%%%%%%%%%%%%%%%%%%%%%%%%%%%%%%%%%%%%%%%%%%%%%%%%%%%%%
\section{Introduction}
%%%%%%%%%%%%%%%%%%%%%%%%%%%%%%%%%%%%%%%%%%%%%%%%%%%%%%%%

It is known that the masses of quarks and leptons ‌span‌ at least 12 orders of magnitude. Furthermore, quarks show small mixing angles, while leptons exhibit two large angles ($\theta_{12}$ and $\theta_{23}$) and one small angle ($\theta_{13}$) which is of the same order of magnitude as the quark Cabibbo mixing angle~\cite{ParticleDataGroup:2024cfk}. The fermion mass hierarchies and contrasting mixing patterns between quark and lepton sectors constitutes the flavor puzzle of the Standard Model (SM), which requires a dynamical explanation. This remains a central mystery in particle physics, driving extensive research into the fundamental principles of the SM flavor sector. It was found that the large lepton mixing angles can be understood by extending the SM with a non-abelian discrete flavor symmetry~\cite{Altarelli:2010gt,Ishimori:2010au,King:2013eh,King:2014nza,King:2017guk,Petcov:2017ggy,Xing:2020ijf,Feruglio:2019ybq,Almumin:2022rml,Ding:2024ozt,Ding:2023htn}. In the models based on traditional discrete flavor symmetry, the vacuum expectation values (VEVs) of the  flavon fields which transform nontrivially under the flavor symmetry group should be oriented along certain directions in flavor space. Dynamically achieving this vacuum alignment is technically challenging, resulting in complex model structures that typically fail to provide quantitative predictions for fermion masses.

Recently modular symmetry has been suggested as the origin of such a discrete flavor symmetry~\cite{Feruglio:2017spp}. This framework avoids the use of flavon fields. Instead, flavor symmetry breaking is uniquely achieved through the VEV of the modulus $\tau$. Here, the Yukawa couplings, which transform non-trivially under modular symmetry, are modular forms which are holomorphic functions of $\tau$. This approach avoids vacuum alignment challenges and yields highly predictive fermion mass models. The minimal lepton model predicts all lepton masses and mixing parameters with just six real parameters~\cite{Ding:2022nzn}, while the unified model describes both quark and lepton sectors using only fourteen~\cite{Ding:2023ydy,Ding:2024pix}. Moreover, the complex modulus $\tau$ potentially bridges particle physics and cosmology: it can drive cosmic inflation~\cite{Ding:2024neh,King:2024ssx,Casas:2024jbw,Kallosh:2024ymt,Ding:2024euc}, be responsible for the reheating of the Universe~\cite{Ding:2024euc}, and generate baryon asymmetry~\cite{Duch:2025abl}. The collapse of modular domain wall could generate gravitational waves~\cite{King:2024lki}. It was shown that the modular symmetry could  provide a solution to the strong CP problem without introducing axion~\cite{Feruglio:2023uof,Petcov:2024vph,Penedo:2024gtb,Feruglio:2024ytl,Feruglio:2025ajb}. Although the predictive power of this framework may be reduced by the K$\ddot{\mathrm{a}}$hler potential which is less  constrained by modular symmetry~\cite{Chen:2019ewa}. The problem can be solved by extending modular symmetry to include a traditional flavor symmetry. This idea gives rise to the so-called eclectic flavor group~\cite{Baur:2019kwi,Baur:2019iai,Nilles:2020tdp,Nilles:2020gvu,Nilles:2020nnc,Nilles:2020kgo,Ding:2023ynd,Li:2023dvm,Li:2024pff} or quasi-eclectic flavor symmetry~\cite{Chen:2021prl}.

In the framework of originally modular symmetry, the supersymmetry (SUSY) is required to enforce that the Yukawa couplings are modular forms of level $N$ as functions of $\tau$~\cite{Lauer:1989ax,Ferrara:1989bc,Ferrara:1989qb,Feruglio:2017spp}.  However, the experimental evidence for low energy SUSY remains elusive~\cite{ParticleDataGroup:2024cfk}, leaving its realization in nature uncertain. This motivates exploring modular invariance independently of SUSY. It was suggested that the framework of automorphic forms provides a non-supersymmetric realization of the modular flavor symmetry~\cite{Ding:2020zxw}. Subsequently, a non-supersymmetric formulation of the modular flavor symmetry was recently proposed in Refs.~\cite{Qu:2024rns,Qu:2025ddz}. The harmonic condition replaces the requirement for holomorphicity, while modularity remains intact. Consequently, the Yukawa couplings emerge as polyharmonic Maa{\ss} forms of level $N$. Similar to the supersymmetric modular flavor symmetry, the generic integer weight polyharmonic Maa{\ss} forms of level $N$ naturally organize into multiplets of the finite modular group $\Gamma^{\prime}_N$~\cite{Qu:2025ddz} up to the automorphic factor, while the even integer weight polyharmonic Maa{\ss} forms can be arranged into multiplets of $\Gamma_N$~\cite{Qu:2024rns}.  In comparison with the holomorphic modular forms which are limited to $k\geq0$, the polyharmonic Maa{\ss} forms extend to negative weights and there exist non-holomorphic polyharmonic Maa{\ss} forms at weights $k=0,1$ and $2$. However, the polyharmonic Maa{\ss} forms of level $N$ coincide with holomorphic modular forms at weights $k\geq3$. This non-holomorphic modular symmetry, enabled by negative weight polyharmonic Maa{\ss} forms, enriches modular invariance, offering a new framework to address fermion masses and flavor mixing. The non-holomorphic modular symmetry can consistently combine with generalized CP (gCP) symmetry, which restricts coupling phases, boosting modular invariant model predictions, as in supersymmetric modular flavor symmetry~\cite{Novichkov:2019sqv}. The non-holomorphic modular flavor symmetry has been exploited to study lepton masses and flavor mixing  by using the inhomogeneous finite modular group $\Gamma_N$ or the homogeneous finite modular group $\Gamma^{\prime}_N$ in the literature. Some phenomenologically viable models based on the finite modular groups $\Gamma_2\cong S_3$~\cite{Okada:2025jjo}, $\Gamma_3\cong A_4$~\cite{Qu:2024rns,Kumar:2024uxn,Nomura:2024ghc,Nomura:2024atp,Nomura:2024ctl,Nomura:2024vus,Kobayashi:2025hnc,Loualidi:2025tgw}, $\Gamma^{\prime}_{3}\cong T^{\prime}$~\cite{Qu:2025ddz}, $\Gamma_4\cong S_4$~\cite{Ding:2024inn}, and  $\Gamma_5\cong A_5$~\cite{Li:2024svh} have been constructed.

The homogeneous finite modular group $\Gamma^{\prime}_5=A^{\prime}_5$, a double cover of the alternating group  $A_{5}$, provides a richer structure of modular forms for constructing predictive models in the bottom-up modular invariance approach. The modular $A^{\prime}_5$ symmetry models with holomorphic modular forms have been widely studied in the literature~\cite{Yao:2020zml,Wang:2020lxk,Behera:2021eut,Behera:2022wco}. In this work, we present a comprehensive analysis of lepton models based on the finite modular group $A^{\prime}_5$, employing the integer weight polyharmonic Maa{\ss} forms of level $5$ in the framework of non-holomorphic modular flavor symmetry. The polyharmonic Maa{\ss} forms of level $N=5$ can be decomposed into multiplets of $A^{\prime}_5$ up to the automorphy factor and were derived in Ref.~\cite{Qu:2025ddz}. The modular symmetry can be extended to combine with gCP symmetry, which requires all coupling constants to be real in our basis. We focus on the most economical modular invariant lepton models, excluding any flavon field except $\tau$. In the neutrino sector, we assume neutrinos are Majorana particles, with their masses generated either by the effective Weinberg operator or via the type-I seesaw mechanism. We aim to seek all minimal non-holomorphic lepton flavor models based on $A^{\prime}_5$ using polyharmonic Maa{\ss} forms with weights from $-5$ to $6$, extending beyond those derived from non-holomorphic modular $A_{5}$ symmetry in Ref.~\cite{Li:2024svh}. After we numerically scan over the parameter space of each minimal model under both normal ordering (NO) and inverted ordering (IO) neutrino masses, we find 6 (4) phenomenologically viable Weinberg operator models and 94 (76) phenomenologically viable seesaw models under NO (IO) cases.  It was found that all these models successfully fit experimental data of lepton sector using only 7 real input parameters, as detailed in tables~\ref{tab:WO_bf}, \ref{tab:SS_best_fit_NO} and \ref{tab:SS_best_fit_IO}.

The rest of the paper is organized as follows. In section~\ref{sec:framework}, we briefly review the framework of non-holomorphic modular flavor symmetry. In section~\ref{sec:model}, we systematically classify non-holomorphic $A^{\prime}_5$ modular lepton models with gCP symmetry and present the phenomenologically viable models with the fewest free parameters along with their numerical fitting results. Two example models are presented and a thorough numerical analysis is performed to discuss their phenomenological implications for both NO and IO neutrino mass spectrums in section~\ref{sec:example_models}. Section~\ref{sec:conclusion} concludes the paper. The group theory of $A^{\prime}_5$ in our working basis are presented in Appendix~\ref{sec:A5p_group_theory}. The explicit expressions of the polyharmonic Maa{\ss} forms of level 5 and weights from $-5$ to $6$ are presented in Appendix~\ref{sec:plo_mass_form}.

%%%%%%%%%%%%%%%%%%%%%%%%%%%%%%%%%%%%%%%%%%%%%%%%%%%%%%%%
\section{\label{sec:framework}Framework}
%%%%%%%%%%%%%%%%%%%%%%%%%%%%%%%%%%%%%%%%%%%%%%%%%%%%%%%%%

We shall briefly review the framework of the non-holomorphic modular invariant theory, in which the Yukawa couplings are generic integer weight polyharmonic Maa{\ss} forms~\cite{Qu:2025ddz}. The full modular group $\Gamma=SL(2,\mathbb{Z})$ can be generated in terms of two generators $S$ and $T$ satisfying the relations:
\begin{equation}
S^2=-\mathbb{1}_{2}\,,\qquad S^4=(ST)^3=\mathbb{1}_{2},\qquad S^2T=TS^2\,,
\end{equation}
where $\mathbb{1}_{2}$ represents the $2\times2$ identity matrix. The matrix representations of the two generators can be written to be
\begin{eqnarray}
S=\begin{pmatrix}
0 & 1 \\
-1 & 0
\end{pmatrix}\,,\qquad
T=\begin{pmatrix}
1 & 1 \\
0 & 1
\end{pmatrix}\,.
\end{eqnarray}
Under the action of a modular transformation, the complex modulus $\tau$ in the upper half plane $\mathcal{H}=\left\{\tau\in\mathbb{C}|\Im(\tau)>0\right\}$ transforms as follows
\begin{equation}
\gamma\tau=\frac{a\tau+b}{c\tau+d}\,,\qquad \gamma=\begin{pmatrix}
a & b \\
c & d
\end{pmatrix}\in SL(2,\mathbb{Z})\,.
\end{equation}
The modular symmetry provides an origin of discrete flavor symmetry through the quotient $\Gamma^{\prime}_N\equiv\Gamma/\Gamma(N)$, where $\Gamma^{\prime}_N$ is the homogeneous finite modular group, and $\Gamma(N)$ denotes the so-called principal congruence subgroup of level $N$ and it is defined as
\begin{equation}
	\Gamma(N)=\left\{ \gamma \in SL(2,\mathbb{Z}) ~~\Big |~~ \gamma \equiv \begin{pmatrix}
		1 &~ 0 \\ 0 &~ 1
	\end{pmatrix} \mod N \right\}\,.
\end{equation}
For $N\leq5$, the homogeneous finite modular group $\Gamma^{\prime}_N$ can be generated by generators $S$ and $T$ satisfying the multiplication rule~\cite{Feruglio:2017spp,Liu:2019khw}
\begin{equation}\label{eq:GammaN_defing}
\Gamma^{\prime}_N=\left\{S,T\Big|S^4=(ST)^3=T^N=1,~S^2T=TS^2\right\}\,.
\end{equation}
Additional relations are needed to render the group finite for $N\geq 6$~\cite{deAdelhartToorop:2011re,Li:2021buv,Ding:2020msi}.

Modular flavor symmetry, first developed in supersymmetric models, requires Yukawa couplings to be holomorphic modular forms through combined modular and supersymmetric constraints~\cite{Feruglio:2017spp}. When extending it to the non-holomorphic regime, the construction of modular invariant Lagrangians necessarily  incorporate both matter fields and polyharmonic Maa{\ss} forms of level $N$. The polyharmonic Maa{\ss} forms $Y(\tau)$ of weight $k$ at level $N$ are functions of the complex modulus $\tau$, exhibiting‌ well-defined transformation properties under the group $\Gamma(N)$:
\begin{equation}
Y(\tau)\mapsto Y(\gamma\tau)=(c\tau+d)^{k}Y(\tau),\qquad \gamma\in\Gamma(N)\,.
\end{equation}
Moreover, they have to satisfy the following Laplacian condition and proper growth condition~\cite{Qu:2024rns},
\begin{eqnarray}
\nonumber&&\left(-4y^2\dfrac{\partial}{\partial\tau}\dfrac{\partial}{\partial \bar{\tau}} +2iky\dfrac{\partial}{\partial\bar{\tau}}\right)Y(\tau)=0\,,\\
&&Y(\tau)=\mathcal{O}(y^\alpha) \qquad \text{as $y\rightarrow +\infty$, uniformly in $x$}\,,
\end{eqnarray}
where $\alpha$ is some real parameter and $\tau\equiv x+i y$.

For weights $k\geq3$  and level $N$, non-holomorphic polyharmonic Maa{\ss} forms do not exist. In fact, such polyharmonic Maa{\ss} forms  coincide precisely  with known holomorphic modular forms, as documented in Ref.~\cite{Ding:2023htn}. At weight $k=2$, the modified Eisenstein series $\widehat{E}_2(\tau)$ is the unique non-holomorphic polyharmonic Maa{\ss} form, distinct from holomorphic modular forms. Its explicit expansion is given by:
\begin{equation}
\widehat{E}_2(\tau)=1-\frac{3}{\pi y}-24\sum^{\infty}_{n=1}\sigma_1(n)q^n=1-\frac{3}{\pi y}-24q-72q^{2}-96q^{3}-168q^{4}-144q^{5} -\ldots\,,
\end{equation}
where $q=e^{2\pi i \tau}$ and $\sigma_1(n)=\sum_{d|n}d$ denotes the divisor function, summing over all positive divisors of $n$. The construction of integer weights polyharmonic Maa{\ss} forms for cases $k\leq1$ was systematically presented in Ref.~\cite{Qu:2025ddz}. They can be derived from non-holomorphic Eisenstein series associated with each cusp of the principal congruence subgroup $\Gamma(N)$, as established in Ref.~\cite{Diamond:2005afc}. The cusps correspond to equivalence classes of $\mathbb{Q}\cup \{i\infty\}$ under the fractional linear action of $\Gamma(N)$, where $\mathbb{Q}$ denoted the set of rational numbers. A cusp is denoted as $\overline{A/C}$ where $A$ and $C$ are two coprime integers. The weight $k$ non-holomorphic Eisenstein series corresponding to the cusp $\overline{A/C}$ is defined by:
\begin{equation}\label{eq:non_holo_Eisenstein}
E_k(N; \tau; s; \overline{A/C}) = \sum_{\substack{(c,d)\equiv (-C, A) \,({\rm mod}\, N) \\ \gcd(c,d)=1} }  \dfrac{y^s}{(c\tau + d)^k |c\tau + d|^{2s}} \,.
\end{equation}
with $\gcd(c, d)$ indicating the greatest common divisor of $c$ and $d$. The condition for two coprime pairs $(A,C)$ and $(A^{\prime},C^{\prime})$  representing the  same cusp of $\Gamma(N)$ is the congruence $(A,C)\equiv \pm (A^{\prime},C^{\prime})\,({\rm mod}\, N)$~\cite{cohen:2017mfa}. For non-positive weights $k\leq 0$, the polyharmonic Maa{\ss} forms are precisely given by non-holomorphic Eisenstein series $\left\{E_k(N; \tau; 1-k; \overline{A/C})\Big|\overline{A/C}\in\mathcal{C}(N)\right\}$~\cite{Qu:2025ddz}, where $\mathcal{C}(N)$ denote a system of representatives of the cusps of $\Gamma(N)$, and their explicit expressions for $N\leq7$ are given in Ref.~\cite{Qu:2025ddz}. The linear space of the weight 1 polyharmonic Maa{\ss} forms is spanned by the first order derivative on $s$ of the non-holomorphic Eisenstein series $\left.\frac{\partial}{\partial s} E_1(N; \tau; s; \overline{A/C}) \right|_{s=0}$, where $\overline{A/C}$ ranges over all cusps of $\Gamma(N)$, up to equivalence.

It has been shown that the polyharmonic Maa{\ss} forms of integer weight $k$ and level $N$ can be arranged into distinct irreducible multiplets of $\Gamma^{\prime}_N$,  apart from the automorphy factor $(c\tau + d)^{k}$~\cite{Qu:2024rns,Qu:2025ddz}. Thus, one can choose a basis in which the polyharmonic Maa{\ss} form multiplet $Y^{(k)}_{\bm{r}}(\tau)$ transforms under an irreducible representation $\rho_{\bm{r}}$ of the finite modular group $\Gamma^{\prime}_N$, i.e.
\begin{equation}
Y^{(k)}_{\bm{r}}(\gamma \tau) = (c\tau + d)^k \rho_{\bm{r}}(\gamma) Y^{(k)}_{\bm{r}}(\tau)\,,\quad \gamma\in \Gamma\,,
\end{equation}
where $\gamma$ is a representative element of $\Gamma^{\prime}_N$. Note that the product of two polyharmonic Maa{\ss} forms of weights $k$ and $k^{\prime}$ generally is not a weight $k+k^{\prime}$ polyharmonic Maa{\ss} form, because the Laplacian condition could be spoiled for $k, k^{\prime}<0$. In the present work, we are interested in finite modular group $\Gamma^{\prime}_{5}\cong A^{\prime}_{5}$. The corresponding polyharmonic Maa{\ss} form multiplets for integer weights $k$ between $-5$ to $6$ were previously computed in Ref~\cite{Qu:2025ddz}, and we summarize their expressions in appendix~\ref{sec:plo_mass_form}.

In the framework of the non-holomorphic modular invariant theory, supersymmetry is not a requisite anymore. Instead, Yukawa couplings manifest as general integer weight polyharmonic Maa{\ss} forms. Under these conditions, the modular invariant Lagrangian for Yukawa interactions takes the form
\begin{eqnarray}
\mathcal{L}_{Y} =- Y^{(k_{Y})}(\tau) \psi^c \psi H + \mathrm{h.c.} \,,
\end{eqnarray}
where the polyharmonic Maa{\ss} form multiplet $Y^{(k_{Y})}(\tau)$, the Higgs field $H$ and the Weyl spinors $\psi$, $\psi^c$ (representing matter fields) transform under the modular group as
\begin{eqnarray}
\nonumber&&Y^{(k_{Y})}(\tau)\mapsto Y^{(k_{Y})}(\gamma\tau)=(c\tau+d)^{k_{Y}}\rho_{Y}(\gamma)Y^{(k_{Y})}(\tau), \qquad H(x) \mapsto (c\tau+d)^{-k_{H}}\rho_{H}(\gamma)H(x)\,,\\
&&\psi(x) \mapsto (c\tau+d)^{-k_{\psi}}\rho_{\psi}(\gamma) \psi(x)\,,\qquad \psi^c(x) \mapsto (c\tau+d)^{-k_{\psi^c}}\rho_{\psi^c}(\gamma)\psi^c(x)\,.
\end{eqnarray}
Here the modular weights $k_Y$, $k_{H}$, $k_{\psi}$ and $k_{\psi^c}$  are integers, while $\rho_{Y}(\gamma)$, $\rho_{H}(\gamma)$, $\rho_{\psi}(\gamma)$ and $\rho_{\psi^c}(\gamma)$ are irreducible representations of $\Gamma^{\prime}_N$. The modular invariance of $\mathcal{L}_{Y}$ requires the following weight and representation balance conditions:
\begin{equation}
k_{Y}=k_{H}+k_{\psi}+k_{\psi^c}\,, \qquad  \rho_{Y}\otimes\rho_{H}\otimes\rho_{\psi}\otimes\rho_{\psi^c} \ni\bm{1}\,,
\end{equation}
where $\bm{1}$ denotes the trivial singlet of $\Gamma^{\prime}_N$.

The gCP symmetry can be naturally incorporated into the framework of non-holomorphic modular flavor symmetry~\cite{Qu:2024rns,Qu:2025ddz}. Remarkably, when the modular generators $S$ and $T$ take unitary symmetric representation matrices, gCP invariance requires all coupling constants associated with each invariant singlet in the Lagrangian to be real~\cite{Novichkov:2019sqv}. This work analyzes the finite modular group $\Gamma^{\prime}_5\cong A^{\prime}_5$ of level $N=5$. The modular invariance constrains the Yukawa couplings to be polyharmonic Maa{\ss} forms of level $5$ given in appendix~\ref{sec:plo_mass_form}. We  work in the $A^{\prime}_5$ basis listed in appendix~\ref{sec:A5p_group_theory}, where the representation matrices of the generators $S$ and $T$ are unitary and symmetric across all irreducible representations. Consequently, gCP symmetry could constrain all coupling constants in the modular invariant Lagrangian to be real, with any CP violation stemming solely from the vacuum expectation value of $\tau$.

%%%%%%%%%%%%%%%%%%%%%%%%%%%%%%%%%%%%%%%%%%%%%%%%%%%%%%%%
\section{\label{sec:model}Model building  based on $A^{\prime}_{5}$ modular symmetry}
%%%%%%%%%%%%%%%%%%%%%%%%%%%%%%%%%%%%%%%%%%%%%%%%%%%%%%%%%

This section establishes a systematic classification for minimal lepton mass models governed by the non-holomorphic finite modular symmetry $\Gamma^{\prime}_{5}\cong A^{\prime}_{5}$ which is the double cover of the alternating group $A_{5}$. The Yukawa couplings in these models originate from the polyharmonic Maa{\ss} forms of level $N=5$ and integer weights~\cite{Qu:2025ddz}. As detailed in table~\ref{tab:MF_summary}, these modular forms decompose into irreducible representations of $A^{\prime}_{5}$, encompassing novel ‌higher-dimensional irreducible representations $\bm{\widehat {2}}$, $\bm{\widehat {2}^{\prime}}$, $\bm{\widehat {4}}$ and $\bm{\widehat {6}}$ absent in $A_{5}$. Neutrinos are assumed to acquire mass through both the Weinberg operator and the type-I seesaw mechanism with three right-handed neutrinos. The analysis concentrates on the most economical scenarios, where modular invariance is achieved without introducing flavon fields beside the complex modulus $\tau$. The flavor symmetry $A^{\prime}_{5}$ is spontaneously broken when $\tau$ acquires a VEV. The Higgs doublet $H$ is assumed to possess a modular weight of zero and  remain invariant under the $A^{\prime}_{5}$ group. The left-handed lepton doublets $L\equiv(L_1, L_2, L_3)^{T}$ with weight $k_{L}$ are assigned to either the triplet $\bm{3}$ or $\bm{3^{\prime}}$. For the three generations of right-handed charged leptons  $E^{c}_{1,2,3}$, their modular weights  $k_{E^{c}_{1,2,3}}$, correspond to representations that include direct sums of singlets $\bm{1}\oplus\bm{1}\oplus\bm{1}$, mixed doublet-singlet combinations ($\bm{\widehat {2}}\oplus\bm{1}$ or $\bm{\widehat {2}^{\prime}}\oplus\bm{1}$), or triplet structures $\bm{3}$ or $\bm{3^{\prime}}$. In type-I seesaw models, the right-handed neutrinos $N^{c}$ with modular weight $k_{N^{c}}$ are allocated to either the irreducible representation $\bm{3}$ or $\bm{3^{\prime}}$. Each representation assignment theoretically permits an infinite parameter space for field weight assignments, while independent coupling terms of the Lagrangian generally increases with the weight of the involved modular forms. To minimize free parameters,  the polyharmonic  Maa{\ss} forms of integer weight $k=-5$ to $k=6$ are employed. In the following, we will establish the definitive expressions of charged lepton and neutrino Yukawa couplings preserving $A^{\prime}_{5}$ symmetry, then write their mass matrices. The systematic exploration of viable representations and weight assignments aims to identify the most predictive and phenomenologically consistent  models. In the present work, the modular group $A^{\prime}_{5}$ is extended to combine with the gCP symmetry. Thus, the gCP invariance enforces the coupling constants accompanying each invariant term in the Lagrangian to be real.

\subsection{\label{sec:Ch_masses}Charged lepton sector }

We proceed to systematically analyze the Lagrangian and mass matrices for charged leptons. Due to ‌space limitations‌, we list ‌below‌ the models for which $\mathcal{L}_e$ contains ‌at most three‌ independent terms.

\begin{description}[labelindent=-0.8em, leftmargin=0.3em]
	
\item[~~(\romannumeral1)]{$L\sim\bm{3}$ and $E^{c}_{1,2,3}\sim\bm{1}$ }

In this case, the three right-handed charged leptons are distinguished from each other by their different modular weights $k_{E^{c}_{i}}$. The Lagrangian for the charged lepton Yukawa coupling reads as
\begin{equation}
\label{eq:Le_1st} C^{(1)}~:~-\mathcal{L}_e=\alpha\left(E^{c}_1LY^{(k_{E^{c}_{1}}+k_{L})}_{\bm{3}}H^{*}\right)_{\bm{1}}+
\beta\left(E^{c}_2LY^{(k_{E^{c}_{2}}+k_{L})}_{\bm{3}}H^{*}\right)_{\bm{1}}+\gamma\left(E^{c}_3LY^{(k_{E^{c}_{3}}+k_{L})}_{\bm{3}}H^{*}\right)_{\bm{1}}\,,
\end{equation}
where the symbol $Y^{(k_{E^{c}_{i}}+k_{L})}_{\bm{3}}$ denotes triplet $\bm{3}$ polyharmonic Maa{\ss} forms of level 5 and weight  $k_{E^{c}_{i}}+k_{L}$ under $A^{\prime}_5$. To ensure minimality, only weights producing a single independent triplet Maa{\ss} form are selected for the charged lepton sector. Thus, the viable values for the sum of modular weights $k_{E^{c}_{i}}+k_{L}$ are  $\pm4,\,\pm2$ and $0$, as can be seen from table~\ref{tab:MF_summary}. The proportional rows in the charged lepton mass matrix would make at least one lepton massless. Row permutation simply redefines right-handed lepton fields without changing physical predictions. Adopting the hierarchy $k_{E^{c}_{1}}+k_{L}<k_{E^{c}_{2}}+k_{L}<k_{E^{c}_{3}}+k_{L}$ without loss of generality, we find 10 distinct charged lepton mass matrices arising from unique assignments of $k_{E^{c}_{i}}+k_{L}$. The three terms in Eq.~\eqref{eq:Le_1st} exhibit analogous mathematical structures, with each corresponding to distinct rows in the charged lepton mass matrix. Consequently, the mass matrices for charged leptons across all ten cases can be systematically derived as outlined below:
\begin{equation}\label{eq:ch_mass1}
m^{(1)}_e(k_{E^{c}_1}+k_{L},k_{E^{c}_2}+k_{L},k_{E^{c}_3}+k_{L})=\begin{pmatrix}
                \alpha Y^{(k_{E^{c}_1}+k_{L})}_{\bm{3}, 1}  ~&~ \alpha Y^{(k_{E^{c}_1}+k_{L})}_{\bm{3}, 3} ~&~ \alpha Y^{(k_{E^{c}_1}+k_{L})}_{\bm{3}, 2}  \\
                \beta Y^{(k_{E^{c}_2}+k_{L})}_{\bm{3}, 1}  ~&~ \beta Y^{(k_{E^{c}_2}+k_{L})}_{\bm{3}, 3} ~&~ \beta Y^{(k_{E^{c}_2}+k_{L})}_{\bm{3}, 2}  \\
                \gamma Y^{(k_{E^{c}_3}+k_{L})}_{\bm{3}, 1}  ~&~\gamma Y^{(k_{E^{c}_3}+k_{L})}_{\bm{3}, 3} ~&~ \gamma Y^{(k_{E^{c}_3}+k_{L})}_{\bm{3}, 2}
        \end{pmatrix}v\,.
\end{equation}
where the charged lepton mass matrix $m^{(1)}_{e}$ is given in the right-left basis $E^{c}m^{(1)}_{e}L$, $v$ is the vacuum expectation value of $H$ and we denote $Y^{(k_{E^{c}_i}+k_{L})}_{\bm{3}}\equiv (Y^{(k_{E^{c}_i}+k_{L})}_{\bm{3},1},Y^{(k_{E^{c}_i}+k_{L})}_{\bm{3},2},Y^{(k_{E^{c}_i}+k_{L})}_{\bm{3},3})^{T}$. All the 10 configurations are systematically derived in Ref.~\cite{Li:2024svh}.

\item[~~(\romannumeral2)]{$L\sim\bm{3^{\prime}}$ and $E^{c}_{1,2,3}\sim\bm{1}$ }

The charged lepton mass terms in the Lagrangian are expressed as
\begin{equation}
C^{(2)}~:~-\mathcal{L}_e=\alpha\left(E^{c}_1LY^{(k_{E^{c}_{1}}+k_{L})}_{\bm{3^{\prime}}}H^{*}\right)_{\bm{1}}+
\beta\left(E^{c}_2LY^{(k_{E^{c}_{2}}+k_{L})}_{\bm{3^{\prime}}}H^{*}\right)_{\bm{1}}+\gamma\left(E^{c}_3LY^{(k_{E^{c}_{3}}+k_{L})}_{\bm{3^{\prime}}}H^{*}\right)_{\bm{1}}\,,
\end{equation}
where $k_{E^{c}_{i}}+k_{L}\in\{-4,\,-2,\,0,\,2,\,4\}$ with $k_{E^{c}_{1}}+k_{L}<k_{E^{c}_{2}}+k_{L}<k_{E^{c}_{3}}+k_{L}$. The ten distinct charged lepton mass matrices corresponding to different assignments of $k_{E^{c}_{i}}+k_{L}$ are
\begin{equation}\label{eq:ch_mass2}
m^{(2)}_e(k_{E^{c}_1}+k_{L},k_{E^{c}_2}+k_{L},k_{E^{c}_3}+k_{L})=\begin{pmatrix}
                \alpha Y^{(k_{E^{c}_1}+k_{L})}_{\bm{3^{\prime}}, 1}  ~&~ \alpha Y^{(k_{E^{c}_1}+k_{L})}_{\bm{3^{\prime}}, 3} ~&~ \alpha Y^{(k_{E^{c}_1}+k_{L})}_{\bm{3^{\prime}}, 2}  \\
                \beta Y^{(k_{E^{c}_2}+k_{L})}_{\bm{3^{\prime}}, 1}  ~&~ \beta Y^{(k_{E^{c}_2}+k_{L})}_{\bm{3^{\prime}}, 3} ~&~ \beta Y^{(k_{E^{c}_2}+k_{L})}_{\bm{3^{\prime}}, 2}  \\
                \gamma Y^{(k_{E^{c}_3}+k_{L})}_{\bm{3^{\prime}}, 1}  ~&~\gamma Y^{(k_{E^{c}_3}+k_{L})}_{\bm{3^{\prime}}, 3} ~&~ \gamma Y^{(k_{E^{c}_3}+k_{L})}_{\bm{3^{\prime}}, 2}
        \end{pmatrix}v\,,
\end{equation}
which are also obtained in Ref.~\cite{Li:2024svh}.

\item[~~(\romannumeral3)]{$L\sim\bm{3}$ and $E^{c}\sim\bm{3}$ }
	
For this assignment, the triplet right-handed charged lepton field with modular weight $k_{E^{c}}$ is defined as $E^{c}=(E^{c}_{1},E^{c}_{2},E^{c}_{3})^{T}$. For the simplest weight assignment, the Lagrangian $\mathcal{L}_e$  for the charged lepton masses is given by :
\begin{equation}\label{eq:Le_C3i}
C^{(3)}_{i}~:~-\mathcal{L}_e=\alpha\left(E^{c}LY^{(k_{i})}_{\bm{1}}H^{*}\right)_{\bm{1}}+
	\beta \left(E^{c}LY^{(k_{i})}_{\bm{3}}H^{*}\right)_{\bm{1}}+\gamma\left(E^{c}LY^{(k_{i})}_{\bm{5}}H^{*}\right)_{\bm{1}}\,,
\end{equation}
where $k_{i}=k_{E^{c}}+k_{L}=-4,-2,0,2$ for  $i=1,2,3,4$, respectively. If the modular group $A^{\prime}_{5}$ is extended to combine with the gCP symmetry, all the couplings $\alpha$, $\beta$ and $\gamma$ are further constrained to be real, and $\alpha$ can be taken to be positive real number. The four distinct charged lepton mass matrices corresponding to different assignments of the weight $k_{E^c}+k_{L}$ can be straightforwardly obtained
\begin{equation}\label{eq:ch_mass3}
\hskip-0.13in m^{(3)}_{e}(k_{i})=
	\begin{pmatrix}
	\alpha  Y^{(k_{i})}_{\bm{1}}+2 \gamma  Y^{(k_{i})}_{\bm{5},1} & -\beta Y^{(k_{i})}_{\bm{3},3}-\sqrt{3} \gamma  Y^{(k_{i})}_{\bm{5},5} & \beta  Y^{(k_{i})}_{\bm{3},2}-\sqrt{3} \gamma  Y^{(k_{i})}_{\bm{5},2} \\
	\beta  Y^{(k_{i})}_{\bm{3},3}-\sqrt{3} \gamma  Y^{(k_{i})}_{\bm{5},5} & \sqrt{6} \gamma  Y^{(k_{i})}_{\bm{5},4} & \alpha  Y^{(k_{i})}_{\bm{1}}-\beta  Y^{(k_{i})}_{\bm{3},1}-\gamma  Y^{(k_{i})}_{\bm{5},1} \\
	-\beta  Y^{(k_{i})}_{\bm{3},2}-\sqrt{3} \gamma  Y^{(k_{i})}_{\bm{5},2} & \alpha  Y^{(k_{i})}_{\bm{1}}+\beta  Y^{(k_{i})}_{\bm{3},1}-\gamma  Y^{(k_{i})}_{\bm{5},1} & \sqrt{6} \gamma  Y^{(k_{i})}_{\bm{5},3} \\
		\end{pmatrix}v\,.
\end{equation}

\item[~~(\romannumeral4)]{$L\sim\bm{3}$ and $E^{c}\sim\bm{3^{\prime}}$ }

From the multiplication rule $\bm{3}\otimes\bm{3^{\prime}}=\bm{4}\oplus\bm{5}$, we find that the Lagrangian $\mathcal{L}_e$  for the charged lepton masses can be written as :
\begin{eqnarray}
\nonumber C^{(4)}_{i}&:&-\mathcal{L}_e=\alpha\left(E^{c}LY^{(k_{i})}_{\bm{5}}H^{*}\right)_{\bm{1}}\,, \quad \text{for}  \quad k_{i}=-4,-2,0,2\,, \\
\tilde{C}^{(4)}&:&-\mathcal{L}_e=\alpha\left(E^{c}LY^{(4)}_{\bm{5}I}H^{*}\right)_{\bm{1}}+\beta\left(E^{c}LY^{(4)}_{\bm{5}II}H^{*}\right)_{\bm{1}}+\gamma\left(E^{c}LY^{(4)}_{\bm{4}}H^{*}\right)_{\bm{1}}\,.
\end{eqnarray}
After the electroweak symmetry breaking by the VEVs of Higgs, we obtain the charged lepton mass matrices
\begin{eqnarray}
\nonumber m^{(4)}_{e}(k_{i})&=&
	\alpha\begin{pmatrix}
	\sqrt{3}   Y^{(k_{i})}_{\bm{5},1} &   Y^{(k_{i})}_{\bm{5},5} &  Y^{(k_{i})}_{\bm{5},2} \\
	 Y^{(k_{i})}_{\bm{5},4} & -\sqrt{2}   Y^{(k_{i})}_{\bm{5},3} & -\sqrt{2}   Y^{(k_{i})}_{\bm{5},5} \\
	  Y^{(k_{i})}_{\bm{5},3} & -\sqrt{2}   Y^{(k_{i})}_{\bm{5},2} & -\sqrt{2}   Y^{(k_{i})}_{\bm{5},4} \\
	\end{pmatrix}v\,, \\
\nonumber 	\tilde{m}^{(4)}_{e}&=&
	\left[\begin{pmatrix}
\sqrt{3} \alpha  Y^{(4)}_{\bm{5}I,1}+	\sqrt{3} \beta  Y^{(4)}_{\bm{5}II,1}  & \alpha  Y^{(4)}_{\bm{5}I,5}+\beta  Y^{(4)}_{\bm{5}II,5} &\alpha  Y^{(4)}_{\bm{5}I,2}+\beta  Y^{(4)}_{\bm{5}II,2} \\
\alpha  Y^{(4)}_{\bm{5}I,4}+\beta  Y^{(4)}_{\bm{5}II,4} & -\sqrt{2} \alpha  Y^{(4)}_{\bm{5}I,3}-\sqrt{2} \beta  Y^{(4)}_{\bm{5}II,3} & -\sqrt{2} \alpha  Y^{(4)}_{\bm{5}I,5}-\sqrt{2} \beta  Y^{(4)}_{\bm{5}II,5} \\
\alpha  Y^{(4)}_{\bm{5}I,3}+\beta  Y^{(4)}_{\bm{5}II,3} & -\sqrt{2} \alpha  Y^{(4)}_{\bm{5}I,2}-\sqrt{2} \beta  Y^{(4)}_{\bm{5}II,2}  & -\sqrt{2} \alpha  Y^{(4)}_{\bm{5}I,4}-\sqrt{2} \beta  Y^{(4)}_{\bm{5}II,4} \\
	\end{pmatrix} \right. \\
\label{eq:ch_mass4}	&&\left.+\gamma\begin{pmatrix}
		0  & \sqrt{2} Y^{(4)}_{\bm{4},4} & \sqrt{2}   Y^{(4)}_{\bm{4},1}\\
		-\sqrt{2}   Y^{(4)}_{\bm{4},3} & - Y^{(4)}_{\bm{4},2} &  Y^{(4)}_{\bm{4},4} \\
		-\sqrt{2}  Y^{(4)}_{\bm{4},2} &  Y^{(4)}_{\bm{4},1} & - Y^{(4)}_{\bm{4},3} \\
	\end{pmatrix}\right]v\,,
\end{eqnarray}
The $C^{(4)}_{i}$ scenario faces a key limitation: a single coupling $\alpha$ proves insufficient to account for the observed hierarchy of three distinct charged lepton masses.

\item[~~(\romannumeral5)]{$L\sim\bm{3^{\prime}}$ and $E^{c}\sim\bm{3}$ }

This case is related to the previous one through interchanging the representation assignments of $L$ and $E^{c}$. As a consequence, the charged lepton mass matrix is the transpose of $m^{(4)}_{e}(k_{i})$ and $\tilde{m}^{(4)}_{e}$.

\item[~~(\romannumeral6)]{$L\sim\bm{3^{\prime}}$ and $E^{c}\sim\bm{3^{\prime}}$ }
		
The  charged lepton Lagrangian  reads as
\begin{equation}
C^{(6)}_{i}~:~-\mathcal{L}_e=\alpha\left(E^{c}LY^{(k_{i})}_{\bm{1}}H^{*}\right)_{\bm{1}}+
	\beta \left(E^{c}LY^{(k_{i})}_{\bm{3^{\prime}}}H^{*}\right)_{\bm{1}}+\gamma\left(E^{c}LY^{(k_{i})}_{\bm{5}}H^{*}\right)_{\bm{1}}\,,
\end{equation}
where  $k_{i}=k_{E^{c}}+k_{L}\in\{-4,\,-2,\,0,\,2\}$, and the corresponding charged lepton  mass matrices take the following form
\begin{equation}\label{eq:ch_mass6}
\hskip-0.13in m^{(6)}_{e}(k_{i})=
		\begin{pmatrix}
	\alpha  Y^{(k_{i})}_{\bm{1}}+2 \gamma  Y^{(k_{i})}_{\bm{5},1} & -\beta  Y^{(k_{i})}_{\bm{3^{\prime}},3}-\sqrt{3} \gamma  Y^{(k_{i})}_{\bm{5},4} & \beta  Y^{(k_{i})}_{\bm{3^{\prime}},2}-\sqrt{3} \gamma  Y^{(k_{i})}_{\bm{5},3} \\
	\beta  Y^{(k_{i})}_{\bm{3^{\prime}},3}-\sqrt{3} \gamma  Y^{(k_{i})}_{\bm{5},4} & \sqrt{6} \gamma  Y^{(k_{i})}_{\bm{5},2} & \alpha  Y^{(k_{i})}_{\bm{1}}-\beta  Y^{(k_{i})}_{\bm{3^{\prime}},1}-\gamma  Y^{(k_{i})}_{\bm{5},1} \\
	-\beta  Y^{(k_{i})}_{\bm{3^{\prime}},2}-\sqrt{3} \gamma  Y^{(k_{i})}_{\bm{5},3} & \alpha  Y^{(k_{i})}_{\bm{1}}+\beta  Y^{(k_{i})}_{\bm{3^{\prime}},1}-\gamma  Y^{(k_{i})}_{\bm{5},1} & \sqrt{6} \gamma  Y^{(k_{i})}_{\bm{5},5} \\
			\end{pmatrix}v\,.
\end{equation}

\item[~~(\romannumeral7)]{$L\sim\bm{3}$, $E^c_{d}\sim\bm{\widehat {2}}$,  $E^c_{3}\sim\bm{1}$ }

Without loss of generality, we consider the first two right-handed charged lepton generations as an $A^{\prime}_{5}$ doublet $E^{c}_d=(E^{c}_1, E^{c}_2)$  with modular weight $k_{E^{c}_{d}}$, while $E^c_{3}$ with modular weight $k_{E^{c}_{3}}$ remains invariant under $A^{\prime}_{5}$. For modular weights allowing up to three distinct terms in the charged lepton Yukawa couplings, the Lagrangian takes the form:
\begin{eqnarray}
\nonumber C^{(7)}_{i}&:&-\mathcal{L}_e=\alpha \left(Y_{\bm{\widehat {4}}}^{(3)} E^{c}_{d}L H^{*} \right)_{\bm{1}}+\beta \left(Y_{\bm{3}}^{(k_{i})} E^{c}_{3}L H^{*} \right)_{\bm{1}}\,,\\
\nonumber \tilde{C}^{(7)}_{i}&:&-\mathcal{L}_e=\alpha \left( Y_{\bm{\widehat {2}}}^{(5)} E^{c}_{d}L H^{*}\right)_{\bm{1}}+\beta \left( Y_{\bm{\widehat {4}}}^{(5)} E^{c}_{d}L H^{*}\right)_{\bm{1}}+\gamma \left( Y_{\bm{3}}^{(k_{i})} E^{c}_{3}L H^{*}\right)_{\bm{1}}\,, \\
 \widehat {C}^{(7)}&:&-\mathcal{L}_e=\alpha \left( Y_{\bm{\widehat {4}}}^{(3)} E^{c}_{d}L H^{*}\right)_{\bm{1}}+\beta \left( Y_{\bm{3}I}^{(6)} E^{c}_{3}L H^{*}\right)_{\bm{1}}+\gamma \left( Y_{\bm{3}II}^{(6)} E^{c}_{3}L H^{*}\right)_{\bm{1}}\,,
\end{eqnarray}
where the parameter $k_{i}=k_{E_{3}^c}+k_L$ takes the values $-4,-2,0,2,4$  for $i=1,2,3,4,5$, respectively. The corresponding charged lepton mass matrices are given by:
\begin{eqnarray}
\nonumber   m^{(7)}_{e}(k_{i}) &=& \left(
\begin{array}{ccc}
 \sqrt{2} \alpha  Y_{\bm{\widehat {4}},3}^{(3)} & -\alpha  Y_{\bm{\widehat {4}},2}^{(3)} & -\sqrt{3} \alpha  Y_{\bm{\widehat {4}},4}^{(3)} \\
 \sqrt{2} \alpha  Y_{\bm{\widehat {4}},2}^{(3)} & -\sqrt{3} \alpha  Y_{\bm{\widehat {4}},1}^{(3)} & \alpha  Y_{\bm{\widehat {4}},3}^{(3)} \\
 \beta  Y_{\bm{3},1}^{(k_{i})} & \beta  Y_{\bm{3},3}^{(k_{i})} & \beta  Y_{\bm{3},2}^{(k_{i})} \\
\end{array}
\right)v\,, \\
\nonumber \tilde{m}^{(7)}_{e}(k_{i}) &=& \left(
\begin{array}{ccc}
 \sqrt{2} \beta  Y_{\bm{\widehat {4}},3}^{(5)}-\alpha  Y_{\bm{\widehat {2}},2}^{(5)} & -\sqrt{2} \alpha  Y_{\bm{\widehat {2}},1}^{(5)}-\beta  Y_{\bm{\widehat {4}},2}^{(5)} & -\sqrt{3} \beta  Y_{\bm{\widehat {4}},4}^{(5)} \\
 \sqrt{2} \beta  Y_{\bm{\widehat {4}},2}^{(5)}-\alpha  Y_{\bm{\widehat {2}},1}^{(5)} & -\sqrt{3} \beta  Y_{\bm{\widehat {4}},1}^{(5)} & \sqrt{2} \alpha  Y_{\bm{\widehat {2}},2}^{(5)}+\beta  Y_{\bm{\widehat {4}},3}^{(5)} \\
 \gamma  Y_{\bm{3},1}^{(k_{i})} & \gamma  Y_{\bm{3},3}^{(k_{i})} & \gamma  Y_{\bm{3},2}^{(k_{i})} \\
\end{array}
\right)v\,, \\
\label{eq:ch_mass7}  \widehat {m}^{(7)}_{e} &=& \left(
\begin{array}{ccc}
 \sqrt{2} \alpha  Y_{\bm{\widehat {4}},3}^{(3)} & -\alpha  Y_{\bm{\widehat {4}},2}^{(3)} & -\sqrt{3} \alpha  Y_{\bm{\widehat {4}},4}^{(3)} \\
 \sqrt{2} \alpha  Y_{\bm{\widehat {4}},2}^{(3)} & -\sqrt{3} \alpha  Y_{\bm{\widehat {4}},1}^{(3)} & \alpha  Y_{\bm{\widehat {4}},3}^{(3)} \\
 \beta  Y_{\bm{3}I,1}^{(6)}+\gamma  Y_{\bm{3}II,1}^{(6)} & \beta  Y_{\bm{3}I,3}^{(6)}+\gamma  Y_{\bm{3}II,3}^{(6)} & \beta  Y_{\bm{3}I,2}^{(6)}+\gamma  Y_{\bm{3}II,2}^{(6)} \\
\end{array}
\right)v\,.
\end{eqnarray}

\item[~~(\romannumeral8)]{$L\sim\bm{3}$, $E^c_{d}\sim\bm{\widehat {2}^{\prime}}$,  $E^c_{3}\sim\bm{1}$ }

The charged lepton Yukawa couplings include up to three terms over 25 allowed modular weight assignments. These assignments lead to the following Lagrangian structure:
\begin{equation}
C^{(8)}_{i,j}~:~ -\mathcal{L}_e=\alpha \left(Y_{\bm{\widehat {6}}I}^{(k_{i})} E^{c}_{d}L H^{*} \right)_{\bm{1}}+\beta \left(Y_{\bm{\widehat {6}}II}^{(k_{i})} E^{c}_{d}L H^{*} \right)_{\bm{1}}+\gamma \left(Y_{\bm{3}}^{(k_{j})} E^{c}_{3}L H^{*} \right)_{\bm{1}}\,,
\end{equation}
with modular weights $k_{i}=k_{E_{d}^c}+k_{L}=-5,-3,-1,1,3$ and $k_{j}=k_{E_{3}^c}+k_{L}=-4,-2,0,2,4$. The corresponding charged lepton mass matrices are expressed as:
 \begin{eqnarray}
\nonumber  m^{(8)}_{e}(k_{i},k_{j})&=&\left(
\begin{array}{ccc}
 \sqrt{2} \alpha  Y_{\bm{\widehat {6}}I,6}^{(k_{i})} & \sqrt{2} \alpha  Y_{\bm{\widehat {6}}I,5}^{(k_{i})} & \alpha  (Y_{\bm{\widehat {6}}I,1}^{(k_{i})}+Y_{\bm{\widehat {6}}I,2}^{(k_{i})}) \\
 \sqrt{2} \alpha  Y_{\bm{\widehat {6}}I,3}^{(k_{i})} & \alpha  (Y_{\bm{\widehat {6}}I,1}^{(k_{i})}-  Y_{\bm{\widehat {6}}I,2}^{(k_{i})}) & \sqrt{2} \alpha  Y_{\bm{\widehat {6}}I,4}^{(k_{i})} \\
 \gamma  Y_{\bm{3},1}^{(k_{j})} & \gamma   Y_{\bm{3},3}^{(k_{j})} & \gamma   Y_{\bm{3},2}^{(k_{j})} \\
\end{array}
\right)v      \\
\label{eq:ch_mass8}&& +\beta\left(
\begin{array}{ccc}
 \sqrt{2}  Y_{\bm{\widehat {6}}II,6}^{(k_{i})} & \sqrt{2}   Y_{\bm{\widehat {6}}II,5}^{(k_{i})} &  Y_{\bm{\widehat {6}}II,1}^{(k_{i})}+Y_{\bm{\widehat {6}}II,2}^{(k_{i})} \\
 \sqrt{2}  Y_{\bm{\widehat {6}}II,3}^{(k_{i})} &   Y_{\bm{\widehat {6}}II,1}^{(k_{i})}-Y_{\bm{\widehat {6}}II,2}^{(k_{i})} & \sqrt{2}  Y_{\bm{\widehat {6}}II,4}^{(k_{i})} \\
 0 &0 & 0 \\
\end{array}
\right)v   \,.
\end{eqnarray}

\item[~~(\romannumeral9)]{$L\sim\bm{3^{\prime}}$, $E^c_{d}\sim\bm{\widehat {2}}$,  $E^c_{3}\sim\bm{1}$ }

Based on the weight assignments of the lepton fields, the charged lepton Lagrangian is demonstrated to be
\begin{equation}
 C^{(9)}_{i,j}~:~-\mathcal{L}_e=\alpha \left(Y_{\bm{\widehat {6}}I}^{(k_{i})} E^{c}_{d}L H^{*} \right)_{\bm{1}}+\beta \left(Y_{\bm{\widehat {6}}II}^{(k_{i})} E^{c}_{d}L H^{*} \right)_{\bm{1}}+\gamma \left(Y_{\bm{3}'}^{(k_{j})} E^{c}_{3}L H^{*} \right)_{\bm{1}}\,,
\end{equation}
with modular weight configurations $k_{i}=k_{E_{d}^c}+k_{L}=-5,-3,-1,1,3$ and $k_{j}=k_{E_{3}^c}+k_{L}=-4,-2,0,2,4$.  The corresponding lepton mass matrices are directly determined from the Lagrangian terms as
\begin{equation}  \label{eq:ch_mass9}
 m^{(9)}_{e}(k_{i},k_{j})=\left(
\begin{array}{ccc}
 \alpha  Y_{\bm{\widehat {6}}I,5}^{(k_{i})}+\beta  Y_{\bm{\widehat {6}}II,5}^{(k_{i})} & -\alpha  Y_{\bm{\widehat {6}}I,3}^{(k_{i})}-\beta  Y_{\bm{\widehat {6}}II,3}^{(k_{i})} & \alpha  Y_{\bm{\widehat {6}}I,2}^{(k_{i})}+\beta  Y_{\bm{\widehat {6}}II,2}^{(k_{i})} \\
 -\alpha  Y_{\bm{\widehat {6}}I,4}^{(k_{i})}-\beta  Y_{\bm{\widehat {6}}II,4}^{(k_{i})} & -\alpha  Y_{\bm{\widehat {6}}I,1}^{(k_{i})}-\beta  Y_{\bm{\widehat {6}}II,1}^{(k_{i})} & -\alpha  Y_{\bm{\widehat {6}}I,6}^{(k_{i})}-\beta  Y_{\bm{\widehat {6}}II,6}^{(k_{i})} \\
 \gamma  Y_{\bm{3}^{\prime},1}^{(k_{j})} & \gamma  Y_{\bm{3}^{\prime},3}^{(k_{j})} & \gamma  Y_{\bm{3}^{\prime},2}^{(k_{j})} \\
\end{array}
\right)v \,.
\end{equation}

\item[~~(\romannumeral10)]{$L\sim\bm{3^{\prime}}$, $E^c_{d}\sim\bm{\widehat {2}^{\prime}}$,  $E^c_{3}\sim\bm{1}$ }

In the same fashion as previous cases, we can read out the Lagrangian for charged lepton masses as follows,
\begin{eqnarray}
\nonumber C^{(10)}_{i}&:&-\mathcal{L}_e=\alpha \left(Y_{\bm{\widehat {4}}}^{(3)} E^{c}_{d}L H^{*} \right)_{\bm{1}}+\beta \left(Y_{\bm{3}^{\prime}}^{(k_{i})} E^{c}_{3}L H^{*} \right)_{\bm{1}}\,, \\
\nonumber \tilde{C}^{(10)}_{i}&:& -\mathcal{L}_e=\alpha \left( Y_{\bm{\widehat {2}^{\prime}}}^{(5)} E^{c}_{d}L H^{*}\right)_{\bm{1}}+\beta \left( Y_{\bm{\widehat {4}}}^{(5)} E^{c}_{d}L H^{*}\right)_{\bm{1}}+\gamma \left( Y_{\bm{3}^{\prime}}^{(k_{i})} E^{c}_{3}L H^{*}\right)_{\bm{1}}\,, \\
\widehat {C}^{(10)}&:&-\mathcal{L}_e=\alpha \left( Y_{\bm{\widehat {4}}}^{(3)} E^{c}_{d}L H^{*}\right)_{\bm{1}}+\beta \left( Y_{\bm{3}^{\prime}I}^{(6)} E^{c}_{3}L H^{*}\right)_{\bm{1}}+\gamma \left( Y_{\bm{3}^{\prime}II}^{(6)} E^{c}_{3}L H^{*}\right)_{\bm{1}}\,,
\end{eqnarray}
where indices satisfy $i=1,2,3,4,5$ for modular weight $k_{i}=k_{E_{3}^c}+k_L$ taking values $-4,-2,0,2,4$. Subsequent to electroweak symmetry breaking via Higgs VEV acquisition, the charged lepton mass matrices are derived as:
\begin{eqnarray}
\nonumber   m^{(10)}_{e}(k_{i}) &=& \left(
\begin{array}{ccc}
 \sqrt{2} \alpha  Y_{\bm{\widehat {4}},4}^{(3)} & \sqrt{3} \alpha  Y_{\bm{\widehat {4}},2}^{(3)} & -\alpha  Y_{\bm{\widehat {4}},1}^{(3)} \\
 -\sqrt{2} \alpha  Y_{\bm{\widehat {4}},1}^{(3)} & -\alpha  Y_{\bm{\widehat {4}},4}^{(3)} & \sqrt{3} \alpha  Y_{\bm{\widehat {4}},3}^{(3)} \\
 \beta  Y_{\bm{3}^{\prime},1}^{(k_{i})} & \beta  Y_{\bm{3}^{\prime},3}^{(k_{i})} & \beta  Y_{\bm{3}^{\prime},2}^{(k_{i})} \\
\end{array}
\right)v\,, \\
\nonumber \tilde{m}^{(10)}_{e}(k_{i}) &=&\left(
\begin{array}{ccc}
 \alpha  Y_{\bm{\widehat {2}^{\prime}},2}^{(5)}+\sqrt{2} \beta  Y_{\bm{\widehat {4}},4}^{(5)} & \sqrt{3} \beta  Y_{\bm{\widehat {4}},2}^{(5)} &   -\sqrt{2} \alpha  Y_{\bm{\widehat {2}^{\prime}},1}^{(5)}-\beta  Y_{\bm{\widehat {4}},1}^{(5)} \\
 \alpha  Y_{\bm{\widehat {2}^{\prime}},1}^{(5)}-\sqrt{2} \beta  Y_{\bm{\widehat {4}},1}^{(5)} & \sqrt{2} \alpha  Y_{\bm{\widehat {2}^{\prime}},2}^{(5)}-\beta  Y_{\bm{\widehat {4}},4}^{(5)} & \sqrt{3} \beta  Y_{\bm{\widehat {4}},3}^{(5)} \\
 \gamma  Y_{\bm{3}^{\prime},1}^{(k_{i})} & \gamma  Y_{\bm{3}^{\prime},3}^{(k_{i})} & \gamma  Y_{\bm{3}^{\prime},2}^{(k_{i})} \\
\end{array}
\right)v \,, \\
\label{eq:ch_mass10}  \widehat {m}^{(10)}_{e}(k_{i}) &=&\left(
\begin{array}{ccc}
 \sqrt{2} \alpha  Y_{\bm{\widehat {4}},4}^{(3)} & \sqrt{3} \alpha  Y_{\bm{\widehat {4}},2}^{(3)} & -\alpha  Y_{\bm{\widehat {4}},1}^{(3)} \\
 -\sqrt{2} \alpha  Y_{\bm{\widehat {4}},1}^{(3)} & -\alpha  Y_{\bm{\widehat {4}},4}^{(3)} & \sqrt{3} \alpha  Y_{\bm{\widehat {4}},3}^{(3)} \\
 \beta  Y_{\bm{3}^{\prime}I,1}^{(6)}+\gamma  Y_{\bm{3}^{\prime}II,1}^{(6)} & \beta  Y_{\bm{3}^{\prime}I,3}^{(6)}+\gamma  Y_{\bm{3}^{\prime}II,3}^{(6)} & \beta  Y_{\bm{3}^{\prime}I,2}^{(6)}+\gamma  Y_{\bm{3}^{\prime}II,2}^{(6)} \\
\end{array}
\right)v \,.
\end{eqnarray}

In summary, under the assumption that the left-handed lepton doublets $L$ are assigned to triplets while the right-handed charged leptons $E^{c}_{1,2,3}$ adopt all possible assignments under the finite modular group $A^{\prime}_{5}$, ten unique representation configurations emerge for these fermionic fields. For each configuration, the charged lepton Lagrangian $\mathcal{L}_e$ is restricted to contain no more than three independent terms in our analysis.  Our investigation focuses on the polyharmonic Maa{\ss} forms of integer weights $k=-5$ to $k=6$, as higher weights modular expansions typically introduce excessive free parameters that compromise model predictability. The comprehensive classification of 110 cases across all possible charged lepton sector configurations is systematically presented in table~\ref{tab:sum_ch}. Crucially, the incorporation of gCP symmetry with the $A^{\prime}_{5}$ modular group imposes stringent constraints, reducing all coupling parameters $\alpha$, $\beta$ and $\gamma$ to be real coefficients.

\end{description}

\begin{table}[t!]
\begin{center}
\renewcommand{\tabcolsep}{1.mm}
\renewcommand{\arraystretch}{1.1}
\begin{tabular}{|c|c|c|c|c|c|}  \hline\hline
\texttt{Cases} & $(\rho_{L},\rho_{E^{c}_{1}},\rho_{E^{c}_{2}},\rho_{E^{c}_{3}})$  & $(k_{L}+k_{E^{c}_{1}},k_{L}+k_{E^{c}_{2}},k_{L}+k_{E^{c}_{3}})$ & $m_{e}$ & inputs  \\  \hline
$C^{(1)}$ & $(\bm{3},\bm{1},\bm{1},\bm{1})$ &  $k_{L}+k_{E^{c}_{1}}<k_{L}+k_{E^{c}_{2}}<k_{L}+k_{E^{c}_{3}}$ & $m^{(1)}_{e}$ in Eq.~\eqref{eq:ch_mass1}  & \multirow{2}{*}{$\alpha$, $\beta$, $\gamma$} \\ \cline{1-2} \cline{4-4}
$C^{(2)}$ & $(\bm{3^{\prime}},\bm{1},\bm{1},\bm{1})$ & $\in\{-4,-2,0,2,4\}$ & $m^{(2)}_{e}$ in Eq.~\eqref{eq:ch_mass2} &  \\ \hline \hline
\texttt{Cases} & $(\rho_{L},\rho_{E^{c}})$  & $k_{L}+k_{E^{c}}$ & $m_{e}$ & inputs \\  \hline
 $C^{(3)}_{i}$ & $(\bm{3},\bm{3})$ & \multirow{2}{*}{$-4$, $-2$, $0$, $2$} & $m^{(3)}_{e}$ in Eq.~\eqref{eq:ch_mass3}  & $\alpha$, $\beta$, $\gamma$   \\ \cline{1-2} \cline{4-5}
 $C^{(4)}_{i}$ & \multirow{2}{*}{$(\bm{3},\bm{3^{\prime}})$} &  & $m^{(4)}_{e}$ in Eq.~\eqref{eq:ch_mass4} & $\alpha$  \\ \cline{1-1} \cline{3-5}
  $\tilde{C}^{(4)}$ &  & $4$ & $\tilde{m}^{(4)}_{e}$ in Eq.~\eqref{eq:ch_mass4} & $\alpha$, $\beta$, $\gamma$ \\ \hline
  $C^{(5)}_{i}$ &  \multirow{2}{*}{$(\bm{3^{\prime}},\bm{3})$} & $-4$, $-2$, $0$, $2$ & $(m^{(4)}_{e})^{T}$ & $\alpha$  \\ \cline{1-1} \cline{3-5}
  $\tilde{C}^{(5)}$ &  & $4$ & $(\tilde{m}^{(4)}_{e})^{T}$ & \multirow{2}{*}{$\alpha$, $\beta$, $\gamma$} \\ \cline{1-4}
  $C^{(6)}_{i}$ & $(\bm{3^{\prime}},\bm{3^{\prime}})$ & $-4$, $-2$, $0$, $2$ & $m^{(6)}_{e}$ in Eq.~\eqref{eq:ch_mass6} &  \\ \hline \hline
  \texttt{Cases} & $(\rho_{L},\rho_{E^{c}_{d}},\rho_{E^{c}_{3}})$  & $(k_{L}+k_{E^{c}_{d}},k_{L}+k_{E^{c}_{3}})$ & $m_{e}$ & inputs \\  \hline
  $C^{(7)}_{i}$ & \multirow{3}{*}{$(\bm{3},\bm{\widehat {2}},\bm{1})$} & $(3,-4)$, $(3,-2)$, $(3,0)$, $(3,2)$, $(3,4)$   & $m^{(7)}_{e}$ in Eq.~\eqref{eq:ch_mass7}  & $\alpha$, $\beta$ \\ \cline{1-1} \cline{3-5}
  $\tilde{C}^{(7)}_{i}$ &  & $(5,-4)$, $(5,-2)$, $(5,0)$, $(5,2)$, $(5,4)$ & $\tilde{m}^{(7)}_{e}$ in Eq.~\eqref{eq:ch_mass7}  & \multirow{4}{*}{$\alpha$, $\beta$, $\gamma$} \\ \cline{1-1} \cline{3-4}
   $\widehat {C}^{(7)}$ &  & $(3,6)$ & $\widehat {m}^{(7)}_{e}$ in Eq.~\eqref{eq:ch_mass7} &   \\ \cline{1-4}
  $C^{(8)}_{i,j}$ & $(\bm{3},\bm{\widehat {2}^{\prime}},\bm{1})$ & \multirow{2}{*}{$\left(\in\{-5,-3,-1,1,3\},\in\{-4,-2,0,2,4\}\right)$} & $m^{(8)}_{e}$ in Eq.~\eqref{eq:ch_mass8} &  \\ \cline{1-2} \cline{4-4}
  $C^{(9)}_{i,j}$ & $(\bm{3^{\prime}},\bm{\widehat {2}},\bm{1})$ & & $m^{(9)}_{e}$ in Eq.~\eqref{eq:ch_mass9} &  \\ \hline
   $C^{(10)}_{i}$ & \multirow{3}{*}{$(\bm{3^{\prime}},\bm{\widehat {2}^{\prime}},\bm{1})$} & $(3,-4)$, $(3,-2)$, $(3,0)$, $(3,2)$, $(3,4)$ & $m^{(10)}_{e}$ in Eq.~\eqref{eq:ch_mass10} & $\alpha$, $\beta$\\ \cline{1-1} \cline{3-5}
  $\tilde{C}^{(10)}_{i}$ &  & $(5,-4)$, $(5,-2)$, $(5,0)$, $(5,2)$, $(5,4)$ & $\tilde{m}^{(10)}_{e}$ in Eq.~\eqref{eq:ch_mass10} & \multirow{2}{*}{$\alpha$, $\beta$, $\gamma$} \\ \cline{1-1} \cline{3-4}
   $\widehat {C}^{(10)}$ &  & $(3,6)$ & $\widehat {m}^{(10)}_{e}$ in Eq.~\eqref{eq:ch_mass10} &  \\ \hline \hline
 \end{tabular}
\caption{\label{tab:sum_ch}Possible assignments for $A^{\prime}_5$ representations and weights of lepton fields. This work extends the finite modular group $A^{\prime}_5$ with gCP symmetry which  enforce all model parameters $\alpha$, $\beta$ and $\gamma$ being real.
}
\end{center}
\end{table}

\subsection{\label{sec:Nu_masses}Neutrino sector}

This work assumes Majorana neutrinos, whose masses are generated via the Weinberg operator  or type-I seesaw mechanism with three right-handed neutrinos. Our analysis focuses on phenomenologically viable models characterized by minimal free parameters, leading to explicit neutrino Lagrangian containing solely two  independent terms.  The possible neutrino models for different assignments of lepton doublets $L$ and right-handed neutrinos $N^{c}$ are listed in table~\ref{tab:sum_nu}.

\begin{table}[t!]
\centering
\renewcommand{\tabcolsep}{2.0mm}
\renewcommand{\arraystretch}{1.1}
\begin{tabular}{|c|c|c|c|c|c|c|c|c|c|c|c|}  \hline\hline
\multicolumn{5}{|c|}{ Weinberg operator} \\ \hline
\texttt{Cases} & $\rho_{L}$  & $k_{L}$ & $m_{\nu}$ & parameters \\  \hline
$W_{i}$& $\bm{3}$ & \multirow{2}{*}{$-2$, $-1$, $0$, $1$} & $m_{\nu}$ in Eq.~\eqref{eq:WO_mnu} &  \multirow{2}{*}{$g_{1}$, $g_{2}$} \\ \cline{1-2} \cline{4-4}
$W^{\prime}_{i}$& $\bm{3^{\prime}}$ &  & $m^{\prime}_{\nu}$ in Eq.~\eqref{eq:WO_mnu} &  \\ \hline \hline
\multicolumn{5}{|c|}{Seesaw } \\ \hline
\texttt{Cases} & $(\rho_{L},\rho_{N^{c}})$  & $(k_{L}+k_{N^{c}},k_{N^{c}})$ & $(m_{D},m_{N})$ & parameters \\  \hline
$S_{i,j}$& $(\bm{3},\bm{3^{\prime}})$ & \multirow{2}{*}{$(\in\{-4,-2,0,2\},\in\{-2,-1,0,1\})$} & $(m_{D},m_{N})$ in Eq.~\eqref{eq:SS_mnu1}& \multirow{2}{*}{$\frac{g^{2}}{M_{1}}$, $\frac{M_{2}}{M_{1}}$}  \\ \cline{1-2} \cline{4-4}

$S^{\prime}_{i,j}$& $(\bm{3^{\prime}},\bm{3})$ & & $(m^{\prime}_{D},m^{\prime}_{N})$ in Eq.~\eqref{eq:SS_mnu2}&   \\ \hline \hline

\end{tabular}
\caption{\label{tab:sum_nu} Summary of the representation and modular weight assignments of the matter fields in the minimal phenomenologically viable neutrino mass models based on $A^{\prime}_{5}$ modular symmetry and gCP symmetry. We analyze Majorana neutrino mass mechanisms, including the Weinberg operator and type-I seesaw with a triplet of right-handed neutrino states. 
}
\end{table}

\subsubsection{Weinberg operator }

In the first case, light neutrino masses are generated by the effective Weinberg operator. When the left-handed lepton doublets $L$ transform as triplets under the $A^{\prime}_5$ flavor symmetry, specifically assigned to either the $\bm{3}$  or $\bm{3^{\prime}}$ representation, two distinct classes of neutrino models arise. This follows the group decomposition:
\begin{equation}
\bm{3}\otimes\bm{3}=\bm{1}\oplus\bm{3}_{A}\oplus\bm{5}_{S}, \qquad \bm{3^{\prime}}\otimes\bm{3^{\prime}}=\bm{1}\oplus\bm{3^{\prime}}_{A}\oplus\bm{5}_{S}\,.
\end{equation}
Consequently, nonzero neutrino masses require the inclusion of either a singlet polyharmonic Maa{\ss} form $Y_{\bm{1}}$ or a quintuplet Maa{\ss} form $Y_{\bm{5}}$. The minimal modular invariant Lagrangian for the two cases are:
\begin{eqnarray}\label{eq:weinberg}
\nonumber W_{i}~&:&-\mathcal{L}_{\nu}=\frac{g_{1}}{2\Lambda}\left(LLHHY^{(2k_{i})}_{\bm{1}}\right)_{\bm{1}}+\frac{g_{2}}{2\Lambda}\left(LLHHY^{(2k_{i})}_{\bm{5}}\right)_{\bm{1}}\,, ~~~\text{for}~~~L\sim \bm{3}\,, \\
W^{\prime}_{i}~&:&-\mathcal{L}_{\nu}=\frac{g_{1}}{2\Lambda}\left(LLHHY^{(2k_{i})}_{\bm{1}}\right)_{\bm{1}}+\frac{g_{2}}{2\Lambda}\left(LLHHY^{(2k_{i})}_{\bm{5}}\right)_{\bm{1}}\,, ~~~\text{for}~~~L\sim \bm{3^{\prime}}\,,
\end{eqnarray}
$k_{i}=k_{L}=-2,-1,0,1$ for  $i=1,2,3,4$, respectively. Applying the $A^{\prime}_{5}$ tensor product decomposition, the corresponding light neutrino mass matrices are derived as:
\begin{eqnarray}
\nonumber && \hskip-0.5in m_{\nu}(k_{L})=\frac{v^2}{2\Lambda}\begin{pmatrix}
	g_{1}Y^{(2k_{L})}_{\bm{1}}+2 g_{2}Y^{(2k_{L})}_{\bm{5},1}& -\sqrt{3} g_{2}Y^{(2k_{L})}_{\bm{5},5}& -\sqrt{3} g_{2}Y^{(2k_{L})}_{\bm{5},2}\\
	-\sqrt{3} g_{2}Y^{(2k_{L})}_{\bm{5},5}& \sqrt{6} g_{2}Y^{(2k_{L})}_{\bm{5},4}& g_{1}Y^{(2k_{L})}_{\bm{1}}-g_{2}Y^{(2k_{L})}_{\bm{5},1}\\
	-\sqrt{3} g_{2}Y^{(2k_{L})}_{\bm{5},2}& g_{1}Y^{(2k_{L})}_{\bm{1}}-g_{2}Y^{(2k_{L})}_{\bm{5},1}& \sqrt{6} g_{2}Y^{(2k_{L})}_{\bm{5},3}\\
\end{pmatrix}\,, \\
\label{eq:WO_mnu} &&\hskip-0.5in m^{\prime}_{\nu}(k_{L})=\frac{v^2}{2\Lambda}\begin{pmatrix}
	g_{1}Y^{(2k_{L})}_{\bm{1}}+2 g_{2}Y^{(2k_{L})}_{\bm{5},1} & -\sqrt{3} g_{2}Y^{(2k_{L})}_{\bm{5},4} & -\sqrt{3} g_{2}Y^{(2k_{L})}_{\bm{5},3} \\
	-\sqrt{3} g_{2}Y^{(2k_{L})}_{\bm{5},4} & \sqrt{6} g_{2}Y^{(2k_{L})}_{\bm{5},2} & g_{1}Y^{(2k_{L})}_{\bm{1}}-g_{2}Y^{(2k_{L})}_{\bm{5},1} \\
	-\sqrt{3} g_{2}Y^{(2k_{L})}_{\bm{5},3} & g_{1}Y^{(2k_{L})}_{\bm{1}}-g_{2}Y^{(2k_{L})}_{\bm{5},1} & \sqrt{6} g_{2}Y^{(2k_{L})}_{\bm{5},5} \\
\end{pmatrix}\,.
\end{eqnarray}
For the neutrino sector, the eight possible light neutrino mass matrices corresponding to distinct assignments of $k_{L}$ are directly extracted from Ref.~\cite{Li:2024svh}.

\subsubsection{Seesaw models }

The Lagrangian for neutrino masses require at least four terms when three right-handed neutrinos are assigned to either $A^{\prime}_5$ singlets or a singlet plus a doublet $\bm{1}\oplus\bm{\widehat{2}^{(\prime)}}$. This inherent complexity significantly diminishes the predictive power of such models. To address this limitation, we propose assigning both the three right-handed neutrinos $N^{c}=(N^{c}_{1},N^{c}_{2},N^{c}_{3})^{T}$ and the lepton doublets $L$ to $A^{\prime}_5$ triplet $\bm{3}$ or $\bm{3^{\prime}}$. Thus, the neutrino mass Lagrangian takes a concise form:
\begin{equation}
-\mathcal{L}_{\nu}=g\left(N^cLY^{(k_{L}+k_{N})}_{\bm{r_{1}}}H\right)_{\bm{1}}+\frac{M}{2}\left(N^cN^cY^{(2k_{N})}_{\bm{r_{2}}}\right)_{\bm{1}}\,,
\end{equation}
where $Y^{(k_{L}+k_{N})}_{\bm{r_{1}}}$ and $Y^{(2k_{N})}_{\bm{r_{2}}}$ denote Maa{\ss} form multiplets that preserve modular invariance. The specific forms of these modular forms are governed by the assigned weights $(k_{L},k_{N^{c}})$ and the assigned representations of $(L,N^{c})$. As shown in Ref.~\cite{Li:2024svh}, the assignments $(L,N^{c})\sim(\bm{3},\bm{3})$ and $(\bm{3^{\prime}},\bm{3^{\prime}})$ require more parameters (at least three for Dirac and two for Majorana mass matrices), making them less efficient. For better parameter economy, we focus on the hybrid assignments $(L,N^{c})\sim(\bm{3},\bm{3^{\prime}})$ and $(L,N^{c})\sim(\bm{3^{\prime}},\bm{3})$, which are simpler while preserving modular symmetry.

\begin{description}[labelindent=-0.8em, leftmargin=0.3em]
\item[~~(\romannumeral1)]{$L\sim\bm{3}$, $N^c\sim\bm{3^{\prime}}$ }

For modular invariance, we identify 16 valid $(k_{L},k_{N^c})$ pairs that maintain the seesaw Lagrangian structure:
\begin{equation}
 S_{i,j}~:~	-\mathcal{L}_{\nu} = g\left(Y_{\bm{5}}^{(k_{i})}N^{c}L H\right)_{\bm{1}}+\frac{M_{1}}{2} \left(Y_{\bm{1}}^{(k_{j})}N^{c}N^{c}\right)_{\bm{1}} +\frac{M_{2}}{2} \left(Y_{\bm{5}}^{(k_{j})}N^{c}N^{c}\right)_{\bm{1}}\,,
\end{equation}
where $k_{i}=k_{L}+k_{N^{c}}\in\{-4,-2,0,2\}$ and $k_{j}=2k_{N^{c}}\in\{-4,-2,0,2\}$ for $i,j=1,2,3,4$. The resulting Dirac and Majorana neutrino mass matrices are explicitly constructed as:
\begin{eqnarray}
\nonumber \hskip-0.13in m_D(k_{i})&=&\left(
	\begin{array}{ccc}
	\sqrt{3} gY^{(k_{i})}_{\bm{5},1} & g Y^{(k_{i})}_{\bm{5},5} & gY^{(k_{i})}_{\bm{5},2}\\
	gY^{(k_{i})}_{\bm{5},4}& -\sqrt{2} gY^{(k_{i})}_{\bm{5},3} & -\sqrt{2} gY^{(k_{i})}_{\bm{5},5} \\
	gY^{(k_{i})}_{\bm{5},3} & -\sqrt{2} gY^{(k_{i})}_{\bm{5},2} & -\sqrt{2} gY^{(k_{i})}_{\bm{5},4}\\
	\end{array}
	\right)v\,, \\
\label{eq:SS_mnu1} \hskip-0.13in m_{N}(k_{j})&=&\left(
	\begin{array}{ccc}
		M_{1}Y^{(k_{j})}_{\bm{1}}+2 M_{2}Y^{(k_{j})}_{\bm{5},1} & -\sqrt{3} M_{2}Y^{(k_{j})}_{\bm{5},4} & -\sqrt{3} M_{2}Y^{(k_{j})}_{\bm{5},3} \\
		-\sqrt{3} M_{2}Y^{(k_{j})}_{\bm{5},4} & \sqrt{6} M_{2}Y^{(k_{j})}_{\bm{5},2} & M_{1}Y^{(k_{j})}_{\bm{1}}-M_{2}Y^{(k_{j})}_{\bm{5},1} \\
		-\sqrt{3} M_{2}Y^{(k_{j})}_{\bm{5},3} & M_{1}Y^{(k_{j})}_{\bm{1}}-M_{2}Y^{(k_{j})}_{\bm{5},1} & \sqrt{6} M_{2}Y^{(k_{j})}_{\bm{5},5} \\
	\end{array}
	\right)\,,
\end{eqnarray}
where parameters $g$, $M_{1}$ and $M_{2}$ are real.

\item[~~(\romannumeral2)]{$L\sim\bm{3^{\prime}}$, $N^c\sim\bm{3}$ }

Similar to previous cases, modular invariance in the three term seesaw Lagrangian permits 16 possible combinations of modular weights $k_{N^c}$ and $k_{L}$, expressed through the Lagrangian:
\begin{equation}
 S^{\prime}_{i,j}~:~	-\mathcal{L}_{\nu} = g\left(Y_{\bm{5}}^{(k_{i})}N^{c}L H\right)_{\bm{1}}+\frac{M_{1}}{2} \left(Y_{\bm{1}}^{(k_{j})}N^{c}N^{c}\right)_{\bm{1}} +\frac{M_{2}}{2} \left(Y_{\bm{5}}^{(k_{j})}N^{c}N^{c}\right)_{\bm{1}}\,,
\end{equation}
where the modular weights satisfy  $k_{i}=k_{L}+k_{N^{c}}\in\{-4,-2,0,2\}$ and $k_{j}=2k_{N^{c}}\in\{-4,-2,0,2\}$, with indices $i,j=1,2,3,4$. The explicit forms of the Dirac and Majorana neutrino mass matrices are derived as:
\begin{eqnarray}
	\nonumber \hskip-0.13in m^{\prime}_D(k_{i})&=&\left(
	\begin{array}{ccc}
	\sqrt{3} gY^{(k_{i})}_{\bm{5},1} & gY^{(k_{i})}_{\bm{5},4} & gY^{(k_{i})}_{\bm{5},3}\\
	gY^{(k_{i})}_{\bm{5},5}& -\sqrt{2} gY^{(k_{i})}_{\bm{5},3} & -\sqrt{2} gY^{(k_{i})}_{\bm{5},2} \\
	gY^{(k_{i})}_{\bm{5},2} & -\sqrt{2} gY^{(k_{i})}_{\bm{5},5} & -\sqrt{2} gY^{(k_{i})}_{\bm{5},4}\\
	\end{array}
	\right)v\,, \\
\label{eq:SS_mnu2} \hskip-0.13in m^{\prime}_{N}(k_{j})&=&\left(
	\begin{array}{ccc}
		M_{1}Y^{(k_{j})}_{\bm{1}}+2 M_{2}Y^{(k_{j})}_{\bm{5},1}& -\sqrt{3} M_{2}Y^{(k_{j})}_{\bm{5},5}& -\sqrt{3} M_{2}Y^{(k_{j})}_{\bm{5},2}\\
		-\sqrt{3} M_{2}Y^{(k_{j})}_{\bm{5},5}& \sqrt{6} M_{2}Y^{(k_{j})}_{\bm{5},4}& M_{1}Y^{(k_{j})}_{\bm{1}}-M_{2}Y^{(k_{j})}_{\bm{5},1}\\
		-\sqrt{3} M_{2}Y^{(k_{j})}_{\bm{5},2}& M_{1}Y^{(k_{j})}_{\bm{1}}-M_{2}Y^{(k_{j})}_{\bm{5},1}& \sqrt{6} M_{2}Y^{(k_{j})}_{\bm{5},3}\\
	\end{array}
	\right)  \,.
\end{eqnarray}

\end{description}

\subsection{Numerical results for lepton masses and mixing in viable models}

\begin{table}[t!]
\centering
\begin{tabular}{|c|c|c||c|c|}\hline \hline
\multirow{2}{*}{Observables}  &  	\multicolumn{2}{c||}{NO}   &      \multicolumn{2}{c|}{IO}       \\ \cline{2-3} \cline{4-5}
	
& $\text{bf}\pm1\sigma$  & $3\sigma$ region & $\text{bf}\pm1\sigma$ & $3\sigma$ region   \\ \hline

&   &  &  &    \\[-0.150in]
	
$\sin^2\theta_{13}$ & $0.02215^{+0.00056}_{-0.00058}$ & $[0.02030,0.02388]$ & $0.02231^{+0.00056}_{-0.00056}$ &  $[0.02060,0.02409]$  \\ [0.050in]
	
$\sin^2\theta_{12}$ & $0.308^{+0.012}_{-0.011}$ & $[0.275,0.345]$ & $0.308^{+0.012}_{-0.011}$ & $[0.275,0.345]$  \\ [0.050in]
	
$\sin^2\theta_{23}$  & $0.470^{+0.017}_{-0.013}$  & $[0.435,0.585]$ & $0.550^{+0.012}_{-0.015}$  & $[0.440,0.584]$   \\ [0.050in]
	
$\delta_{CP}/\pi$  & $1.178^{+0.144}_{-0.228}$ & $[0.689,2.022]$  & $1.522^{+0.122}_{-0.139}$ & $[1.117,1.861]$  \\ [0.050in]

$\frac{\Delta m^2_{21}}{10^{-5}\text{eV}^2}$ & $7.49^{+0.19}_{-0.19}$ & $[6.92,8.05]$ & $7.49^{+0.19}_{-0.19}$ & $[6.92,8.05]$  \\ [0.050in]

$\frac{\Delta m^2_{3\ell}}{10^{-3}\text{eV}^2}$ & $2.513^{+0.021}_{-0.019}$ & $[2.451,2.578]$ & $-2.484^{+0.020}_{-0.020}$ & $[-2.547,-2.421]$ \\ [0.050in]

$\Delta m^2_{21}/\Delta m^2_{3\ell}$  &  $0.0298^{+0.00079}_{-0.00079}$  & $[0.0268,0.0328]$ & $-0.0302^{+0.00080}_{-0.00080}$ & $[-0.0333,-0.0272]$  \\ [0.050in]

$m_e/m_{\mu}$  & $0.004737$ & --- & $0.004737$ & --- \\ [0.050in]

$m_{\mu}/m_{\tau}$  & $0.05882$ & --- & $0.05882$ & --- \\ [0.050in]
 $m_{e}/\text{MeV}$ & $0.469652$  &  --- & $0.469652$  &  ---  \\ [0.050in] \hline \hline
	
\end{tabular}
\caption{\label{tab:bf_13sigma_data}
The best fit values, $1\sigma$ and $3\sigma$ ranges for the mixing parameters and lepton mass ratios are presented, where the experimental data and uncertainties for both the NO and IO neutrino mass spectrums are sourced from NuFIT 6.0 with Super-Kamiokande atmospheric data~\cite{Esteban:2024eli}. It is important to note that $\Delta m^2_{3\ell}=\Delta m^2_{31}>0$ for NO and $\Delta m^2_{3\ell}=\Delta m^2_{32}<0$ for IO. The $1\sigma$ uncertainties for the charged lepton mass ratios are considered to be $0.1\%$ of their central values in the $\chi^2$ analysis.}
\end{table}

As discussed in sections~\ref{sec:Ch_masses} and~\ref{sec:Nu_masses}, the mass matrices and associated symmetry transformations for charged leptons and neutrinos have been analyzed in detail. A key requirement in constructing lepton models is ensuring consistency between the representation assignments and modular weight distributions for the lepton doublets $L$ across both sectors. In this work, we focus on the scenario in which the lepton doublets $L$ are assigned to a triplet representation of $A^{\prime}_{5}$. Additionally, we consider all viable representation assignments for the three generators of right-handed charged leptons $E^{c}$. This leads to the identification of ten distinct representation assignments, labeled (\romannumeral1) $\sim$ (\romannumeral10) for the charged leptons. For modular forms with integer weights in the range $-5\leq k\leq6$, we construct 110 charged lepton mass matrices, each incorporating no more than three real parameters up to the complex modulus $\tau$. A complete summary is provided in table~\ref{tab:sum_ch}. Note that the 20 charged lepton matrices of the first two cases were previously analyzed in Ref.~\cite{Li:2024svh}.  Turning to the neutrino sector, we find 40 different mass matrices in total: 8 of them arise from the Weinberg operator formalism, and 32 of them are generated by the type-I seesaw mechanism. Unlike the charged lepton matrices, each light neutrino mass matrix depends on only two real parameters up to the modulus $\tau$, as detailed in table~\ref{tab:sum_nu}.

Building upon this classification,  we focus on analyzing the ``minimal'' models which are defined as those with the fewest input parameters. By combining all viable charged lepton structures from table~\ref{tab:sum_ch} with neutrino sector constructions from table~\ref{tab:sum_nu}, we find 2200 possible lepton models with fewer than seven real free parameters. However, we find that it proves challenging to reproduce the experimentally observed lepton masses and mixing parameters using only one or two Yukawa couplings in the charged lepton mass matrix. Thus, all viable models require seven independent real input parameters, as models with fewer parameters fail to satisfy experimental  constraints. Note that the 400 lepton models labeled $C^{(1)}-W_{j}$, $C^{(2)}-W^{\prime}_{j}$, $C^{(1)}-S_{j,k}$ and $C^{(2)}-S^{\prime}_{j,k}$ ($j,k=1,2,3,4$), which are invariant under the action of the finite modular group $A_5$, have been discussed in detail in Ref.~\cite{Li:2024svh}. We therefore obtained 1440 new minimal non-supersymmetric models under $A^{\prime}_{5}$ finite modular symmetry labeled as follows:
\begin{eqnarray}
\nonumber \hskip-0.28in \text{WO models} &:& C^{(3)}_{i}-W_{j},  \quad \tilde{C}^{(4)}-W_{j},  \quad \tilde{C}^{(5)}-W^{\prime}_{j}, \quad  C^{(6)}_{i}-W^{\prime}_{j}, \quad  \tilde{C}^{(7)}_{m}-W_{j},  \\
\nonumber \hskip-0.28in && \widehat {C}^{(7)}-W_{j}, \quad C^{(8)}_{m,n}-W_{j}, \quad C^{(9)}_{m,n}-W^{\prime}_{j}, \quad  \tilde{C}^{(10)}_{m}-W^{\prime}_{j}, \quad \widehat {C}^{(10)}-W^{\prime}_{j}\,, \\
\nonumber  \hskip-0.18in \text{SS models} &:& C^{(3)}_{i}-S_{j,k}, \quad  \tilde{C}^{(4)}-S_{j,k}, \quad  \tilde{C}^{(5)}-S^{\prime}_{j,k}, \quad  C^{(6)}_{i}-S^{\prime}_{j,k}, \quad  \tilde{C}^{(7)}_{m}-S_{j,k},  \\
\label{eq:models_7pars}\hskip-0.18in && \widehat {C}^{(7)}-S_{j,k}, \quad C^{(8)}_{m,n}-S_{j,k}, \quad  C^{(9)}_{m,n}-S^{\prime}_{j,k}, \quad  \tilde{C}^{(10)}_{m}-S^{\prime}_{j,k}, \quad \widehat {C}^{(10)}-S^{\prime}_{j,k}\,,
\end{eqnarray}
with the indices $i,j,k=1,2,3,4$ and $m,n=1,2,3,4,5$. Note that all models feature uniquely determined modular weight assignments (omitted here for brevity). Each model in Eq.~\eqref{eq:models_7pars} features a charged lepton mass matrix with three real couplings $\alpha$, $\beta$ and $\gamma$. For model with neutrino masses derived from the Weinberg operator, the mass matrices depend on a single independent parameter $g_{2}/g_{1}$ along with the overall scale $g_{1}v^2/\Lambda$ and the modulus $\tau$. In seesaw models, neutrino masses are governed by $M_{2}/M_{1}$ (plus the overall scale $g^2v^2/M_{1}$ and the modulus $\tau$).

\begin{figure}[t!]
\centering
\begin{tabular}{c}
\includegraphics[width=0.485\linewidth]{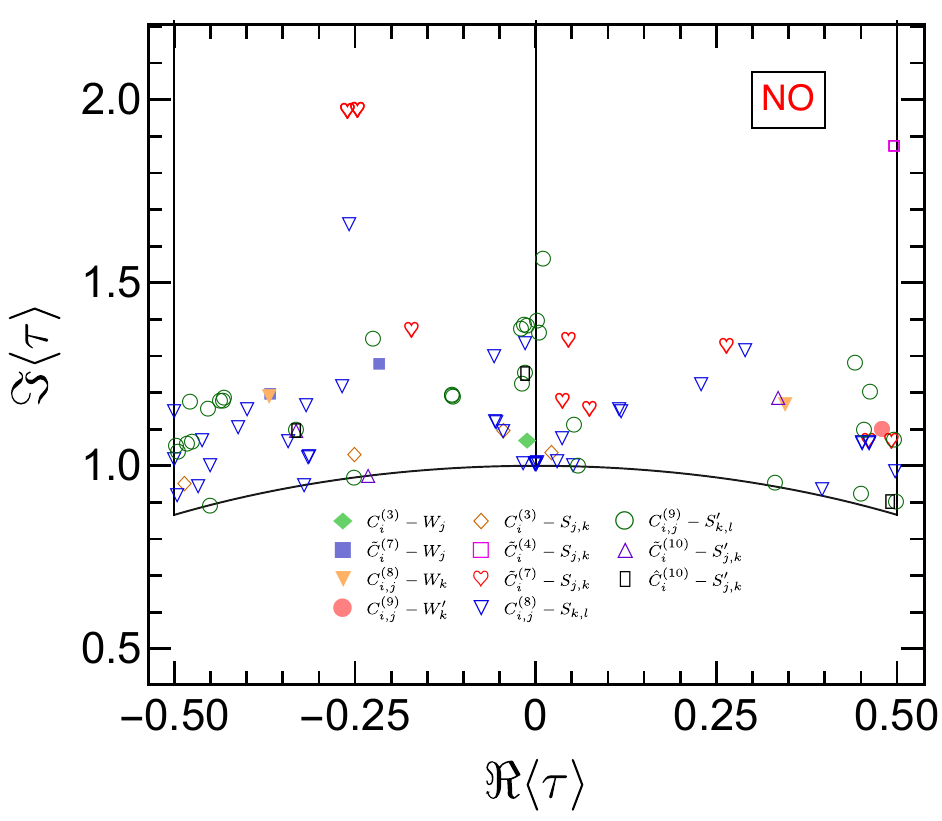}~~
\includegraphics[width=0.485\linewidth]{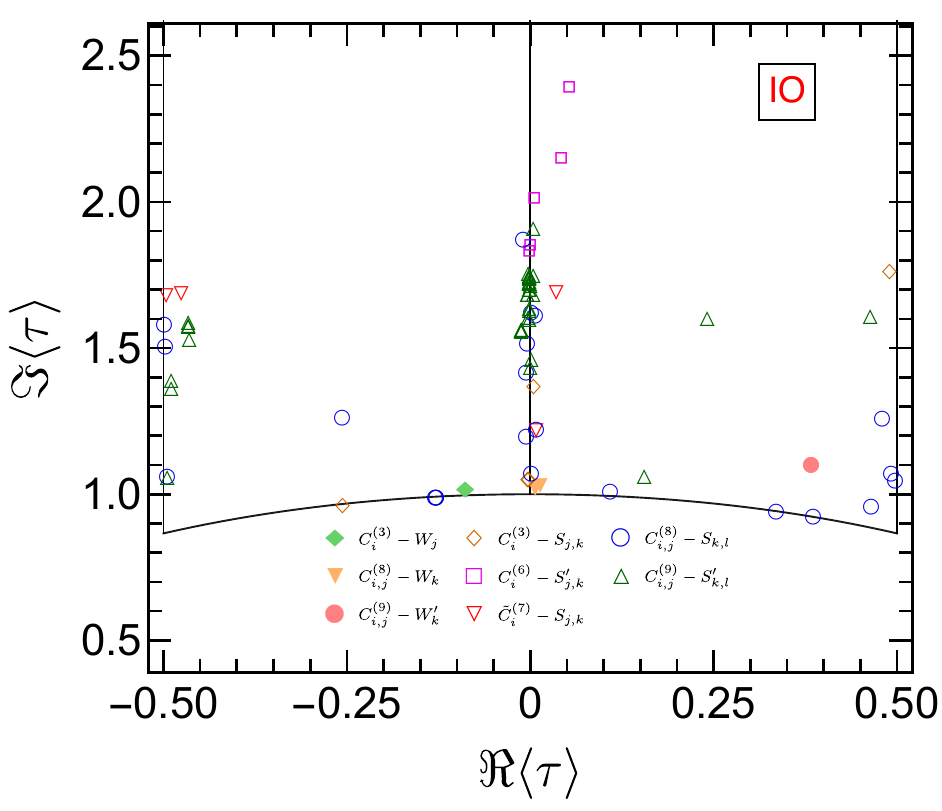}
\end{tabular}
\caption{\label{fig:bf_tau} The best fit values of $\tau$ for the 6 (4) viable Weinberg operator models and the 94 (76) seesaw models under the NO (IO) neutrino mass spectrum. }
\end{figure}

\begin{figure}[t!]
\centering
\begin{tabular}{c}
\includegraphics[width=0.92\linewidth]{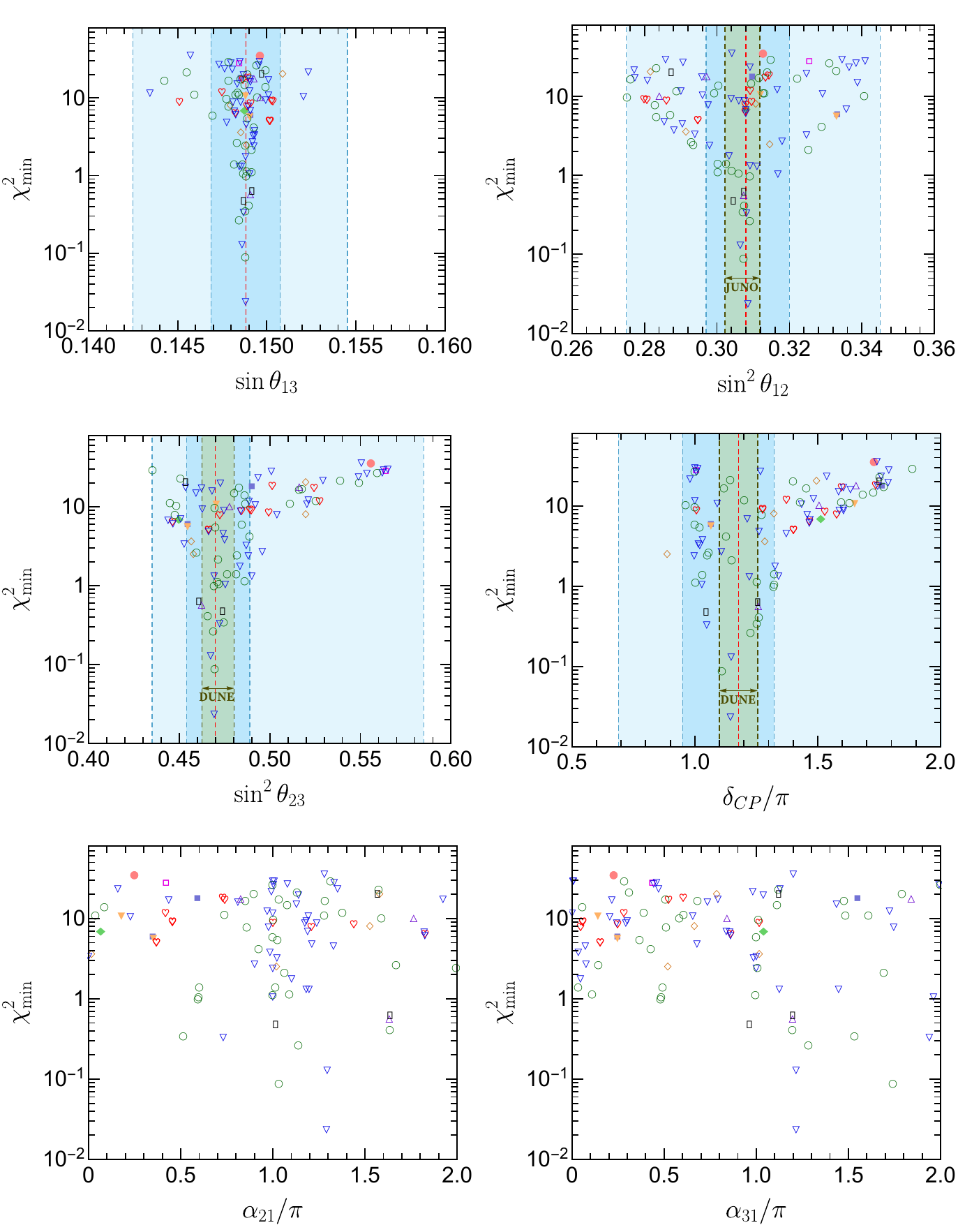}
\end{tabular}
\caption{\label{fig:bf_mixing_NO}  Best fit values of the three lepton mixing angles and tree CP violation phases at minimum $\chi^2$ for all 100 viable models with NO neutrino mass spectrum. The red dashed lines show best fit values, light blue bands mark $1\sigma$ and $3\sigma$ ranges from NuFIT 6.0 with Super-Kamiokande atmospheric data~\cite{Esteban:2024eli}. The lighter green band in the $\sin^{2}\theta_{12}$ panel shows the prospective $3\sigma$ range after 6 years of JUNO running~\cite{JUNO:2022mxj}, and those in the $\sin^{2}\theta_{23}$ and $\delta_{CP}$ panels display the resolution  after 15 years of DUNE running~\cite{DUNE:2020ypp}, calculated using NuFIT 6.0 best fit values. }
\end{figure}

\begin{figure}[t!]
\centering
\begin{tabular}{c}
\includegraphics[width=1.\linewidth]{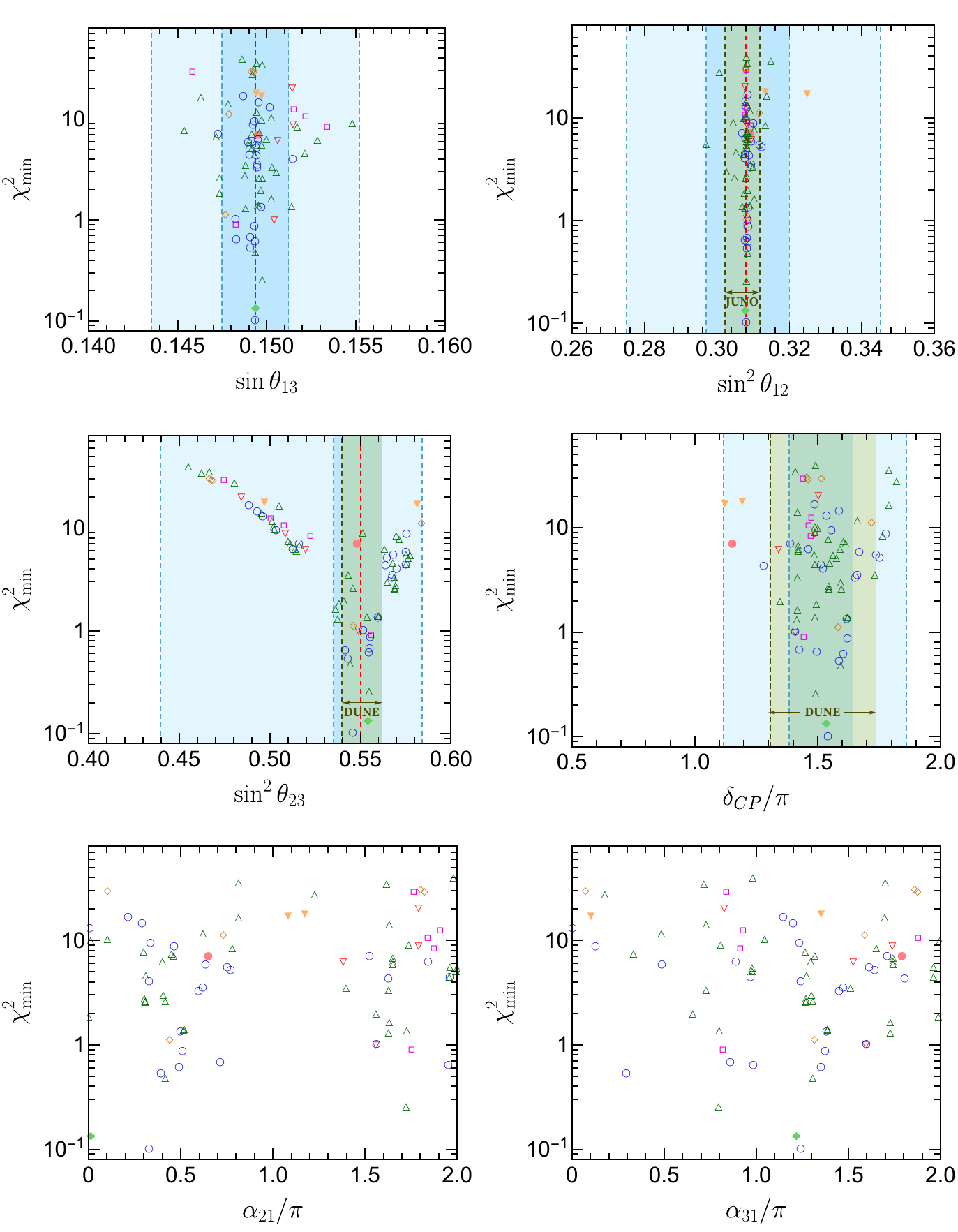}
\end{tabular}
\caption{\label{fig:bf_mixing_IO}  The the best fit $\chi^2$ for all 80 viable models with IO neutrino mass spectrum, covering the three lepton mixing angles and CP violation phases.  }
\end{figure}

This work conducts a systematic numerical investigation of models with gCP symmetry under both NO and IO neutrino mass hierarchies.  To quantitatively evaluate the compatibility of theoretical predictions with current experimental data on lepton masses and mixing parameters, we perform a conventional $\chi^2$ minimization procedure defined as:
\begin{equation}\label{eq:chisq}
  \chi^2 = \sum_{i=1}^7 \left( \frac{P_i-O_i}{\sigma_i}\right)^2\,,
\end{equation}
where $O_{i}$ and $\sigma_{i}$ denote the central experimental values and their  $1\sigma$ uncertainties for seven key dimensionless observables, please see table~\ref{tab:bf_13sigma_data}. It should be noted that the contribution from the Dirac phase $\delta_{CP}$ is incorporated into the $\chi^{2}$ function. The theoretical predictions $P_i$ derive from five dimensionless input parameters:
\begin{equation}\label{eq:five_inputs}
\Re{\tau},  \qquad \Im{\tau}, \qquad \beta /\alpha, \qquad \gamma /\alpha, \qquad
g_2/g_1 ~\left(\text{or}~M_{2}/M_{1}\right)\,,
\end{equation}
for Weinberg operator models (type-I seesaw models). Crucially, the overall scale of the charged lepton mass matrices, set by $\alpha v$, is determined by the electron mass, while the scale for the light neutrino mass matrices, given by $g_{1}v^2/\Lambda$ (or $g^2v^2/M_{1}$ for seesaw models), is constrained by the measured best fit value of the solar mass square difference  $\Delta m^{2}_{21}$.

\begin{figure}[t!]
\centering
\begin{tabular}{c}
\includegraphics[width=0.99\linewidth]{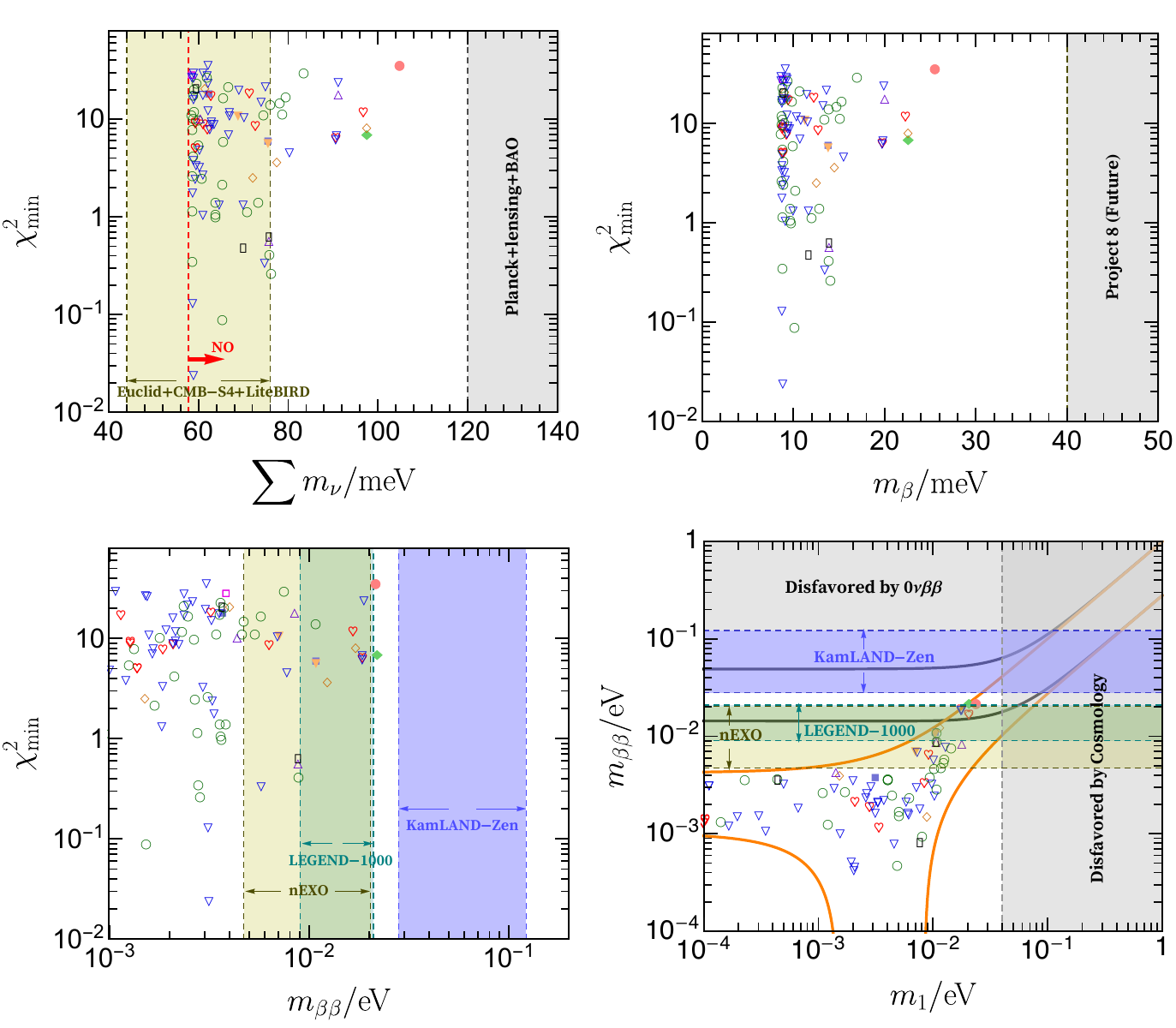}
\end{tabular}
\caption{\label{fig:bf_mass_NO}  The best fit values of the  minimum value of $\chi^2$, the effective mass $m_{\beta\beta}$ in $0\nu\beta\beta$-decay, the kinematical mass $m_{\beta}$ in beta decay $m_{\beta}$ and total neutrino mass sum $\sum m_{\nu}$ for all 100 viable NO models. In neutrino mass panel, we show current limit $\sum m_{\nu}<120\,\text{meV}$ from the Planck $+$ lensing $+$ BAO~\cite{Planck:2018vyg}, future sensitivity $\sum m_{\nu}<(44-76)\,\text{meV}$ of Euclid+CMB-S4+LiteBIRD~\cite{Euclid:2024imf} and normal ordering threshold ($\sum m_{\nu}\geq 57.75\,\text{meV}$). In the panel of the kinematical mass $m_{\beta}$ in beta decay, the gray region represents Project 8 future bound ($m_{\beta}<0.04\,\text{meV}$)~\cite{Project8:2022wqh}. In the panel of the effective Majorana mass $m_{\beta\beta}$, we display the latest result $m_{\beta\beta}<(28-122)\,\text{meV}$ of KamLAND-Zen~\cite{KamLAND-Zen:2024eml}, and the next generation experiments sensitivity ranges $m_{\beta\beta}<(9-21)\,\text{meV}$ from LEGEND-1000~\cite{LEGEND:2021bnm} and $m_{\beta\beta}<(4.7-20.3)\,\text{meV}$ from nEXO~\cite{nEXO:2021ujk}. }
\end{figure}

\begin{figure}[t!]
\centering
\begin{tabular}{c}
\includegraphics[width=0.99\linewidth]{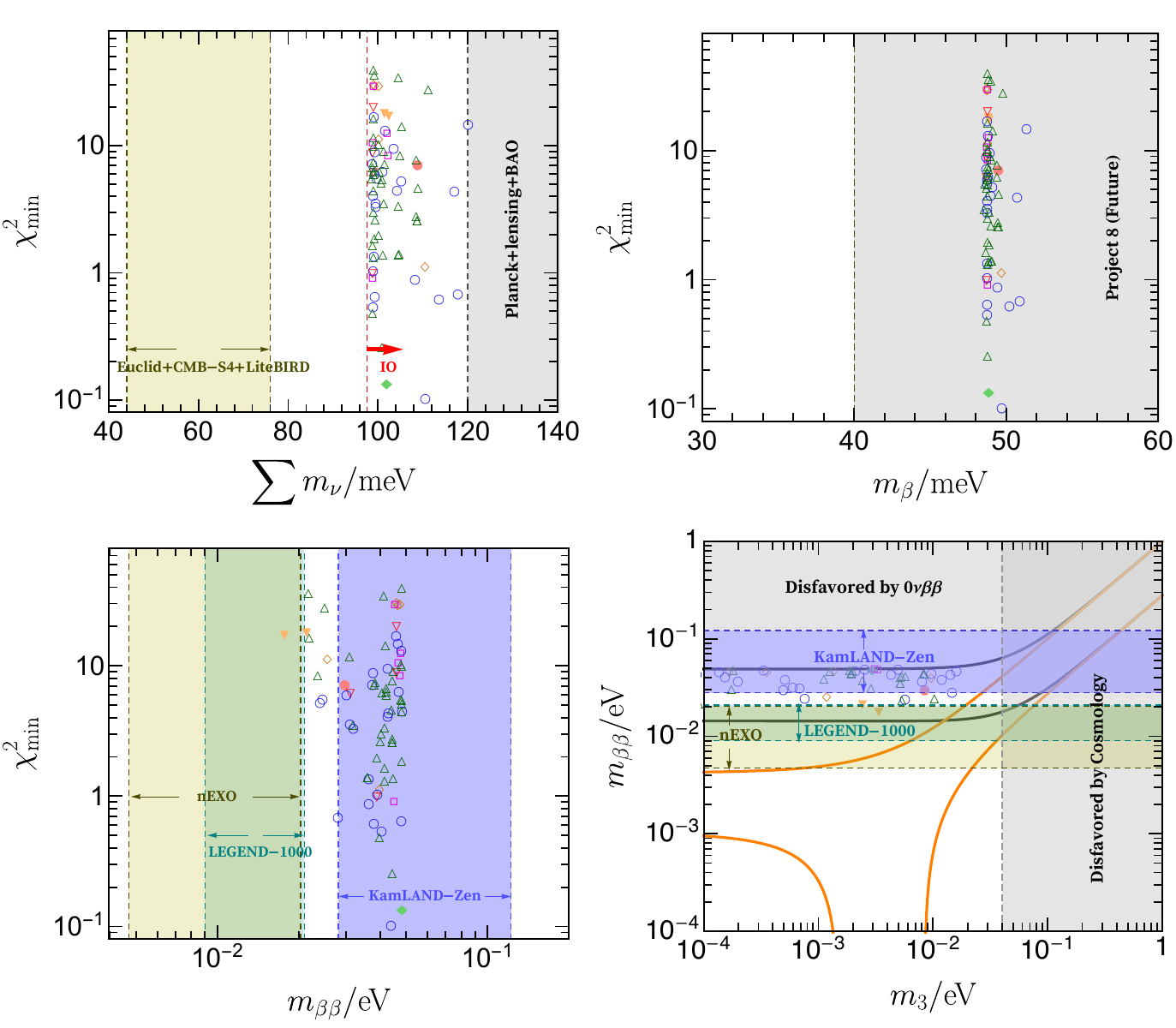}
\end{tabular}
\caption{\label{fig:bf_mass_IO}  The best fit values of the  minimum value of $\chi^2$, the effective mass in $0\nu\beta\beta$-decay  $m_{\beta\beta}$, the kinematical mass in beta decay $m_{\beta}$ and the three neutrino mass sum $\sum m_{\nu}$ for all 80 viable IO models. }
\end{figure}

To evaluate lepton flavor models, we test their predictions against experimental results. For each model in Eq.~\eqref{eq:models_7pars}, we perform a parameter scan and compute the minimum $\chi^2$ value. The absolute values of all dimensionless input parameters are uniformly sampled in the range $[0, 10^6]$, while the vacuum expectation value (VEV) of the modulus $\tau$ is restricted to the fundamental domain $\mathcal{D}=\left\{\tau\in\mathcal{H}\big||\Re(\tau)|\leq1/2, \Im(\tau)>0, |\tau|\geq1\right\}$. Our analysis selects models whose predictions satisfy the following constraints: the three lepton mixing angles ($\theta_{13}$, $\theta_{12}$, $\theta_{23}$), the Dirac CP phase $\delta_{CP}$, and the squared mass ratio $|\Delta m^{2}_{21}/\Delta m^{2}_{3l}|$ must lie within their experimentally allowed $3\sigma$ ranges given in table~\ref{tab:bf_13sigma_data}. Simultaneously, the total neutrino mass $\sum m_{\nu}$ must be below the Planck $+$ lensing $+$ BAO~ limit of 120 meV~\cite{Planck:2018vyg}. Furthermore, the charged lepton mass ratios $m_{e}/m_{\mu}$ and $m_{\mu}/m_{\tau}$ must align with experimental values within a maximum deviation of $0.3\%$. We comprehensively analyze 288 Weinberg operator models and 1152 seesaw models from Eq.~\eqref{eq:models_7pars} using $\chi^2$ fitting. For the NO (IO) mass spectrum, we find that 6 (4) viable Weinberg models and 94 (76) viable seesaw models with 7 real input parameters can accommodate the experimental data at $3\sigma$ level. Table~\ref{tab:WO_bf} provides detailed best fit values for all viable Weinberg models, including input parameters, mixing parameters, neutrino masses, the effective mass $m_{\beta\beta}$ in neutrinoless double beta decay ($0\nu\beta\beta$-decay) and the kinematical mass $m_{\beta}$  in beta decay. For the viable seesaw models, the best fit values of input parameters and derived physical quantities are summarized in  table~\ref{tab:SS_best_fit_NO} for NO case and table~\ref{tab:SS_best_fit_IO} for IO case.

In all models, the modulus $\tau$ is treated as a free parameter within the fundamental domain of $SL(2,\mathbb{Z})$ and adjusted to improve agreement with data. We displays the best fit values of the complex modulus $\tau$ for 100 viable models corresponding to the NO neutrino mass spectrum and 80 models for the IO case in figure~\ref{fig:bf_tau}. We find that the VEVs of $\tau$ in most viable models cluster near the regions $\Re\tau=0$ and $\Re\tau=\pm0.5$ for both mass orderings. For each model, the minimum value of $\chi^2$ can be determined with respect to key observables: the three lepton mixing angles, three CP violation phases, the sum of neutrino masses $\sum m_\nu$, the effective Majorana mass $m_{\beta\beta}$ in $0\nu\beta\beta$-decay and  the kinematical mass in beta decay $m_{\beta}$. The NO results are presented in figures~\ref{fig:bf_mixing_NO} and~\ref{fig:bf_mass_NO}, while the IO results are shown in figures~\ref{fig:bf_mixing_IO} and~\ref{fig:bf_mass_IO}. From figures~\ref{fig:bf_mixing_NO} and~\ref{fig:bf_mixing_IO}, we find that over half of the viable models predict mixing angles and the Dirac CP phase within experimentally allowed $1\sigma$ ranges~\cite{Esteban:2024eli}, demonstrating strong consistency with current data.  Future higher precision measurements of lepton mixing parameters and neutrino masses from neutrino oscillation experiments and cosmological surveys, combined with improved determinations of $m_{\beta\beta}$, could critically distinguish between these modular models. The JUNO experiment~\cite{JUNO:2022mxj} is expected to significantly refine measurements of the solar angle $\theta_{12}$. In figures~\ref{fig:bf_mixing_NO} and~\ref{fig:bf_mixing_IO}, we display the $3\sigma$ confidence interval for $\sin^{2}\theta_{12}$ after six years of JUNO data collection.  The improved precision for measuring $\theta_{23}$ and $\delta_{CP}$ at DUNE~\cite{DUNE:2020ypp} and T2HK~\cite{Hyper-Kamiokande:2018ofw} will further distinguish between models. The figures also show the projected constraints on $\theta_{23}$ and $\delta_{CP}$ after 15 years of DUNE running. These results, combined with future data from JUNO and T2HK, will enable the exclusion of a large fraction of those viable modular  models.

From figures~\ref{fig:bf_mass_NO} and~\ref{fig:bf_mass_IO}, we find that the projected neutrino mass sum $\sum m_{\nu}$ across majority of feasible models falls within the detectable range of upcoming cosmological surveys. Specifically, these predictions align with the anticipated sensitivity threshold $\sum m_{\nu}<(44-76)\,\text{meV}$ expected from the combined Euclid+CMB-S4+LiteBIRD~\cite{Euclid:2024imf}. For the NO  (IO) case, the results indicate  that all viable models predict values of $m_{\beta}$  below (above) the projected detection limit of  $0.04\text{eV}$ from Project 8~\cite{Project8:2022wqh}. The predictions for $m_{\beta\beta}$ in all viable NO models remain consistent with the latest KamLAND-Zen result of $m_{\beta\beta}<(28-122)\,\text{meV}$~\cite{KamLAND-Zen:2024eml}. However, this bound is anticipated to validate the majority of viable IO models.  The next generation detectors like LEGEND-1000 (targeting $m_{\beta\beta} < (9 \sim 21)\,$meV)~\cite{LEGEND:2021bnm} and nEXO (projected to reach $m_{\beta\beta} < (4.7 \sim 20.3)\,$meV)~\cite{nEXO:2021ujk} will possess sufficient sensitivity to scrutinize some NO models and all IO models predictions. The correlation between the lightest neutrino mass ($m_{1}$ or $m_{3}$) and $m_{\beta\beta}$ for each viable scenario is shown in the figures~\ref{fig:bf_mass_NO} and~\ref{fig:bf_mass_IO}, illustrating parameter dependencies across mass orderings.

%%%%%%%%%%%%%%%%%%%%%%%%%%%%%%%%%%%%%%%%%%%%%%%%%%%%%%%%
%\section{\label{sec:example_models}The predictions for lepton masses and mixing of example models}
\section{\label{sec:example_models}Two example models}
%%%%%%%%%%%%%%%%%%%%%%%%%%%%%%%%%%%%%%%%%%%%%%%%%%%%%%%%

\begin{figure}[t!]
\centering
\begin{tabular}{c}
\includegraphics[width=1\linewidth]{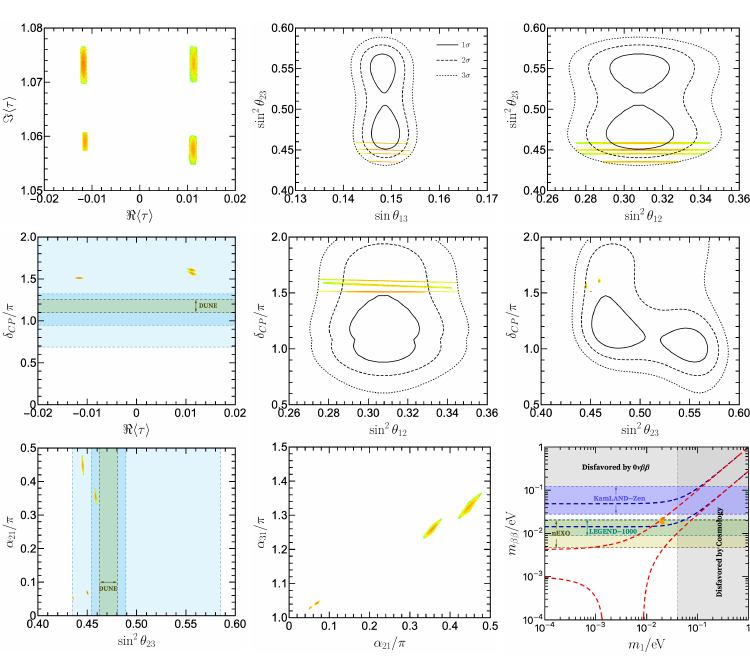}\\
\includegraphics[width=0.48\linewidth]{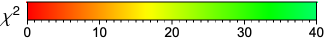}
\end{tabular}
\caption{\label{fig:corr_C33_W1_NO}
The correlations among the neutrino parameters predicted by the example model $C^{(3)}_{3}-W_{1}$ under the NO neutrino mass spectrum. The black lines in the mixing angles and Dirac CP panels show the $1\sigma$, $2\sigma$, and $3\sigma$ bounds (as solid, dashed, and dotted, respectively) for the NO neutrino mass spectrum. These bounds are derived from NuFIT 6.0 with Super-Kamiokande atmospheric data.
}
\end{figure}

\begin{figure}[t!]
\centering
\begin{tabular}{c}
\includegraphics[width=1\linewidth]{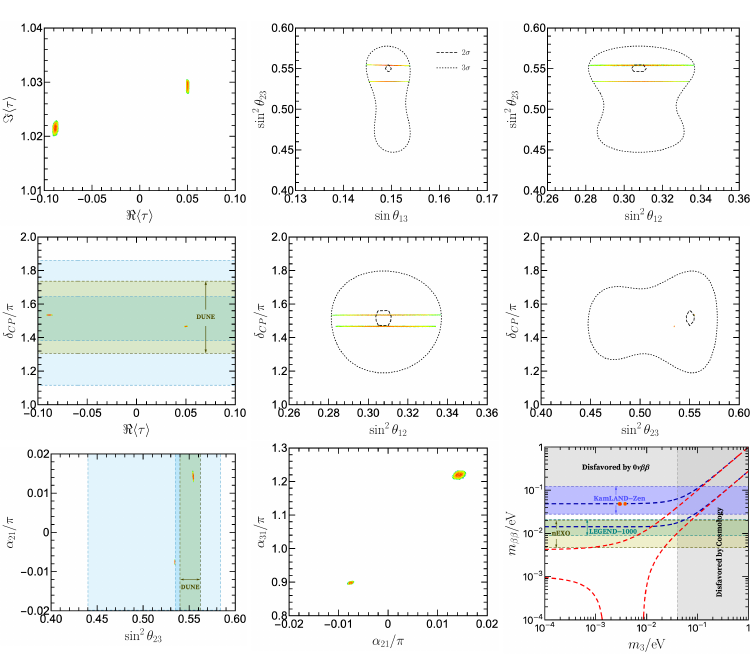}\\
\includegraphics[width=0.48\linewidth]{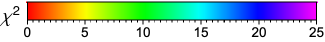}
\end{tabular}
\caption{\label{fig:corr_C33_W1_IO}
The correlations among the neutrino parameters, as established by the specific model $C^{(3)}_{3}-W_{1}$, are explicitly demonstrated within the IO neutrino mass spectrum. }
\end{figure}

Following the discussion above, we find 100 (80) minimal phenomenologically viable lepton flavor models for  NO (IO) neutrino mass spectrum by using the polyharmonic Maa{\ss} forms of level 5 and integer weights, associated with the finite modular group $\Gamma^{\prime}_5\cong A^{\prime}_5$. These non-holomorphic $A^{\prime}_5$ modular models are highly predictive, as they contain only 7 real input parameters which is fewer than the number of observables. This naturally correlates input variables with mixing angles and masses through constrained relationships. A detailed exploration of all viable cases and a full graphical analysis of all viable models are beyond our current scope.   Instead, we focus on two representative models (one Weinberg operator model and one seesaw model) that clearly demonstrate the quality of the results. These examples clearly show the limited parameter range and visible connections in non-holomorphic $A^{\prime}_5$ modular symmetric models, while allowing thorough analysis of their predictions.

For the first example model, we consider that the light neutrino masses arise from the effective Weinberg operator.  The modular weights and representation assignments for the lepton fields are specified as follows:
\begin{equation}
\rho_{L}= \rho_{E^{c}}=\bm{3}, \qquad k_{L}=-2, \qquad k_{E^{c}}=2\,,
\end{equation}
which correspond to the lepton model $C^{(3)}_{3}-W_{1}$ where the definition of $C^{(3)}_{3}$ and $W_{1}$ are given in table~\ref{tab:sum_ch} and table~\ref{tab:sum_nu}, respectively. The resulting mass matrices derived from Eqs.~\eqref{eq:ch_mass3} and~\eqref{eq:WO_mnu} take the form:
\begin{eqnarray}
\nonumber m_e&=&	\begin{pmatrix}
	\alpha  Y^{(0)}_{\bm{1}}+2 \gamma  Y^{(0)}_{\bm{5},1} & -\beta Y^{(0)}_{\bm{3},3}-\sqrt{3} \gamma  Y^{(0)}_{\bm{5},5} & \beta  Y^{(0)}_{\bm{3},2}-\sqrt{3} \gamma  Y^{(0)}_{\bm{5},2} \\
	\beta  Y^{(0)}_{\bm{3},3}-\sqrt{3} \gamma  Y^{(0)}_{\bm{5},5} & \sqrt{6} \gamma  Y^{(0)}_{\bm{5},4} & \alpha  Y^{(0)}_{\bm{1}}-\beta  Y^{(0)}_{\bm{3},1}-\gamma  Y^{(0)}_{\bm{5},1} \\
	-\beta  Y^{(0)}_{\bm{3},2}-\sqrt{3} \gamma  Y^{(0)}_{\bm{5},2} & \alpha  Y^{(0)}_{\bm{1}}+\beta  Y^{(0)}_{\bm{3},1}-\gamma  Y^{(0)}_{\bm{5},1} & \sqrt{6} \gamma  Y^{(0)}_{\bm{5},3} \\
		\end{pmatrix}v\,,\\
m_{\nu}&=&\frac{v^2}{2\Lambda}\begin{pmatrix}
	g_{1}Y^{(-4)}_{\bm{1}}+2 g_{2}Y^{(-4)}_{\bm{5},1}& -\sqrt{3} g_{2}Y^{(-4)}_{\bm{5},5}& -\sqrt{3} g_{2}Y^{(-4)}_{\bm{5},2}\\
	-\sqrt{3} g_{2}Y^{(-4)}_{\bm{5},5}& \sqrt{6} g_{2}Y^{(-4)}_{\bm{5},4}& g_{1}Y^{(-4)}_{\bm{1}}-g_{2}Y^{(-4)}_{\bm{5},1}\\
	-\sqrt{3} g_{2}Y^{(-4)}_{\bm{5},2}& g_{1}Y^{(-4)}_{\bm{1}}-g_{2}Y^{(-4)}_{\bm{5},1}& \sqrt{6} g_{2}Y^{(-4)}_{\bm{5},3}\\
\end{pmatrix}\,,
\end{eqnarray}
with explicit matrix elements determined by polyharmonic  Maa{\ss} forms of respective weights.  The numerical best fit values of input parameters and observable quantities for both mass orderings are presented in table~\ref{tab:WO_bf}. After scanning the parameter space of this model and requiring all observables to lie in their $3\sigma$ experimental ranges, we find several intriguing correlations between input parameters and physical observables for both mass orderings. For NO (IO) case, four (two) viable parameter regions emerge, with correlations shown in figure~\ref{fig:corr_C33_W1_NO} (figure~\ref{fig:corr_C33_W1_IO}). The imposed gCP symmetry restricts all couplings to be real, making $\Re \tau$ to be the only source of CP violation. This strongly constrains both CP phases and parameter distributions, as seen from the narrow bands in figures~\ref{fig:corr_C33_W1_NO} and \ref{fig:corr_C33_W1_IO}. Notably, reversing $\Re(\tau)$ yield identical predictions except for a sign reversal in the CP phase. This explains the broken mirror symmetry in figures~\ref{fig:corr_C33_W1_NO} and \ref{fig:corr_C33_W1_IO}, resulting from the explicit inclusion of $\delta_{CP}$ in the $\chi^2$ minimization.

For NO neutrino mass spectrum, the atmospheric mixing parameter $\sin^2\theta_{23}$ falls within four narrow intervals $[0.435,0.436]\cup[0.445,0.446]\cup[0.449,0.451]\cup[0.458,0.460]$, while the Dirac CP violation phase $\delta_{CP}$ is predicted to lie in narrow intervals $[1.505\pi,1.509\pi]\cup[1.511,1,517]\cup[1.542\pi,1.626\pi]$. These theoretical predictions could be experimentally tested through upcoming projects including DUNE~\cite{DUNE:2020ypp} and T2HK~\cite{Hyper-Kamiokande:2018ofw}, and at the discussed ESS$\nu$SB experiment~\cite{Alekou:2022emd}. The two Majorana CP violation phases are predicted to be in narrow regions $\alpha_{21}\in[0.0489\pi,0.0560\pi]\cup[0.0622\pi,0.0754\pi]\cup[0.334\pi,0.379\pi]\cup[0.420\pi,0.478\pi]$ and $\alpha_{31}\in[1.026\pi,1.033\pi]\cup[1.036\pi,1.050\pi]\cup[1.232\pi,1.290\pi]\cup[1.291\pi,1.366\pi]$. Our analysis shows that the total neutrino mass $\sum m_{\nu}$ lies in $[93.4,\text{meV},99.3,\text{meV}]$, well below the current Planck $+$ lensing $+$ BAO upper limit of $120,\text{meV}$~\cite{Planck:2018vyg}. These predictions will be testable through the next generation experiments Euclid+CMB-S4+ LiteBIRD, which are projected to achieve sensitivities of $\sum m_{\nu}<(44-76)\,\text{meV}$~\cite{Euclid:2024imf}. The effective Majorana mass $m_{\beta\beta}$ is constrained to $[17.21\,\text{meV},22.50\,\text{meV}]$, a range potentially detectable by the KamLAND-Zen experiment~\cite{KamLAND-Zen:2024eml} and the next generation $0\nu\beta\beta$-decay experiments like LEGEND-1000~\cite{LEGEND:2021bnm} and nEXO~\cite{nEXO:2021ujk}.

For the IO neutrino mass spectrum, our analysis shows excellent agreement with experimental data, with $\chi^2_{\text{min}}=0.135$. Through extensive parameter scanning, we find that the predicted values of $\theta_{13}$ and $\theta_{12}$ fully cover their experimentally allowed $3\sigma$ ranges. The model generates precise predictions for other mixing parameters:
\begin{eqnarray}
\nonumber && \sin^2\theta_{23}\in[0.5341,0.5344]\cup[0.5536,0.5548], \qquad \delta_{CP}/\pi\in[1.465,1.471]\cup[1.533,1,538], \\
\nonumber && \alpha_{21}/\pi\in[-0.00823,-0.00694]\cup[0.0129,0.0157], \qquad  \alpha_{31}/\pi\in[0.893,0.904]\cup[1.206,1.233], \\
&& \sum m_{\nu}\in[100.7\,\text{meV},104.3\,\text{meV}], \qquad m_{\beta\beta}\in[47.66\,\text{meV},49.12\,\text{meV}]\,.
\end{eqnarray}
From figure~\ref{fig:corr_C33_W1_IO}, we find that there are strong correlations among the mixing angles and CP violation phases. The model's validity depends on future precision measurements of $\theta_{23}$ and $\delta_{CP}$, especially by the next generation experiments such as DUNE~\cite{DUNE:2020ypp} and T2HK~\cite{Hyper-Kamiokande:2018ofw}, which will provide essential tests for the predicted parameter interdependencies and numerical ranges.

\begin{figure}[t!]
\centering
\begin{tabular}{c}
\includegraphics[width=1\linewidth]{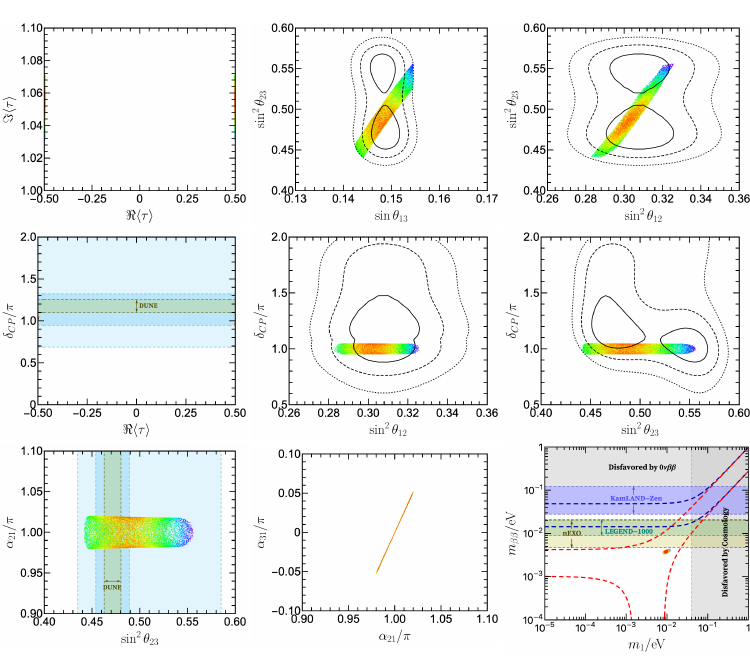}\\
\includegraphics[width=0.48\linewidth]{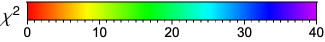}
\end{tabular}
\caption{\label{fig:corr_C951_Sp11_NO}
The correlations between the neutrino parameters predicted in the example model $C^{(9)}_{5,1}-S^{\prime}_{1,1}$ in the NO neutrino mass spectrum. }
\end{figure}

\begin{figure}[t!]
\centering
\begin{tabular}{c}
\includegraphics[width=1\linewidth]{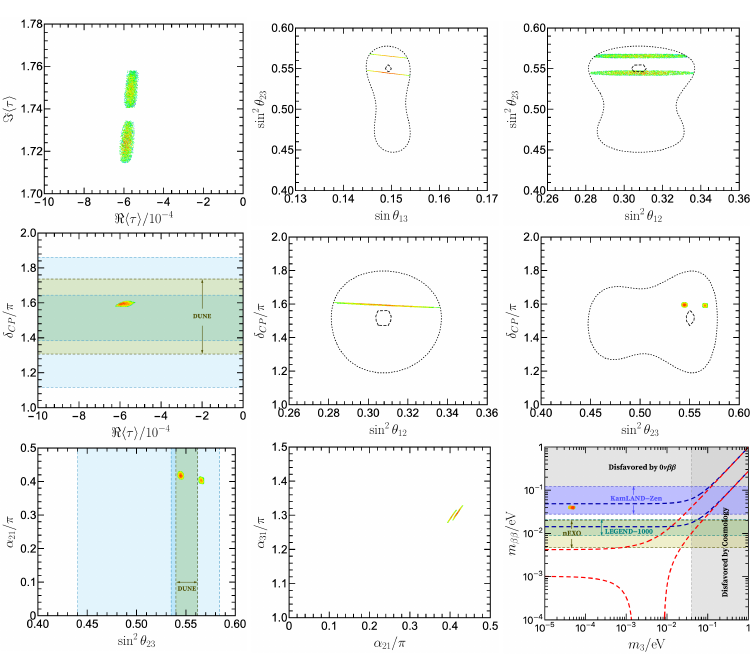}\\
\includegraphics[width=0.48\linewidth]{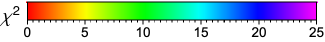}
\end{tabular}
\caption{\label{fig:corr_C951_Sp11_IO}
The correlations between the neutrino parameters predicted in the example model $C^{(9)}_{5,1}-S^{\prime}_{1,1}$ in the IO neutrino mass spectrum. }
\end{figure}

As a benchmark model for the case of light neutrino masses generated via the type-I seesaw mechanism, we assign the following $A^{\prime}_{5}$ representations and modular weights:
\begin{equation}
L\sim \bm{3^{\prime}}, \quad E^{c}_{d}\sim\bm{\widehat {2}},  \quad E^{c}_{3}\sim\bm{1}, \quad N^{c}\sim\bm{3}, \quad  k_{L}=k_{E^{c}_{3}}=k_{N}=-2,  \quad k_{E^{c}_{d}}=5\,.
\end{equation}
which corresponds to the combination $C^{(9)}_{5,1}-S^{\prime}_{1,1}$, where $C^{(9)}_{5,1}$ and $S^{\prime}_{1,1}$  are defined in table~\ref{tab:sum_ch} and table~\ref{tab:sum_nu}. Then the charged lepton mass matrix $m_{e}$, the Dirac neutrino mass matrix $m_D$, and the heavy Majorana neutrino mass matrix $m_{N}$ are derived as follows:
\begin{eqnarray}
\nonumber m_{e}&=&\left(
\begin{array}{ccc}
 \alpha  Y^{(3)}_{\bm{\widehat {6}}I,5}+\beta  Y^{(3)}_{\bm{\widehat {6}}II,5} & -\alpha  Y^{(3)}_{\bm{\widehat {6}}I,3}-\beta  Y^{(3)}_{\bm{\widehat {6}}II,3} & \alpha  Y^{(3)}_{\bm{\widehat {6}}I,2}+\beta  Y^{(3)}_{\bm{\widehat {6}}II,2} \\
 -\alpha  Y^{(3)}_{\bm{\widehat {6}}I,4}-\beta  Y^{(3)}_{\bm{\widehat {6}}II,4} & -\alpha  Y^{(3)}_{\bm{\widehat {6}}I,1}-\beta  Y^{(3)}_{\bm{\widehat {6}}II,1} & -\alpha  Y^{(3)}_{\bm{\widehat {6}}I,6}-\beta  Y^{(3)}_{\bm{\widehat {6}}II,6} \\
 \gamma  Y^{(-4)}_{\bm{3^{\prime}},1} & \gamma  Y^{(-4)}_{\bm{3^{\prime}},3} & \gamma  Y^{(-4)}_{\bm{3^{\prime}},2} \\
\end{array}
\right)v\,,  \\
\nonumber  m_D&=&g\left(
\begin{array}{ccc}
 \sqrt{3} Y^{(-4)}_{\bm{5},1} & Y^{(-4)}_{\bm{5},4} & Y^{(-4)}_{\bm{5},3} \\
 Y^{(-4)}_{\bm{5},5} & -\sqrt{2} Y^{(-4)}_{\bm{5},3} & -\sqrt{2} Y^{(-4)}_{\bm{5},2} \\
 Y^{(-4)}_{\bm{5},2} & -\sqrt{2} Y^{(-4)}_{\bm{5},5} & -\sqrt{2} Y^{(-4)}_{\bm{5},4} \\
\end{array}
\right)v\,, \\
 m_{N}&=&\left(
\begin{array}{ccc}
 M_{1} Y^{(-4)}_{\bm{1}}+2 M_{2} Y^{(-4)}_{\bm{5},1} & -\sqrt{3} M_{2} Y^{(-4)}_{\bm{5},5} & -\sqrt{3} M_{2} Y^{(-4)}_{\bm{5},2} \\
 -\sqrt{3} M_{2} Y^{(-4)}_{\bm{5},5} & \sqrt{6} M_{2} Y^{(-4)}_{\bm{5},4} & M_{1} Y^{(-4)}_{\bm{1}}-M_{2} Y^{(-4)}_{\bm{5},1} \\
 -\sqrt{3} M_{2} Y^{(-4)}_{\bm{5},2} & M_{1} Y^{(-4)}_{\bm{1}}-M_{2} Y^{(-4)}_{\bm{5},1} & \sqrt{6} M_{2} Y^{(-4)}_{\bm{5},3} \\
\end{array}
\right)  \,.
\end{eqnarray}
In order to show the viability and predictions of this model for both NO and IO cases, we scan over the parameter space of the model.  This allows us to evaluate the predictions for lepton masses and mixing parameters, retaining only points where flavor observables fall within their experimentally allowed $3\sigma$ regions. The best fit parameters and associated lepton flavor observables are given in tables~\ref{tab:SS_best_fit_NO} and \ref{tab:SS_best_fit_IO}.  We present correlations between observables and free parameters for the NO and IO cases in figures~\ref{fig:corr_C951_Sp11_NO} and~\ref{fig:corr_C951_Sp11_IO}, respectively. Note that the atmospheric mixing angle and the three CP violation phase parameters are constrained within narrow ranges for both mass hierarchies:
\begin{eqnarray}
\nonumber   \text{NO :} && \sin^2\theta_{12}\in[0.283,0.326], \qquad \sin^2\theta_{23} \in[0.442,0.556], \qquad  \delta_{CP}/\pi\in[0.952,1.048],     \\
\nonumber && \alpha_{21}/\pi\in[0.980,1.020], \qquad \alpha_{31}/\pi\in[-0.0522,0.0522],  \\
\nonumber &&  \sum m_{\nu}\in [70.36\,\text{meV},77.12\,\text{meV}], \qquad m_{\beta\beta}\in [3.563\,\text{meV},4.101\,\text{meV}]\,, \\
\nonumber \text{IO :} && \sin^2\theta_{12}\in[0.283,0.336], \qquad \sin^2\theta_{23}\in[0.542,0.548]\cup[0.563,0.568],   \\
\nonumber && \delta_{CP}/\pi\in[1.579,1.611], \qquad \alpha_{21}/\pi\in[0.393,0.431], \qquad  \alpha_{31}/\pi\in[1.278,1.329],  \\
&&  \sum m_{\nu}\in [97.69\,\text{meV},100.2\,\text{meV}], \qquad m_{\beta\beta}\in[39.13\,\text{meV},40.95\,\text{meV}]\,.
\end{eqnarray}
These precise constraints indicate that the model can be experimentally verified through upcoming long-baseline neutrino facilities such as DUNE~\cite{DUNE:2020ypp} and T2HK~\cite{Hyper-Kamiokande:2018ofw}, which possess enhanced sensitivity to measure CP violation effects and mixing parameters. Furthermore, the atmospheric angle $\sin^2\theta_{23}$ and the Dirac CP phase $\delta_{CP}$ are strongly correlated with the solar angle  $\sin^2\theta_{12}$. These correlations could be tested by JUNO experiment~\cite{JUNO:2022mxj} in combination with DUNE~\cite{DUNE:2020ypp} or T2HK~\cite{Hyper-Kamiokande:2018ofw}.  The neutrino mass parameters $\sum m_{\nu}$ and $m_{\beta\beta}$ are constrained across mass orderings. For the NO case, $\sum m_{\nu}$ lies within the sensitivity range of upcoming cosmological surveys such as Euclid+CMB-S4+LiteBIRD~\cite{Euclid:2024imf}, while $m_{\beta\beta}$ remains below the current KamLAND-Zen limit~\cite{KamLAND-Zen:2024eml} but may be reachable by next generation experiments like LEGEND-1000~\cite{LEGEND:2021bnm} and nEXO~\cite{nEXO:2021ujk}. In contrast, both parameters under the IO case fall entirely within projected detection ranges of future experiments, allowing decisive tests of this mass ordering.

\begin{table}[t!]
\begin{center}
\small
\renewcommand{\tabcolsep}{0.5mm}
\renewcommand{\arraystretch}{1.1}
\caption{\label{tab:WO_bf}Best fit parameters from $\chi^{2}$ minimization in the finite modular symmetry $A^{\prime}_{5}$ and gCP invariant Weinberg operator models for both neutrino mass orderings (NO/IO), including: mixing angles ($\theta_{12}$, $\theta_{13}$, $\theta_{23}$), CP violation phases ($\delta_{CP}$, $\alpha_{21}$, $\alpha_{31}$), neutrino masses ($m_{1,2,3}$), and measurable quantities ($m_{\beta\beta}$, $m_{\beta}$). Charged lepton ratios use global best fit values $m_{e}/m_{\mu}=0.004737$ and $m_{\mu}/m_{\tau}=0.05882$. Derived parameters $\sum m_{\nu}$ and $\Delta m^{2}_{21}/\Delta m^{2}_{31}$ are calculable from masses and thus omitted.
}
\begin{tabular}{|c|c|c|c|c|c|c|c|c|c|c|c|c|c|c|c|c|c|c|c|c|c|c|c|c|c|c|c|c|c|c|c|c|c|c|c|c|c|c|c|c|c|c|c|c|c|c|c|c|c|c|c|c|c|c|c|c|c|c|c|c|c|c|c|c|c|c|c|c|c|c|c|c|c|
c|c|c|c|c|c|c|c|c|c|c|c|c|c|c|}  \hline\hline
\multicolumn{89}{|c|}{NO neutrino mass spectrum} \\ \hline
\texttt{Models} & \multicolumn{11}{c|}{$\Re\langle \tau \rangle$} & \multicolumn{11}{c|}{$\Im\langle \tau \rangle$}  & \multicolumn{11}{c|}{$\beta/\alpha$} & \multicolumn{11}{c|}{$\gamma/\alpha$} & \multicolumn{11}{c|}{$g_{2}/g_{1}$} & \multicolumn{11}{c|}{$\alpha v$/MeV} & \multicolumn{11}{c|}{$(g_{1}v^2/\Lambda)/$meV} &  \multicolumn{11}{c|}{$\chi^2_{\mathrm{min}}$}\\ \hline
%\rowcolor{LightCyan}

$C^{(3)}_{3}-W_{1}$    & \multicolumn{11}{c|}{ $ -$0.0118} &  \multicolumn{11}{c|}{1.07} &  \multicolumn{11}{c|}{$-$0.606} &  \multicolumn{11}{c|}{$-$0.843} &  \multicolumn{11}{c|}{$-$1.51} &  \multicolumn{11}{c|}{594.8} &  \multicolumn{11}{c|}{44.3} &  \multicolumn{11}{c|}{6.96} \\ \hline

$\tilde{C}^{(7)}_{1}-W_{2}$  & \multicolumn{11}{c|}{  $-$0.216 } & \multicolumn{11}{c|}{  1.27 } & \multicolumn{11}{c|}{  0.999 } & \multicolumn{11}{c|}{  19.6 } & \multicolumn{11}{c|}{  0.497 } & \multicolumn{11}{c|}{  14.6 } & \multicolumn{11}{c|}{  42.9 } & \multicolumn{11}{c|}{  17.6 } \\ \hline
$\tilde{C}^{(7)}_{2}-W_{1}$   & \multicolumn{11}{c|}{  $-$0.367 } & \multicolumn{11}{c|}{  1.19 } & \multicolumn{11}{c|}{  $-$0.493 } & \multicolumn{11}{c|}{  1716. } & \multicolumn{11}{c|}{  1.63 } & \multicolumn{11}{c|}{  1.71 } & \multicolumn{11}{c|}{  24.8 } & \multicolumn{11}{c|}{  5.85 } \\ \hline

$C^{(8)}_{4,2}-W_{1}$  & \multicolumn{11}{c|}{  0.346 } & \multicolumn{11}{c|}{  1.17 } & \multicolumn{11}{c|}{  $-$3.30 } & \multicolumn{11}{c|}{  133.1 } & \multicolumn{11}{c|}{  0.998 } & \multicolumn{11}{c|}{  23.0 } & \multicolumn{11}{c|}{  37.6 } & \multicolumn{11}{c|}{  10.8 } \\
\hline
 \hline

 \end{tabular}

\end{center}
\end{table}

\newpage

\begin{center}
\renewcommand{\tabcolsep}{0.5mm}
\renewcommand{\arraystretch}{1.15}
\begin{small}
\setlength\LTcapwidth{\textwidth}
\setlength\LTleft{0.0in}
\setlength\LTright{0pt}
\begin{longtable}{|c|c|c|c|c|c|c|c|c|c|c|c|c|c|c|c|c|c|c|c|c|c|c|c|c|c|c|c|c|c|c|c|c|c|c|c|c|c|c|c|c|c|c|c|c|c|c|c|c|c|c|c|c|c|c|c|c|c|c|c|c|c|c|c|c|c|c|c|c|c|c|c|c|c|c|c|c|c|c|c|c|c|c|c|c|c|c|c|c|c|}
\caption{\label{tab:SS_best_fit_NO}
Best fit lepton mixing and mass parameters at $\chi^{2}$ minimum for viable NO seesaw models.
} \\
\midrule
\specialrule{0em}{1.0pt}{1.0pt}

\endfirsthead

\multicolumn{89}{c}
{{\bfseries \tablename\ \thetable{} -- continued from previous page}} \\
        \hline

\endhead

                \caption[]{continues on next page}\\
                        \endfoot

        \endlastfoot

\hline
\texttt{Models} & \multicolumn{11}{c|}{$\Re\langle \tau \rangle$} & \multicolumn{11}{c|}{$\Im\langle \tau \rangle$}  & \multicolumn{11}{c|}{$\beta/\alpha$} & \multicolumn{11}{c|}{$\gamma/\alpha$} & \multicolumn{11}{c|}{$M_{2}/M_{1}$}  &  \multicolumn{11}{c|}{ $\alpha v$/MeV} & \multicolumn{11}{c|}{$(g^2v^2/M_{1})/$meV} &\multicolumn{11}{c|}{$\chi^2_{\mathrm{min}}$} \\ \hline

%C3i-S_{j,k}
$C^{(3)}_{3}-S_{1,3}$ & \multicolumn{11}{c|}{ $-$0.250 } & \multicolumn{11}{c|}{ 1.04 } & \multicolumn{11}{c|}{ $-$0.734 } & \multicolumn{11}{c|}{ $-$1.04 } & \multicolumn{11}{c|}{ 0.932 } & \multicolumn{11}{c|}{ 528.7 } & \multicolumn{11}{c|}{ 186.7 } & \multicolumn{11}{c|}{ 2.55 } \\ \hline
 $C^{(3)}_{3}-S_{2,1}$ & \multicolumn{11}{c|}{ $-$0.0434 } & \multicolumn{11}{c|}{ 1.10 } & \multicolumn{11}{c|}{ $-$0.725 } & \multicolumn{11}{c|}{ $-$0.975 } & \multicolumn{11}{c|}{ 0.878 } & \multicolumn{11}{c|}{ 528.9 } & \multicolumn{11}{c|}{ 63.8 } & \multicolumn{11}{c|}{ 20.8 } \\ \hline
 $C^{(3)}_{3}-S_{3,2}$ & \multicolumn{11}{c|}{ $-$0.486 } & \multicolumn{11}{c|}{ 0.957 } & \multicolumn{11}{c|}{ $-$0.743 } & \multicolumn{11}{c|}{ $-$1.13 } & \multicolumn{11}{c|}{ $-$0.582 } & \multicolumn{11}{c|}{ 529.0 } & \multicolumn{11}{c|}{ 40.0 } & \multicolumn{11}{c|}{ 8.16 } \\ \hline
 $C^{(3)}_{3}-S_{3,3}$ & \multicolumn{11}{c|}{ 0.0218 } & \multicolumn{11}{c|}{ 1.04 } & \multicolumn{11}{c|}{ $-$0.766 } & \multicolumn{11}{c|}{ $-$0.866 } & \multicolumn{11}{c|}{ 0.907 } & \multicolumn{11}{c|}{ 528.4 } & \multicolumn{11}{c|}{ 39.7 } & \multicolumn{11}{c|}{ 3.68 } \\ \hline

 %C4pi-S_{j,k}
 $\tilde{C}^{(4)}_{1}-S_{4,2}$  & \multicolumn{11}{c|}{ 0.497 }  & \multicolumn{11}{c|}{ 1.87 }  & \multicolumn{11}{c|}{ $-$5.01 }  & \multicolumn{11}{c|}{ $-$1.31 }  & \multicolumn{11}{c|}{ 0.889 }  & \multicolumn{11}{c|}{ 223.7 }  & \multicolumn{11}{c|}{ 122.8 }  & \multicolumn{11}{c|}{ 27.4 } \\
\end{longtable}
 \end{small}
\end{center}

\newpage

\begin{center}
\renewcommand{\tabcolsep}{0.5mm}
\renewcommand{\arraystretch}{1.15}
\begin{small}
\setlength\LTcapwidth{\textwidth}
\setlength\LTleft{0.0in}
\setlength\LTright{0pt}
\begin{longtable}{|c|c|c|c|c|c|c|c|c|c|c|c|c|c|c|c|c|c|c|c|c|c|c|c|c|c|c|c|c|c|c|c|c|c|c|c|c|c|c|c|c|c|c|c|c|c|c|c|c|c|c|c|c|c|c|c|c|c|c|c|c|c|c|c|c|c|c|c|c|c|c|c|c|c|c|c|c|c|c|c|c|c|c|c|c|c|c|c|c|c|c|c|c|c|}
\caption{\label{tab:SS_best_fit_IO}
Best fit lepton mixing and mass parameters at $\chi^{2}$ minimum for viable IO seesaw models. } \\
\midrule
\specialrule{0em}{1.0pt}{1.0pt}

\endfirsthead

\multicolumn{89}{c}
{{\bfseries \tablename\ \thetable{} -- continued from previous page}} \\
        \hline

\endhead

                \caption[]{continues on next page}\\
                        \endfoot

        \endlastfoot

\hline
\texttt{Models} & \multicolumn{11}{c|}{$\Re\langle \tau \rangle$} & \multicolumn{11}{c|}{$\Im\langle \tau \rangle$}  & \multicolumn{11}{c|}{$\beta/\alpha$} & \multicolumn{11}{c|}{$\gamma/\alpha$} & \multicolumn{11}{c|}{$M_{2}/M_{1}$}  &  \multicolumn{11}{c|}{ $\alpha v$/MeV} & \multicolumn{11}{c|}{$(g^2v^2/M_{1})/$meV} &\multicolumn{11}{c|}{$\chi^2_{\mathrm{min}}$} \\ \hline

%C3i-S_{j,k}
$C^{(3)}_{1}-S_{3,3}$  & \multicolumn{11}{c|}{ 0.00532 } & \multicolumn{11}{c|}{ 1.37 } & \multicolumn{11}{c|}{ 1.39 } & \multicolumn{11}{c|}{ $-$0.534 } & \multicolumn{11}{c|}{ 0.828 } & \multicolumn{11}{c|}{ 568.0 } & \multicolumn{11}{c|}{ 65.8 } & \multicolumn{11}{c|}{ 1.14 } \\ \hline
$C^{(3)}_{2}-S_{3,3}$  & \multicolumn{11}{c|}{ 0.491 } & \multicolumn{11}{c|}{ 1.77 } & \multicolumn{11}{c|}{ $-$1.39 } & \multicolumn{11}{c|}{ $-$0.424 } & \multicolumn{11}{c|}{ 1.12 } & \multicolumn{11}{c|}{ 351.2 } & \multicolumn{11}{c|}{ 45.5 } & \multicolumn{11}{c|}{ 29.8 } \\ \hline
$C^{(3)}_{3}-S_{1,1}$  & \multicolumn{11}{c|}{ $-$0.00137 } & \multicolumn{11}{c|}{ 1.06 } & \multicolumn{11}{c|}{ $-$0.765 } & \multicolumn{11}{c|}{ $-$0.864 } & \multicolumn{11}{c|}{ $-$1.79 } & \multicolumn{11}{c|}{ 528.6 } & \multicolumn{11}{c|}{ 246.8 } & \multicolumn{11}{c|}{ 29.5 } \\ \hline
$C^{(3)}_{3}-S_{1,2}$  & \multicolumn{11}{c|}{ $-$0.256 } & \multicolumn{11}{c|}{ 0.967 } & \multicolumn{11}{c|}{ 0.774 } & \multicolumn{11}{c|}{ $-$0.912 } & \multicolumn{11}{c|}{ $-$0.0181 } & \multicolumn{11}{c|}{ 529.0 } & \multicolumn{11}{c|}{ 30.5 } & \multicolumn{11}{c|}{ 11.4 } \\ \hline
$C^{(3)}_{3}-S_{1,3}$  & \multicolumn{11}{c|}{ $-$0.00342 } & \multicolumn{11}{c|}{ 1.06 } & \multicolumn{11}{c|}{ $-$0.764 } & \multicolumn{11}{c|}{ $-$0.862 } & \multicolumn{11}{c|}{ $-$1.89 } & \multicolumn{11}{c|}{ 529.0 } & \multicolumn{11}{c|}{ 296.9 } & \multicolumn{11}{c|}{ 30.8 } \\
\end{longtable}
 \end{small}
\end{center}

\newpage

%%%%%%%%%%%%%%%%%%%%%%%%%%%%%%%%%%%%%%%%%%%%%%%%%%%%%%%%
\section{\label{sec:conclusion} Conclusion }
%%%%%%%%%%%%%%%%%%%%%%%%%%%%%%%%%%%%%%%%%%%%%%%%%%%%%%%%

In this work, we have investigated the application of modular symmetry in lepton flavor models, using polyharmonic Maa{\ss} forms at level $N=5$. In this framework, the Yukawa couplings and mass matrices are constructed from  polyharmonic Maa{\ss} forms of level $N=5$ and depend on a limited set of free parameters. Unlike holomorphic modular forms, polyharmonic Maa{\ss} forms exist at any integer (positive, negative, or zero) weight. They offer greater flexibility than the standard modular forms which are restricted to non-negative weights. This generalization opens a new way for model building and enhances our understanding of the origin of fermion mass hierarchies and flavor mixing.

We present a systematic analysis of lepton models based on $\Gamma^{\prime}_{5} \cong A^{\prime}_{5}$ modular group in combination with gCP symmetry in  the framework of non-holomorphic modular flavor symmetry~\cite{Qu:2024rns,Qu:2025ddz}, and all Yukawa couplings are constrained to be real. In this study, we have explored both scenarios that neutrino masses are generated by the Weinberg operator and the type-I seesaw mechanism. We have explicitly constructed all minimal models requiring only the modulus $\tau$ and no additional flavons, where the Yukawa couplings are derived from polyharmonic Maa{\ss} forms of level 5 with weights $k$ in the range $-5 \leq k \leq 6$. In these models, the left-handed lepton doublets $L$ are assigned to a triplet representation of the $A^{\prime}_{5}$ modular group, while the right-handed charged leptons $E^c$ are assigned to either a triplet or a combination of a singlet and a doublet. Notably, configurations with $E^{c}$ as a singlet $\bm{1}$ align with the baseline models in Ref.~\cite{Li:2024svh}, which established minimal lepton models using non-holomorphic $A_{5}$ modular symmetry. For the right-handed neutrinos $N^c$ in the seesaw mechanism, they transform as either a triplet $\bm{3}$ or $\bm{3^{\prime}}$ under $A^{\prime}_{5}$. This systematic classification yields 288 distinct minimal models for the Weinberg operator and produces another 1152 viable realizations in the seesaw mechanism. The modular weights and representation assignments for charged leptons and neutrinos, along with their respective mass matrices, are comprehensively summarized in tables~\ref{tab:sum_ch} and~\ref{tab:sum_nu}. We show that all the 1440 lepton models depend on five dimensionless parameters (as specified in Eq.~\eqref{eq:five_inputs}) and two overall scale factors.

Through a comprehensive numerical analysis of all 1440 models, we have identified phenomenologically viable candidates for NO and IO neutrino mass spectrums. For the Weinberg operator, we find 6 viable models for NO and 4 models for IO. In the seesaw framework, we obtain 94 viable models for NO and 76 models for IO. In tables~\ref{tab:WO_bf},  \ref{tab:SS_best_fit_NO} and \ref{tab:SS_best_fit_IO}, we present the best fit values of input parameters, lepton mixing angles, CP violation phases, neutrino masses, the effective Majorana mass of $0\nu\beta\beta$-decay and the beta decay kinematic mass for all phenomenologically viable models. Our analysis shows that nearly half of the viable models fit current data well, with all mixing parameters within the $1\sigma$ experimental bounds. Next generation neutrino facilities will transform our ability to discriminate between physics models. The future medium baseline reactor experiments like JUNO will substantially improve the precision of the $\sin^{2}\theta_{12}$ measurement, which will significantly reduce the number of viable models for the NO case.  Simultaneously, next generation long baseline detectors including DUNE and T2HK are projected to achieve unprecedented accuracy in determining atmospheric mixing parameters $\sin^{2}\theta_{23}$ and the Dirac CP phase $\delta_{CP}$. This high precision data will rule out a large fraction of currently viable models for both NO and IO cases, further constraining the theory space. The synergy analysis of data from JUNO, combined with measurements from DUNE or T2HK, will be highly effective in testing viable modular models, especially for those compatible with the NO case. Additional important tests of the models will be provided by precision measurements of the sum of the neutrino masses $\sum m_{\nu}$ and the effective Majorana mass $m_{\beta\beta}$. The complementary data from cosmological observations through the Euclid+CMB-S4+LiteBIRD collaborations will precisely measure the neutrino mass sum $\sum m_{\nu}$, while next generation $0\nu\beta\beta$-decay experiments like LEGEND-1000 and nEXO will probe the effective Majorana mass $m_{\beta\beta}$. Together, these efforts will severely constrain viable neutrino mass models. In particular, all viable models compatible with IO could be conclusively tested through either cosmological mass measurements or next generation $0\nu\beta\beta$-decay experiments, potentially resolving the current mass ordering ambiguity.

To demonstrate our findings more comprehensively, we have presented a very detailed description and statistical analyses of two representative viable benchmark models: the model $C^{(3)}_{3}-W_{1}$ where neutrino masses arise through the Weinberg operator, and the model $C^{(9)}_{5,1}-S^{\prime}_{1,1}$ in which the neutrino masses are generated by the type-I seesaw mechanism. Both models were numerically analyzed for NO and IO neutrino masses. We present predictions for key parameters including lepton flavor mixing angles, CP violation phases, the absolute neutrino masses, and the effective mass parameter relevant for $0\nu\beta\beta$-decay processes. The correlations between key parameters and observables, testable in upcoming experiments, are presented in figures ~\ref{fig:corr_C33_W1_NO}, \ref{fig:corr_C33_W1_IO}, \ref{fig:corr_C951_Sp11_NO} and \ref{fig:corr_C951_Sp11_IO}. The model $C^{(3)}_{3}-W_{1}$ can be critically tested by upcoming cosmological surveys Euclid+CMB-S4+LiteBIRD and neutrinoless double-beta decay experiments LEGEND-1000 and nEXO. For the $C^{(9)}_{5,1}-S^{\prime}_{1,1}$ model, if neutrino masses are NO,  the prediction for $\sum m_{\nu}$ within the sensitivity range of cosmological surveys, but $m_{\beta\beta}$ below the detection threshold of current experiments. In contrast, in the case of IO, both $\sum m_{\nu}$ and $m_{\beta\beta}$ are  within the projected sensitivity of next generation experiments, making it highly testable.

\section*{Acknowledgements}

CCL is supported by Natural Science Basic Research Program of Shaanxi (Program No. 2024JC-YBQN-0004), and the National Natural Science Foundation of China under Grant No. 12247103.  GJD is supported by the National Natural Science Foundation of China under Grant Nos.~12375104. We acknowledgement Bu-Yao Qu for providing the expressions of polyharmonic Maa{\ss} forms of level $N=5$.

\newpage

\section*{Appendix}

\begin{appendix}

\section{\label{sec:A5p_group_theory}Finite modular group $A^{\prime}_{5}$}

The finite modular group $\Gamma^{\prime}_{5} \cong A^{\prime}_5$ is the double cover of the icosahedral group $A_5$ and is defined by generators $S$ and $T$ satisfying the multiplication rules below~\cite{Liu:2019khw,Yao:2020zml}.
\begin{equation}
S^4=T^5=(ST)^3=1,\quad S^2T=TS^2\,.
\end{equation}
The group $A^{\prime}_5$ has 120 elements which is twice as many elements as $A_5$. It has nine irreducible representations: a single one-dimensional representation $\bm{1}$, two doublet representations $\bm{\widehat {2}}$ and $\bm{\widehat {2}^{\prime}}$, paired three-dimensional representations $\bm{3}$ and $\bm{3^{\prime}}$, two four-dimensional representations $\bm{4}$ and $\bm{\widehat {4}}$, along with unique five-dimensional and six-dimensional representations $\bm{5}$ and $\bm{\widehat {6}}$. Notably, the representations $\bm{1}$, $\bm{3}$, $\bm{3^{\prime}}$, $\bm{4}$ and $\bm{5}$ correspond to the five inequivalent irreducible representations of the $A_5$ group. The explicit matrix realizations of the generators $S$ and $T$ across these representations are systematically provided in table~\ref{tab:rep-matrices}.  Then one can straightforwardly calculate the multiplication rules of the irreducible representations as follows,
\begin{eqnarray}
\nonumber &&\hskip-0.2in \bm{\widehat {2}}\otimes \bm{\widehat {2}}=\bm{1_a}\oplus \bm{3_s}\,,\quad \bm{\widehat {2}}\otimes \bm{\widehat {2}^{\prime}}=\bm{4}\,,\quad \bm{\widehat {2}}\otimes \bm{3}=\bm{\widehat {2}}\oplus \bm{\widehat {4}}\,,\quad \bm{\widehat {2}}\otimes \bm{3^{\prime}}=\bm{\widehat {2}^{\prime}}\otimes \bm{3}=\bm{\widehat {6}}\,, \quad \bm{\widehat {2}}\otimes \bm{4}=\bm{\widehat {2}^{\prime}}\oplus \bm{\widehat {6}}\,,\\
\nonumber &&\hskip-0.2in \bm{\widehat {2}}\otimes \bm{\widehat {4}}=\bm{3}\oplus \bm{5},\quad \bm{\widehat {2}}\otimes \bm{5}=\bm{\widehat {2}^{\prime}}\otimes \bm{5}=\bm{\widehat {4}}\oplus \bm{\widehat {6}}, \quad \bm{\widehat {2}}\otimes \bm{\widehat {6}}=\bm{3}\otimes \bm{4}=\bm{3^{\prime}}\oplus \bm{4}\oplus \bm{5}, \quad \bm{\widehat {2}^{\prime}}\otimes \bm{\widehat {2}^{\prime}}=\bm{1_a}\oplus \bm{3^{\prime}_s}\,,\\
\nonumber &&\hskip-0.2in \bm{\widehat {2}^{\prime}}\otimes \bm{3^{\prime}}=\bm{\widehat {2}^{\prime}}\oplus \bm{\widehat {4}}\,, \quad
 \bm{\widehat {2}^{\prime}}\otimes \bm{4}=\bm{\widehat {2}}\oplus \bm{\widehat {6}}\,,\quad \bm{\widehat {2}^{\prime}}\otimes \bm{\widehat {4}}=\bm{3^{\prime}}\oplus \bm{5}\,,\quad \bm{\widehat {2}^{\prime}}\otimes \bm{\widehat {6}}=\bm{3^{\prime}}\otimes \bm{4}=\bm{3}\oplus \bm{4}\oplus \bm{5}\,,\\
\nonumber &&\hskip-0.2in \bm{3}\otimes \bm{3}=\bm{1_s}\oplus \bm{3_a}\oplus \bm{5_s}\,,\quad \bm{3}\otimes \bm{3^{\prime}}=\bm{4}\oplus \bm{5}\,,\quad \bm{3}\otimes \bm{\widehat {4}}=\bm{\widehat {2}}\oplus \bm{\widehat {4}}\oplus \bm{\widehat {6}}\,,\\
\nonumber &&\hskip-0.2in \bm{3}\otimes \bm{5}=\bm{3^{\prime}}\otimes \bm{5}=\bm{3}\oplus \bm{3^{\prime}}\oplus \bm{4}\oplus \bm{5}\,,\quad \bm{3}\otimes \bm{\widehat {6}}=\bm{\widehat {2}^{\prime}}\oplus \bm{\widehat {4}}\oplus \bm{\widehat {6}_{1}}\oplus \bm{\widehat {6}_{2}}\,,\\
\nonumber &&\hskip-0.2in \bm{3^{\prime}}\otimes \bm{3^{\prime}}=\bm{1_s}\oplus \bm{3^{\prime}_a}\oplus \bm{5_s}\,,\quad \bm{3^{\prime}}\otimes \bm{\widehat {4}}=\bm{\widehat {2}^{\prime}}\oplus \bm{\widehat {4}}\oplus \bm{\widehat {6}}\,,\quad \bm{3^{\prime}}\otimes \bm{\widehat {6}}=\bm{\widehat {2}}\oplus \bm{\widehat {4}}\oplus \bm{\widehat {6}_{1}}\oplus \bm{\widehat {6}_{2}}\,,\\
\nonumber &&\hskip-0.2in \bm{4}\otimes \bm{4}=\bm{1_s}\oplus \bm{3_a}\oplus \bm{3^{\prime}_a}\oplus \bm{4_s}\oplus \bm{5_s}\,,\quad \bm{4}\otimes \bm{\widehat {4}}=\bm{\widehat {4}}\oplus \bm{\widehat {6}_{1}}\oplus \bm{\widehat {6}_{2}}\,,\\
\nonumber &&\hskip-0.2in \bm{4}\otimes \bm{5}=\bm{3}\oplus \bm{3^{\prime}}\oplus \bm{4}\oplus \bm{5_1}\oplus \bm{5_2}\,,\quad \bm{4}\otimes \bm{\widehat {6}}=\bm{\widehat {2}}\oplus \bm{\widehat {2}^{\prime}}\oplus \bm{\widehat {4}_1}\oplus \bm{\widehat {4}_2}\oplus \bm{\widehat {6}_{1}}\oplus \bm{\widehat {6}_{2}}\,,\\
\nonumber &&\hskip-0.2in \bm{\widehat {4}}\otimes \bm{\widehat {4}}=\bm{1_a}\oplus \bm{3_s}\oplus \bm{3^{\prime}_s}\oplus \bm{4_s}\oplus \bm{5_a}\,,\quad \bm{\widehat {4}}\otimes \bm{5}=\bm{\widehat {2}}\oplus \bm{\widehat {2}^{\prime}}\oplus \bm{\widehat {4}}\oplus \bm{\widehat {6}_{1}}\oplus \bm{\widehat {6}_{2}}\,,\\
\nonumber &&\hskip-0.2in \bm{\widehat {4}}\otimes \bm{\widehat {6}}=\bm{3}\oplus \bm{3^{\prime}}\oplus \bm{4_1}\oplus \bm{4_2}\oplus \bm{5_1}\oplus \bm{5_2}\,,\quad \bm{5}\otimes \bm{5}=\bm{1_s}\oplus \bm{3_a}\oplus \bm{3^{\prime}_a}\oplus \bm{4_s}\oplus \bm{4_a}\oplus \bm{5_{1,s}}\oplus \bm{5_{2,s}}\,,\\
\nonumber &&\hskip-0.2in \bm{5}\otimes \bm{\widehat {6}}=\bm{\widehat {2}}\oplus \bm{\widehat {2}^{\prime}}\oplus \bm{\widehat {4}_1}\oplus \bm{\widehat {4}_2}\oplus \bm{\widehat {6}_{1}}\oplus \bm{\widehat {6}_{2}}\oplus \bm{\widehat {6}_3}\,,\\
\label{eq:kronecker-Gamma5p}&&\hskip-0.2in \bm{\widehat {6}}\otimes \bm{\widehat {6}}=\bm{1_a}\oplus \bm{3_{1,s}}\oplus \bm{3_{2,s}}\oplus \bm{3'_{1,s}}\oplus \bm{3'_{2,s}}\oplus \bm{4_s}\oplus \bm{4_a}\oplus \bm{5_{1,s}}\oplus \bm{5_{2,a}}\oplus \bm{5_{3,a}}\,.
\end{eqnarray}
In this notation, the indices $\bm{s}$ and $\bm{a}$ represent symmetric and anti-symmetric combinations respectively, and $\bm{4_1}$ and $\bm{4_2}$ stand for the two $\bm{4}$ representations that appear in the Kronecker products. This labeling convention extends systematically to other tensor product decompositions. In our working basis, all Clebsch-Gordan (CG) coefficients for the $A^{\prime}_5$ group are real and they have already been systematically calculated and tabulated in Ref.~\cite{Yao:2020zml}. Consequently, these coefficients are not reproduced in the present investigation.

\newpage
\begin{table}[t!]
\begin{center}
\renewcommand{\tabcolsep}{2.8mm}
\renewcommand{\arraystretch}{1.2}
\begin{tabular}{|c|c|c|c|c|c|}\hline\hline
  $\bm{r}$ & $\rho_{\bm{r}}(S)$ & $\rho_{\bm{r}}(T)$ \\ \hline
    $\bm{1}$ & 1 & 1  \\ \hline
	& &  \\[-0.16in]
    $\bm{\widehat {2}}$ & $i\sqrt{\frac{1}{\sqrt{5}\phi}}\begin{pmatrix}
 \phi  & 1 \\
 1 & -\phi  \\
\end{pmatrix}$ & $\begin{pmatrix}
 \omega _5^2 & 0 \\
 0 & \omega _5^3 \\
\end{pmatrix}$  \\ [0.18in]  \hline
	& & \\[-0.16in]
    $\bm{\widehat {2}^{\prime}}$ & $i\sqrt{\frac{1}{\sqrt{5}\phi}}\begin{pmatrix}
 1 & \phi  \\
 \phi  & -1 \\
\end{pmatrix}$ & $\begin{pmatrix}
 \omega _5 & 0 \\
 0 & \omega _5^4 \\
\end{pmatrix}$  \\[0.18in] \hline
	& & \\[-0.16in]
    $\bm{3}$ & $\frac{1}{\sqrt{5}}\begin{pmatrix}
 1 & -\sqrt{2} & -\sqrt{2} \\
 -\sqrt{2} & -\phi  & \frac{1}{\phi } \\
 -\sqrt{2} & \frac{1}{\phi } & -\phi  \\
\end{pmatrix}$ & $\begin{pmatrix}
 1 & 0 & 0 \\
 0 & \omega _5 & 0 \\
 0 & 0 & \omega _5^4 \\
\end{pmatrix}$ \\ [0.18in] \hline
	& & \\[-0.16in]
  $\bm{3^{\prime}}$ & $\frac{1}{\sqrt{5}}\begin{pmatrix}
 -1 & \sqrt{2} & \sqrt{2} \\
 \sqrt{2} & -\frac{1}{\phi } & \phi  \\
 \sqrt{2} & \phi  & -\frac{1}{\phi } \\
\end{pmatrix}$ & $\begin{pmatrix}
 1 & 0 & 0 \\
 0 & \omega _5^2 & 0 \\
 0 & 0 & \omega _5^3 \\
\end{pmatrix}$  \\ [0.18in] \hline
	& & \\[-0.16in]
    $\bm{4}$ & $\frac{1}{\sqrt{5}}\begin{pmatrix}
 1 & \frac{1}{\phi } & \phi  & -1 \\
 \frac{1}{\phi } & -1 & 1 & \phi  \\
 \phi  & 1 & -1 & \frac{1}{\phi } \\
 -1 & \phi  & \frac{1}{\phi } & 1 \\
\end{pmatrix}$ & $\begin{pmatrix}
 \omega _5 & 0 & 0 & 0 \\
 0 & \omega _5^2 & 0 & 0 \\
 0 & 0 & \omega _5^3 & 0 \\
 0 & 0 & 0 & \omega _5^4 \\
\end{pmatrix}$  \\ [0.18in] \hline
	& & \\[-0.16in]
    $\bm{\widehat {4}}$ & $i\sqrt{\frac{1}{5\sqrt{5}\phi}}\begin{pmatrix}
 -\phi ^2 & \sqrt{3} \phi  & -\sqrt{3} & -\frac{1}{\phi } \\
 \sqrt{3} \phi  & \frac{1}{\phi } & -\phi ^2 & -\sqrt{3} \\
 -\sqrt{3} & -\phi ^2 & -\frac{1}{\phi } & -\sqrt{3} \phi  \\
 -\frac{1}{\phi } & -\sqrt{3} & -\sqrt{3} \phi  & \phi ^2 \\
\end{pmatrix}$ & $\begin{pmatrix}
 \omega _5 & 0 & 0 & 0 \\
 0 & \omega _5^2 & 0 & 0 \\
 0 & 0 & \omega _5^3 & 0 \\
 0 & 0 & 0 & \omega _5^4 \\
\end{pmatrix}$  \\  [0.18in] \hline
	& & \\[-0.16in]
    $\bm{5}$& $\frac{1}{5}\begin{pmatrix}
 -1 & \sqrt{6} & \sqrt{6} & \sqrt{6} & \sqrt{6} \\
 \sqrt{6} & \frac{1}{\phi ^2} & -2 \phi  & \frac{2}{\phi } & \phi ^2 \\
 \sqrt{6} & -2 \phi  & \phi ^2 & \frac{1}{\phi ^2} & \frac{2}{\phi } \\
 \sqrt{6} & \frac{2}{\phi } & \frac{1}{\phi ^2} & \phi ^2 & -2 \phi  \\
 \sqrt{6} & \phi ^2 & \frac{2}{\phi } & -2 \phi  & \frac{1}{\phi ^2} \\
\end{pmatrix}$ & $\begin{pmatrix}
 1 & 0 & 0 & 0 & 0 \\
 0 & \omega _5 & 0 & 0 & 0 \\
 0 & 0 & \omega _5^2 & 0 & 0 \\
 0 & 0 & 0 & \omega _5^3 & 0 \\
 0 & 0 & 0 & 0 & \omega _5^4 \\
\end{pmatrix}$   \\ [0.18in] \hline
	& & \\[-0.16in]
    $\bm{\widehat {6}}$ & $i\sqrt{\frac{1}{5\sqrt{5}\phi}}\begin{pmatrix}
 -1 & \phi  & \frac{1}{\phi } & \sqrt{2} \phi  & \sqrt{2} & \phi ^2 \\
 \phi  & 1 & \phi ^2 & \sqrt{2} & -\sqrt{2} \phi  & -\frac{1}{\phi } \\
 \frac{1}{\phi } & \phi ^2 & 1 & -\sqrt{2} & \sqrt{2} \phi  & -\phi  \\
 \sqrt{2} \phi  & \sqrt{2} & -\sqrt{2} & -\phi  & -1 & \sqrt{2} \phi  \\
 \sqrt{2} & -\sqrt{2} \phi  & \sqrt{2} \phi  & -1 & \phi  & \sqrt{2} \\
 \phi ^2 & -\frac{1}{\phi } & -\phi  & \sqrt{2} \phi  & \sqrt{2} & -1 \\
\end{pmatrix}$ & $\begin{pmatrix}
 1 & 0 & 0 & 0 & 0 & 0 \\
 0 & 1 & 0 & 0 & 0 & 0 \\
 0 & 0 & \omega _5 & 0 & 0 & 0 \\
 0 & 0 & 0 & \omega _5^2 & 0 & 0 \\
 0 & 0 & 0 & 0 & \omega _5^3 & 0 \\
 0 & 0 & 0 & 0 & 0 & \omega _5^4 \\
\end{pmatrix}$  \\ [0.18in] \hline\hline
\end{tabular}
\caption{\label{tab:rep-matrices}The representation matrices of the generators $S$ and $T$ in different irreducible representations of $A^{\prime}_5$, where $\phi=\frac{1+\sqrt{5}}{2}$  is the golden ratio and $\omega_5$ is the quintic unit root $\omega_5=e^{2\pi i/5}$. }
\end{center}
\end{table}

\section{\label{sec:plo_mass_form}Polyharmonic Maa{\ss} form multiplets of level $5$}

In the present work,  non-holomorphic modular symmetry requires Yukawa couplings to be the so-called polyharmonic Maa{\ss} forms of level $5$, which combine holomorphic and non-holomorphic modular forms. These forms decompose into representations of finite modular groups: integer weight forms under $\Gamma^{\prime}_5\cong A^{\prime}_{5}$ (up to the automorphy factor)~\cite{Qu:2025ddz} and even weight forms under $\Gamma_5\cong A_{5}$~\cite{Qu:2024rns}. The tensor product of two level 5 modular form multiplets (weights $k$, $k^{\prime}$) is still a modular form multiplet (weight $k+k^{\prime}$). However, this does not hold for polyharmonic Maa{\ss} forms, as their product often loses harmonicity, particularly in the case of negative weights. Thus, higher weight holomorphic forms can be built from tensor products of weight 1 modular forms. Research shows that the polyharmonic Maa{\ss} forms  of weight 1 and level 5 organize into two irreducible six-dimensional representations $Y^{(1)}_{\bm{\widehat {6}}I}$ and $Y^{(1)}_{\bm{\widehat {6}}II}$ which  correspond  to the holomorphic and non-holomorphic modular form realizations of the $A^{\prime}_{5}$ group, respectively. The explicit construction of $Y^{(1)}_{\bm{\widehat {6}}I}$ , derived in Ref.~\cite{Yao:2020zml}, is expressed as:
\begin{equation}\label{eq:W1_polyharmonic form}
Y^{(1)}_{\bm{\widehat {6}}I}(\tau)= \begin{pmatrix} F_1^5 + 2 F_2^5  \\  2 F_1^5 - F_2^5 \\ 5 F_1^4 F_2 \\ 5\sqrt{2}F_1^3 F_2^2 \\-5\sqrt{2} F_1^2 F_2^3 \\ 5F_1 F_2^4 \end{pmatrix}\equiv \begin{pmatrix}
Y_1(\tau) \\
Y_2(\tau) \\
Y_3(\tau) \\
Y_4(\tau) \\
Y_5(\tau) \\
Y_6(\tau)
\end{pmatrix}\,,
\end{equation}
where the functions $F_1(\tau)$ and $F_2(\tau)$ are defined through theta constants and Dedekind eta functions:
\begin{equation}
F_1(\tau)= e^{-\frac{\pi i}{10}}\dfrac{\theta_{(\frac{1}{10},\frac{1}{2})}(5\tau)}{\eta(\tau)^{3/5}},~~~F_2(\tau)= e^{-\frac{3\pi i}{10}}\dfrac{\theta_{(\frac{3}{10},\frac{1}{2})}(5\tau)}{\eta(\tau)^{3/5}}\,,
\end{equation}
where $\theta_{(m^{\prime},m^{\prime\prime})}(\tau)$ represents the theta constant with characteristic indices $(m^{\prime},m^{\prime\prime})$, and $\eta(\tau)$ is the Dedekind eta function. These fundamental modular forms are universally expressed as:
\begin{equation}
\theta_{(m^{\prime},m^{\prime\prime})}(\tau)=\sum_{m\in \mathbb{Z}} e^{2\pi i [\frac{1}{2}(m+m^{\prime})^2\tau + (m+m^{\prime})m^{\prime\prime}]}\,, \quad  \eta(\tau)=q^{1/24}\prod_{n=1}^\infty \left(1-q^n \right),\qquad
        q\equiv e^{2 \pi i\tau}\,.
\end{equation}
Theta constants sum over integer lattices, while the Dedekind eta function manifests as an infinite product in terms of $q$ series. Then we can obtain that the $q$-expansion of the modular forms $Y_1(\tau),\cdots, Y_{6}(\tau)$ are given by
\begin{eqnarray}
\nonumber \hskip-0.3in Y_1(\tau) &=& 1+5 q+10 q^3-5 q^4+5 q^5+10 q^6+5 q^9 + \cdots , \\
\nonumber \hskip-0.3in Y_2(\tau) &=& 2+5 q+10 q^2+5 q^4+5 q^5+10 q^6+10 q^7-5 q^9+10 q^{10} + \cdots  , \\
\nonumber \hskip-0.3in Y_3(\tau) &=& 5 q^{1/5} (1+2 q+2 q^2+q^3+2 q^4+2 q^5+2 q^6+q^7+2 q^8+2 q^9 + \cdots  ), \\
\nonumber \hskip-0.3in Y_4(\tau) &=& 5\sqrt{2} q^{2/5} (1+q+q^2+q^3+2 q^4+q^6+q^7+2 q^8+q^9 + \cdots  ), \\
\nonumber \hskip-0.3in Y_5(\tau) &=& -5\sqrt{2} q^{3/5} (1+q^2+q^3+q^4-q^5+2 q^6+q^8+q^9+ \cdots  ), \\
\hskip-0.3in Y_6(\tau) &=& 5q^{4/5} (1-q+2 q^2+2 q^6-2 q^7+2 q^8+q^9 + \cdots  )\,.
\end{eqnarray}
The expressions for linearly independent higher weight modular form multiplets can be derived from the tensor products of lower weight modular multiplets. The independent modular multiplets for level $5$ up to weight 6 are:
\begin{equation}\label{eq:Yw2to6}
	\begin{array}{lll}
Y^{(2)}_{\bm{3}}
		=\left(Y^{(1)}_{\bm{\widehat {6}}I}Y^{(1)}_{\bm{\widehat {6}}I}\right)_{\bm{3_{1,s}}}\,, & ~
Y^{(2)}_{\bm{3^{\prime}}}
		=\left(Y^{(1)}_{\bm{\widehat {6}}I}Y^{(1)}_{\bm{\widehat {6}}I}\right)_{\bm{3^{\prime}_{1,s}}}\,, & ~
Y^{(2)}_{\bm{5}}
		=\left(Y^{(1)}_{\bm{\widehat {6}}I}Y^{(1)}_{\bm{\widehat {6}}I}\right)_{\bm{5_{1,s}}}\,, \\
Y^{(3)}_{\bm{\widehat {4}}}
		=\left(Y^{(1)}_{\bm{\widehat {6}}I}Y^{(2)}_{\bm{3}}\right)_{\bm{\widehat {4}}}\,, & ~
Y^{(3)}_{\bm{\widehat {6}}I}
		=\left(Y^{(1)}_{\bm{\widehat {6}}I}Y^{(2)}_{\bm{3}}\right)_{\bm{\widehat{6}_{1}}}\,, & ~
Y^{(3)}_{\bm{\widehat {6}}II}
		=\left(Y^{(1)}_{\bm{\widehat {6}}I}Y^{(2)}_{\bm{5}}\right)_{\bm{\widehat{6}_{1}}}\,, \\
Y^{(4)}_{\bm{1}}
		=\left(Y^{(2)}_{\bm{3}}Y^{(2)}_{\bm{3}}\right)_{\bm{1}} \,, & ~
Y^{(4)}_{\bm{3}}
		=\left(Y^{(2)}_{\bm{3}}Y^{(2)}_{\bm{5}}\right)_{\bm{3}} \,, & ~
Y^{(4)}_{\bm{3^{\prime}}}
		=\left(Y^{(2)}_{\bm{3}}Y^{(2)}_{\bm{5}}\right)_{\bm{3^{\prime}}}\,, \\
Y^{(4)}_{\bm{4}}
		=\left(Y^{(2)}_{\bm{3}}Y^{(2)}_{\bm{3^{\prime}}}\right)_{\bm{4}} \,, & ~
Y^{(4)}_{\bm{5}I}
		=\left(Y^{(2)}_{\bm{3}}Y^{(2)}_{\bm{3}}\right)_{\bm{5}} \,, & ~
Y^{(4)}_{\bm{5}II}
		=\left(Y^{(2)}_{\bm{3}}Y^{(2)}_{\bm{3^{\prime}}}\right)_{\bm{5}}\,, \\
Y^{(5)}_{\bm{\widehat {2}}}
		=\left(Y^{(2)}_{\bm{3}}Y^{(3)}_{\bm{\widehat {4}}}\right)_{\bm{\widehat {2}}}\,, & ~
Y^{(5)}_{\bm{\widehat {2}^{\prime}}}
		=\left(Y^{(2)}_{\bm{3}}Y^{(3)}_{\bm{\widehat {6}}I}\right)_{\bm{\widehat {2}^{\prime}}} \,,& ~
Y^{(5)}_{\bm{\widehat {4}}}
		=\left(Y^{(2)}_{\bm{3}}Y^{(3)}_{\bm{\widehat {4}}}\right)_{\bm{\widehat {4}}}  \,,\\
Y^{(5)}_{\bm{\widehat {6}}I}
		=\left(Y^{(2)}_{\bm{3}}Y^{(3)}_{\bm{\widehat {4}}}\right)_{\bm{\widehat {6}}}\,, & ~
Y^{(5)}_{\bm{\widehat {6}}II}
		=\left(Y^{(2)}_{\bm{3}}Y^{(3)}_{\bm{\widehat {6}}I}\right)_{\bm{\widehat {6}_{1}}} \,,& ~
Y^{(5)}_{\bm{\widehat {6}}III}
		=\left(Y^{(2)}_{\bm{3}}Y^{(3)}_{\bm{\widehat {6}}II}\right)_{\bm{\widehat {6}_{1}}}  \,,\\
Y^{(6)}_{\bm{1}}
		=\left(Y^{(3)}_{\bm{\widehat {6}}I}Y^{(3)}_{\bm{\widehat {6}}II}\right)_{\bm{1}} \,, & ~
Y^{(6)}_{\bm{3}I}
		=\left(Y^{(3)}_{\bm{\widehat{4}}}Y^{(3)}_{\bm{\widehat {6}}I}\right)_{\bm{3}} \,, & ~
Y^{(6)}_{\bm{3}II}
		=\left(Y^{(3)}_{\bm{\widehat {6}}I}Y^{(3)}_{\bm{\widehat {6}}I}\right)_{\bm{3_{1,s}}}\,, \\
Y^{(6)}_{\bm{3^{\prime}}I}
		=\left(Y^{(3)}_{\bm{\widehat {4}}}Y^{(3)}_{\bm{\widehat {4}}}\right)_{\bm{3^{\prime}}} \,, & ~
Y^{(6)}_{\bm{3^{\prime}}II}
		=\left(Y^{(3)}_{\bm{\widehat {6}}I}Y^{(3)}_{\bm{\widehat {6}}I}\right)_{\bm{3^{\prime}_{1}}} \,, & ~
Y^{(6)}_{\bm{4}I}
		=\left(Y^{(3)}_{\bm{\widehat {4}}}Y^{(3)}_{\bm{\widehat {4}}}\right)_{\bm{4}}\,, \\
Y^{(6)}_{\bm{4}II}
		=\left(Y^{(3)}_{\bm{\widehat {4}}}Y^{(3)}_{\bm{\widehat {6}}I}\right)_{\bm{4_{1}}} \,, & ~
Y^{(6)}_{\bm{5}I}
		=\left(Y^{(3)}_{\bm{\widehat {4}}}Y^{(3)}_{\bm{\widehat {6}}I}\right)_{\bm{5_{1}}} \,, & ~
Y^{(6)}_{\bm{5}II}
		=\left(Y^{(3)}_{\bm{\widehat {6}}I}Y^{(3)}_{\bm{\widehat {6}}I}\right)_{\bm{5_{1}}}\,.
	\end{array}
\end{equation}
Here, the explicit forms of these holomorphic modular multiplets can be constructed by using the CG coefficients of $A^{\prime}_{5}$, which have been provided in Ref.~\cite{Yao:2020zml}.

The weight 1 non-holomorphic modular sextet $Y^{(1)}_{\bm{\widehat {6}}II}$ was first derived in Ref.~\cite{Qu:2025ddz}, with its Fourier series expansion exhibiting the following component structure:
\begin{eqnarray}
\nonumber Y^{(1)}_{\bm{\widehat {6}}II,1} &=& a_1 + 2 \ln y - \dfrac{5 \Gamma(0,4\pi y)}{q} - \dfrac{10 \Gamma(0,8\pi y)}{q^2} - \dfrac{5 \Gamma(0,16\pi y)}{q^4} - \dfrac{5 \Gamma(0,20\pi y)}{q^5} + \cdots \\
\nonumber &&\hskip-0.4in - 5 \left( q \ln 5 + 2 q^2 \ln 5 + 2 q^3 \ln 3 + q^4 \ln 80 + 2 q^5 \ln 5 + 2 q^6 \ln \dfrac{15}{2} + \cdots  \right)  \,, \\
\nonumber Y^{(1)}_{\bm{\widehat {6}}II,2} &=& a_2- \ln y + \dfrac{5 \Gamma(0,4\pi y)}{q} + \dfrac{10 \Gamma(0,12\pi y)}{q^3} - \dfrac{5 \Gamma(0,16\pi y)}{q^4} + \dfrac{5\Gamma(0,20\pi y)}{q^5} + \cdots  \\
\nonumber &&\hskip-0.4in + 5 \left( q \ln 5 + 2 q^2 \ln 2  + 2 q^3 \ln 5 + q^4 \ln \dfrac{16}{5} + 2 q^5 \ln 5 + 2 q^6 \ln \dfrac{10}{3} + \cdots \right) \,, \\
\nonumber Y^{(1)}_{\bm{\widehat {6}}II,3} &=&- 5 q^{1/5} \left( \dfrac{\Gamma(0,16\pi y/5)}{q} - \dfrac{\Gamma(0,36\pi y/5)}{q^2} + \dfrac{2\Gamma(0,56\pi y/5)}{q^3} + \dfrac{2\Gamma(0,136\pi y/5)}{q^7} + \cdots \right)  \\
\nonumber &&\hskip-0.4in - 10 q^{1/5} \left( q \ln \dfrac{3}{2} + 2 q^3 \ln 2 - q^4 \ln \dfrac{7}{3} + q^5 \ln \dfrac{13}{2} + 2 q^7 \ln \dfrac{3}{2} + q^9 \ln \dfrac{23}{2} + \cdots \right) \,, \\
\nonumber Y^{(1)}_{\bm{\widehat {6}}II,4} &=& 5\sqrt{2}q^{2/5} \left( \dfrac{\Gamma(0,12\pi y/5)}{q} + \dfrac{\Gamma(0,52\pi y/5)}{q^3} + \dfrac{\Gamma(0,72\pi y/5)}{q^4} + \dfrac{\Gamma(0,92\pi y/5)}{q^5} + \cdots \right)  \\
\nonumber &&\hskip-0.4in +5\sqrt{2}q^{2/5} \left( \ln 2 + q \ln        7 + q^2 \ln \dfrac{16}{3} + q^3 \ln 17 + 2 q^4 \ln 2  + 4 q^5 \ln 81 + q^6 \ln 2 + \cdots  \right) \,,  \\
\nonumber Y^{(1)}_{\bm{\widehat {6}}II,5} &=& 5\sqrt{2} q^{3/5} \left( \dfrac{\Gamma(0,8\pi y/5)}{q} + \dfrac{\Gamma(0,28\pi y/5)}{q^2} + \dfrac{\Gamma(0,48\pi y/5)}{q^3} + \dfrac{\Gamma(0,68\pi y/5)}{q^4} + \cdots \right) \\
\nonumber &&\hskip-0.4in +5\sqrt{2} q^{3/5} \left( \ln 3 + 4 q \ln 2 + q^2 \ln 13 + q^3 \ln \dfrac{81}{2} + q^4 \ln 23 + q^5 \ln 112 + 2 q^6 \ln 3 + \cdots \right)  \,,  \\
\nonumber Y^{(1)}_{\bm{\widehat {6}}II,6} &=& 5 q^{4/5} \left( \dfrac{\Gamma(0,4\pi y/5)}{q} + \dfrac{2 \Gamma(0,24\pi y/5)}{q^2} + \dfrac{2\Gamma(0,44\pi y/5)}{q^3} + \dfrac{\Gamma(0,64\pi y/5)}{q^4} + \cdots \right) \\
&&\hskip-0.4in +10 q^{4/5} \left( 2 \ln 2 + 2 q \ln 3 + q^2 \ln 14 + q^3 \ln 19 + 4 q^4 \ln 2 + q^5 \ln 29 + q^6 \ln 34 + \cdots \right) \,,
\end{eqnarray}
where the constants $a_1$ and $a_2$ are
\begin{eqnarray}
\nonumber a_1 &=& 2 \gamma + 2\ln 20\pi - 5 \left[\ln \Gamma\left( \dfrac{1}{5} \right) + \ln \Gamma\left( \dfrac{2}{5} \right) - \ln \Gamma\left( \dfrac{3}{5} \right) - \ln \Gamma\left( \dfrac{4}{5} \right) \right] \simeq 0.5831\,, \\
a_2 &=& - \gamma - \ln 20 \pi + 5 \left[ \ln \Gamma\left( \dfrac{1}{5} \right) - \ln \Gamma\left( \dfrac{2}{5} \right) + \ln \Gamma\left( \dfrac{3}{5} \right) - \ln \Gamma\left( \dfrac{4}{5} \right) \right] \simeq 0.1501 \,.~~~~\quad~~
\end{eqnarray}
where $\gamma=0.577216$  is Euler’s constant, while the gamma function $\Gamma(z)$ and  incomplete gamma function $\Gamma(a,z)$ are respectively defined by the integrals:
\begin{eqnarray}\label{eq:gamma_func}
\Gamma(z)=\int_0^{\infty} t^{z-1}e^{-t}\,dt\,, \qquad \Gamma(a,z)=\int_z^{\infty} t^{a-1} e^{-t}\,dt\,,
\end{eqnarray}
which satisfies the following recurssion relation,
\begin{eqnarray}\label{eq:Gamma_recursion}
\Gamma(z+1)=z\Gamma(z), \qquad \Gamma(a+1,z)=a\Gamma(a,z) + z^a\,e^{-z}\,.
\end{eqnarray}
For weight $k=2$ and level $5$, the polyharmonic Maa{\ss} forms contain not only the modular multiplets from Eq.~\eqref{eq:Yw2to6}, but also the modified Eisenstein series $\widehat {E}_{2}(\tau)$, which transforms as an $A^{\prime}_{5}$ singlet with the $q$-expansion
\begin{equation}
 Y^{(2)}_{\bm{1}}=\widehat {E}_{2}(\tau)= 1 - \frac{3}{\pi y} - 24 q - 72 q^2 - 96 q^3 - 168 q^4 - 144 q^5 + \cdots \,.
\end{equation}
At weight $k=0$ and level $5$, there are four linearly independent modular multiplets of polyharmonic Maa{\ss} forms, corresponding to the representations $\bm{1}$, $\bm{3}$, $\bm{3^{\prime}}$ and $\bm{5}$: $Y^{(0)}_{\bm{1}}$,  $Y^{(0)}_{\bm{3}}$, $Y^{(0)}_{\bm{3^{\prime}}}$ and $Y^{(0)}_{\bm{5}}$  whose Fourie expansions are given by
 \begin{eqnarray}
\nonumber Y_{\bm{1}}^{(0)} &=& 1\,, \\
\nonumber Y_{\bm{3},1}^{(0)} &=& y + \frac{15 e^{-4 \pi y}}{2 \pi}\left( \frac{1}{ q}+ \frac{e^{-4 \pi
y}}{3 q^2}+ \frac{4\, e^{-8 \pi y}}{9 q^3}+ \frac{3\, e^{-12 \pi y}}{4 q^4} + \cdots  \right)  \\
\nonumber &&+\frac{5\sqrt{5}\ln \phi}{2\pi} + \frac{15 q}{2 \pi}\left( 1 + \frac{ q}{3}+ \frac{4 q^2}{9}+ \frac{3 q^3}{4}+ \frac{13 q^4}{15} + \cdots \right) \,,   \\
\nonumber Y_{\bm{3},2}^{(0)} &=& - \frac{75 q^{1/5}e^{-16 \pi y/5}}{8 \sqrt{2} \pi}\left( \frac{1}{ q}+ \frac{28\, e^{-4\pi y}}{27 q^2}+ \frac{4\, e^{-8 \pi y}}{7 q^3}+ \frac{80\, e^{-12 \pi y}}{57 q^4} + \cdots  \right)   \\
\nonumber &&- \frac{25 q^{1/5}}{2 \sqrt{2} \pi}\left( 1 + \frac{ q}{3}+ \frac{12 q^2}{11}+ \frac{11 q^3}{16}+ \frac{4 q^4}{7}+ \frac{6 q^5}{13} + \cdots \right) \,,   \\
\nonumber Y_{\bm{3},3}^{(0)} &=& - \frac{25 q^{4/5}e^{-4 \pi y/5}}{2 \sqrt{2} \pi}\left( \frac{1}{ q}+ \frac{e^{-4 \pi y}}{3 q^2}+ \frac{12\, e^{-8 \pi y}}{11 q^3}+ \frac{11\, e^{-12 \pi y}}{16 q^4} + \cdots  \right)  \\
\nonumber &&- \frac{75 q^{4/5}}{8 \sqrt{2} \pi}\left( 1 + \frac{28 q}{27}+ \frac{4 q^2}{7}+ \frac{80 q^3}{57}+ \frac{5 q^4}{9}+ \frac{40 q^5}{29} + \cdots \right)\,,  \\
\nonumber Y_{\bm{3}',1}^{(0)} &=& y - \frac{5e^{-4 \pi y} }{\pi}\left( \frac{1}{ q}+ \frac{3\, e^{-4 \pi y}}{4 q^2}+ \frac{e^{-8 \pi y}}{ q^3}+ \frac{3\, e^{-12 \pi y}}{4 q^4} + \cdots  \right)   \\
\nonumber &&-\frac{5\sqrt{5}\ln \phi}{2\pi}- \frac{5q }{\pi}\left( 1 + \frac{3 q}{4}+ q^2+ \frac{3 q^3}{4}+ \frac{6 q^4}{5} + \cdots \right)  \,, \\
\nonumber Y_{\bm{3}',2}^{(0)} &=& \frac{25 q^{2/5}e^{-12 \pi y/5}}{3 \sqrt{2} \pi}\left( \frac{1}{ q}+ \frac{15\, e^{-4 \pi y}}{16 q^2}+ \frac{18\, e^{-8 \pi y}}{13 q^3}+ \frac{7\, e^{-12 \pi y}}{12 q^4} + \cdots  \right)   \\
\nonumber && + \frac{25 q^{2/5}}{4 \sqrt{2} \pi}\left( 1 + \frac{12 q}{7}+  q^2+ \frac{32 q^3}{17}+ \frac{12 q^4}{11}+ \frac{40 q^5}{27} + \cdots \right) \,, \\
\nonumber Y_{\bm{3}',3}^{(0)} &=& \frac{25 q^{3/5}e^{-8 \pi y/5}}{4 \sqrt{2} \pi}\left( \frac{1}{ q}+ \frac{12\, e^{-4 \pi y}}{7 q^2}+ \frac{e^{-8 \pi y}}{ q^3}+ \frac{32\, e^{-12 \pi y}}{17 q^4} + \cdots  \right)  \\
\nonumber && + \frac{25 q^{3/5}}{3 \sqrt{2} \pi}\left( 1 + \frac{15 q}{16}+ \frac{18 q^2}{13}+ \frac{7 q^3}{12}+ \frac{33 q^4}{23}+ \frac{27 q^5}{28} + \cdots \right)\,,  \\
\nonumber Y_{\bm{5},1}^{(0)} &=& y  - \frac{3 e^{-4 \pi y}}{2 \pi}\left( \frac{1}{ q}+ \frac{3\, e^{-4 \pi y}}{2 q^2}+ \frac{4\, e^{-8 \pi y}}{3 q^3}+ \frac{7\, e^{-12 \pi y}}{4 q^4} + \cdots  \right) \\
\nonumber &&-\frac{5\ln 5}{4\pi} - \frac{3q }{2 \pi}\left( 1 + \frac{3 q}{2}+ \frac{4 q^2}{3}+ \frac{7 q^3}{4}+ \frac{ q^4}{5} + \cdots \right)\,,   \\
\nonumber Y_{\bm{5},2}^{(0)} &=& \frac{35 \sqrt{3} q^{1/5}e^{-16 \pi y/5}}{8 \sqrt{2} \pi}\left( \frac{1}{ q}+ \frac{52\, e^{-4\pi y}}{63 q^2}+ \frac{48\, e^{-8 \pi y}}{49 q^3}+ \frac{80\, e^{-12 \pi y}}{133 q^4} + \cdots  \right)   \\
\nonumber && + \frac{5 \sqrt{3} q^{1/5}}{2 \sqrt{2} \pi}\left( 1 + 2 q+ \frac{12 q^2}{11}+ \frac{31 q^3}{16}+ \frac{32 q^4}{21}+ \frac{21 q^5}{13} + \cdots \right) \,, \\
\nonumber Y_{\bm{5},3}^{(0)} &=& \frac{5 \sqrt{2} q^{2/5}e^{-12 \pi y/5}}{\sqrt{3} \pi}\left( \frac{1}{ q}+ \frac{45\, e^{-4 \pi y}}{32 q^2}+ \frac{21\, e^{-8 \pi y}}{26 q^3}+ \frac{13\, e^{-12 \pi y}}{8 q^4} + \cdots  \right)  \\
\nonumber && + \frac{15 \sqrt{3} q^{2/5}}{4 \sqrt{2} \pi}\left( 1 + \frac{16 q}{21}+ \frac{14 q^2}{9}+ \frac{12 q^3}{17}+ \frac{12 q^4}{11}+ \frac{80 q^5}{81} + \cdots \right)\,,  \\
\nonumber Y_{\bm{5},4}^{(0)} &=& \frac{15 \sqrt{3} q^{3/5}e^{-8 \pi y/5}}{4 \sqrt{2} \pi}\left( \frac{1}{ q}+ \frac{16\, e^{-4 \pi y}}{21 q^2}+ \frac{14\, e^{-8 \pi y}}{9 q^3}+ \frac{12\, e^{-12 \pi y}}{17 q^4} + \cdots  \right)  \\
\nonumber&& + \frac{5 \sqrt{2} q^{3/5}}{\sqrt{3} \pi}\left( 1 + \frac{45 q}{32}+ \frac{21 q^2}{26}+ \frac{13 q^3}{8}+ \frac{18 q^4}{23}+ \frac{3 q^5}{2} + \cdots \right) \,, \\
\nonumber Y_{\bm{5},5}^{(0)} &=& \frac{5 \sqrt{3} q^{4/5}e^{-4 \pi y/5}}{2 \sqrt{2} \pi}\left( \frac{1}{ q}+ \frac{2\, e^{-4 \pi y}}{ q^2}+ \frac{12\, e^{-8 \pi y}}{11 q^3}+ \frac{31\, e^{-12 \pi y}}{16 q^4} + \cdots  \right)  \\
&& + \frac{35 \sqrt{3} q^{4/5}}{8 \sqrt{2} \pi}\left( 1 + \frac{52 q}{63}+ \frac{48 q^2}{49}+ \frac{80 q^3}{133}+ \frac{10 q^4}{7}+ \frac{120 q^5}{203} + \cdots \right)\,,
\end{eqnarray}
The weight $k=-1$ polyharmonic Maa{\ss} forms of level $5$ can be arranged into two linearly independent sextuplets $\bm{\widehat {6}}$ of $A^{\prime}_5$ as follow,
{\small
\begin{eqnarray}
\nonumber Y_{\bm{\widehat {6}}I,1}^{(-1)} &=& \dfrac{y^2}{2} +  \dfrac{27 }{16 \pi^2}\left( \dfrac{\Gamma(2, 4 \pi y)}{ q}+ \dfrac{13 \Gamma(2, 8 \pi y)}{9 q^2}+ \dfrac{56 \Gamma(2, 12 \pi y)}{81 q^3}+ \dfrac{61 \Gamma(2, 16 \pi y)}{48 q^4} + \cdots  \right)  \\
\nonumber &&\hskip-0.4in + c_{\bm{\widehat {6}}I,1}^{(-1)} + \dfrac{23 }{16 \pi^2}\left( q + \dfrac{36 q^2}{23}+ \dfrac{182 q^3}{207}+ \dfrac{601 q^4}{368}+ \dfrac{623 q^5}{575}+ \dfrac{553 q^6}{414} + \cdots \right) \,,  \\
\nonumber Y_{\bm{\widehat {6}}I,2}^{(-1)} &=& - \dfrac{51 }{16 \pi^2}\left( \dfrac{\Gamma(2, 4 \pi y)}{ q}+ \dfrac{14 \Gamma(2, 8 \pi y)}{17 q^2}+ \dfrac{494 \Gamma(2, 12 \pi y)}{459 q^3}+ \dfrac{209 \Gamma(2, 16 \pi y)}{272 q^4} + \cdots  \right)   \\
\nonumber &&\hskip-0.4in + c_{\bm{\widehat {6}}I,2}^{(-1)} - \dfrac{49 }{16 \pi^2}\left( q + \dfrac{13 q^2}{14}+ \dfrac{152 q^3}{147}+ \dfrac{673 q^4}{784}+ \dfrac{1249 q^5}{1225}+ \dfrac{839 q^6}{882} + \cdots \right) \,,\\
\nonumber Y_{\bm{\widehat {6}}I,3}^{(-1)} &=& - \dfrac{75 q^{1/5}}{256 \pi^2}\left( \dfrac{\Gamma(2, \frac{16 \pi y}{5})}{ q}+ \dfrac{1712 \Gamma(2, \frac{36 \pi y}{5})}{243 q^2}+ \dfrac{48 \Gamma(2, \frac{56 \pi y}{5})}{49 q^3}+ \dfrac{1920 \Gamma(2, 76 \pi y/5)}{361 q^4} + \cdots  \right)   \\
\nonumber &&\hskip-0.4in - \dfrac{25 q^{1/5}}{16 \pi^2}\left( 1 + \dfrac{11 q}{18}+ \dfrac{122 q^2}{121}+ \dfrac{61 q^3}{256}+ \dfrac{562 q^4}{441}+ \dfrac{7 q^5}{26}+ \dfrac{962 q^6}{961} + \cdots \right) \,, \\
\nonumber Y_{\bm{\widehat {6}}I,4}^{(-1)} &=& - \dfrac{475 q^{2/5}}{72 \sqrt{2} \pi^2}\left( \dfrac{\Gamma(2, \frac{12 \pi y}{5})}{ q}+ \dfrac{945 \Gamma(2,\frac{32 \pi y}{5})}{1216 q^2}+ \dfrac{3051 \Gamma(2,\frac{52 \pi y}{5})}{3211 q^3}+ \dfrac{307 \Gamma(2, \frac{72 \pi y}{5})}{342 q^4} + \cdots  \right)   \\
\nonumber &&\hskip-0.4in - \dfrac{175 q^{2/5}}{32 \sqrt{2} \pi^2}\left( 1 + \dfrac{388 q}{343}+ \dfrac{257 q^2}{252}+ \dfrac{2308 q^3}{2023}+ \dfrac{122 q^4}{121}+ \dfrac{6080 q^5}{5103}+ \dfrac{241 q^6}{256} + \cdots \right) \,, \\
\nonumber Y_{\bm{\widehat {6}}I,5}^{(-1)} &=& - \dfrac{75 q^{3/5}}{16 \sqrt{2} \pi^2}\left( \dfrac{\Gamma(2, \frac{8 \pi y}{5})}{ q}+ \dfrac{34 \Gamma(2, \frac{28 \pi y}{5})}{49 q^2}+ \dfrac{181 \Gamma(2, \frac{48 \pi y}{5})}{216 q^3}+ \dfrac{194 \Gamma(2, 68 \pi y/5)}{289 q^4} + \cdots  \right)   \\
\nonumber &&\hskip-0.4in - \dfrac{175 q^{3/5}}{72 \sqrt{2} \pi^2}\left( 1 + \dfrac{405 q}{224}+ \dfrac{1503 q^2}{1183}+ \dfrac{139 q^3}{84}+ \dfrac{4743 q^4}{3703}+ \dfrac{10377 q^5}{5488}+ \dfrac{122 q^6}{121} + \cdots \right) \,, \\
\nonumber Y_{\bm{\widehat {6}}I,6}^{(-1)} &=& - \dfrac{75 q^{4/5}}{16 \pi^2}\left( \dfrac{\Gamma(2, 4 \pi y/5)}{ q}+ \dfrac{29 \Gamma(2, 24 \pi y/5)}{27 q^2}+ \dfrac{122 \Gamma(2, 44 \pi y/5)}{121 q^3}+ \dfrac{261 \Gamma(2, 64 \pi y/5)}{256 q^4} + \cdots  \right)   \\
\nonumber &&\hskip-0.4in  - \dfrac{1225 q^{4/5}}{256 \pi^2}\left( 1 + \dfrac{176 q}{189}+ \dfrac{2552 q^2}{2401}+ \dfrac{17280 q^3}{17689}+ \dfrac{145 q^4}{147}+ \dfrac{5760 q^5}{5887}+ \dfrac{15032 q^6}{14161} + \cdots \right)\,, \\
\nonumber Y_{\bm{\widehat {6}}II,1}^{(-1)} &=& - \dfrac{49 }{16 \pi^2}\left( \dfrac{\Gamma(2, 4 \pi y)}{ q}+ \dfrac{13 \Gamma(2, 8 \pi y)}{14 q^2}+ \dfrac{152 \Gamma(2, 12 \pi y)}{147 q^3}+ \dfrac{673 \Gamma(2, 16 \pi y)}{784 q^4} + \cdots  \right)   \\
\nonumber &&\hskip-0.4in + c_{\bm{\widehat {6}}I,2}^{(-1)} - \dfrac{51 }{16 \pi^2}\left(+ \dfrac{ q}{1}+ \dfrac{14 q^2}{17}+ \dfrac{494 q^3}{459}+ \dfrac{209 q^4}{272}+ \dfrac{417 q^5}{425}+ \dfrac{443 q^6}{459} + \cdots \right)\,,  \\
\nonumber Y_{\bm{\widehat {6}}II,2}^{(-1)} &=& \dfrac{y^2}{2} - \dfrac{23 }{16 \pi^2}\left( \dfrac{\Gamma(2, 4 \pi y)}{ q}+ \dfrac{36 \Gamma(2, 8 \pi y)}{23 q^2}+ \dfrac{182 \Gamma(2, 12 \pi y)}{207 q^3}+ \dfrac{601 \Gamma(2, 16 \pi y)}{368 q^4} + \cdots  \right)  \\
\nonumber &&\hskip-0.4in - c_{\bm{\widehat {6}}I,1}^{(-1)} - \dfrac{27 }{16 \pi^2}\left(+ \dfrac{ q}{1}+ \dfrac{13 q^2}{9}+ \dfrac{56 q^3}{81}+ \dfrac{61 q^4}{48}+ \dfrac{209 q^5}{225}+ \dfrac{311 q^6}{243} + \cdots \right) \,, \\
\nonumber Y_{\bm{\widehat {6}}II,3}^{(-1)} &=& - \dfrac{1225 q^{1/5}}{256 \pi^2}\bigg( \dfrac{\Gamma(2, \frac{16 \pi y}{5})}{ q}+ \dfrac{176 \Gamma(2, \frac{36 \pi y}{5})}{189 q^2}+ \dfrac{2552 \Gamma(2, \frac{56 \pi y}{5})}{2401 q^3} + \dfrac{17280 \Gamma(2, \frac{76 \pi y}{5})}{17689 q^4} + \cdots  \bigg)  \\
\nonumber &&\hskip-0.4in - \dfrac{75 q^{1/5}}{16 \pi^2}\left( 1 + \dfrac{29 q}{27}+ \dfrac{122 q^2}{121}+ \dfrac{261 q^3}{256}+ \dfrac{1286 q^4}{1323}+ \dfrac{183 q^5}{169} + \cdots \right)\,,  \\
\nonumber Y_{\bm{\widehat {6}}II,4}^{(-1)} &=& - \dfrac{175 q^{2/5}}{72 \sqrt{2} \pi^2}\left( \dfrac{\Gamma(2, \frac{12 \pi y}{5})}{ q}+ \dfrac{405 \Gamma(2,\frac{32 \pi y}{5})}{224 q^2}+ \dfrac{1503 \Gamma(2,\frac{52 \pi y}{5})}{1183 q^3}+ \dfrac{139 \Gamma(2, \frac{72 \pi y}{5})}{84 q^4} + \cdots  \right)   \\
\nonumber &&\hskip-0.4in - \dfrac{75 q^{2/5}}{16 \sqrt{2} \pi^2}\left( 1 + \dfrac{34 q}{49}+ \dfrac{181 q^2}{216}+ \dfrac{194 q^3}{289}+ \dfrac{122 q^4}{121}+ \dfrac{1120 q^5}{2187} + \cdots \right) \,, \\
\nonumber Y_{\bm{\widehat {6}}II,5}^{(-1)} &=& \dfrac{175 q^{3/5}}{32 \sqrt{2} \pi^2}\left( \dfrac{\Gamma(2, \frac{8 \pi y}{5})}{ q}+ \dfrac{388 \Gamma(2, \frac{28 \pi y}{5})}{343 q^2}+ \dfrac{257 \Gamma(2, \frac{48 \pi y}{5})}{252 q^3}+ \dfrac{2308 \Gamma(2, 68 \pi y/5)}{2023 q^4} + \cdots  \right)   \\
\nonumber &&\hskip-0.4in + \dfrac{475 q^{3/5}}{72 \sqrt{2} \pi^2}\left( 1 + \dfrac{945 q}{1216}+ \dfrac{3051 q^2}{3211}+ \dfrac{307 q^3}{342}+ \dfrac{9531 q^4}{10051}+ \dfrac{11259 q^5}{14896} + \cdots \right)\,,  \\
\nonumber Y_{\bm{\widehat {6}}II,6}^{(-1)} &=& \dfrac{25 q^{4/5}}{16 \pi^2}\left( \dfrac{\Gamma(2, 4 \pi y/5)}{ q}+ \dfrac{11 \Gamma(2, 24 \pi y/5)}{18 q^2}+ \dfrac{122 \Gamma(2, 44 \pi y/5)}{121 q^3}+ \dfrac{61 \Gamma(2, 64 \pi y/5)}{256 q^4} + \cdots  \right)   \\
&&\hskip-0.4in + \dfrac{75 q^{4/5}}{256 \pi^2}\left( 1 + \dfrac{1712 q}{243}+ \dfrac{48 q^2}{49}+ \dfrac{1920 q^3}{361}+ \dfrac{55 q^4}{18}+ \dfrac{4480 q^5}{841}+ \dfrac{368 q^6}{289} + \cdots \right)\,,
\end{eqnarray}}
where the constant terms are
\begin{equation}
 c_{\bm{\widehat {6}}I,1}^{(-1)} =  0.197869\,,  \qquad c_{\bm{\widehat {6}}I,2}^{(-1)} = -0.280512\,.
\end{equation}
At weight $k=-2$ and level $5$,  the four linearly independent polyharmonic Maa{\ss} form  multiplets $Y^{(-2)}_{\bm{1}}$,  $Y^{(-2)}_{\bm{3}}$, $Y^{(-2)}_{\bm{3^{\prime}}}$ and $Y^{(-2)}_{\bm{5}}$  have the following forms
{\small
\begin{eqnarray}
\nonumber Y_{\mathbf{1}}^{(-2)}&=& \frac{y^3}{3} - \frac{5}{\pi^3}\left(\frac{3\Gamma(3,4\pi y)}{4 q} + \frac{27\Gamma(3,8\pi y)}{32 q^2} + \frac{7\Gamma(3,12\pi y)}{9 q^3} + \cdots\right) \\
\nonumber&&-\frac{\pi}{12}\frac{\zeta(3)}{\zeta(4)} -\frac{q}{\pi^3}\left( \frac{15 }{2} +\frac{135 q}{16} + \frac{70 q^2}{9} + \frac{1095 q^3}{128} + \frac{189 q^4}{25} + \frac{35 q^5}{4} + \cdots \right)\,, \\
\nonumber  Y_{\mathbf{3},1}^{(-2)} &=& \frac{y^3}{3} - \frac{63 }{32 \pi^3}\left( \frac{\Gamma(3, 4 \pi y)}{ q}+ \frac{31 \Gamma(3, 8 \pi y)}{36 q^2}+ \frac{1612 \Gamma(3, 12 \pi y)}{1701 q^3}+ \frac{57 \Gamma(3, 16 \pi y)}{64 q^4} + \cdots  \right) \\
\nonumber &&-\frac{\pi L(3,\chi_3)}{12\sqrt{5}L(4,\chi_3)} - \frac{63q }{16 \pi^3}\left( 1 + \frac{31 q}{36}+ \frac{1612 q^2}{1701}+ \frac{57 q^3}{64}+ \frac{7813 q^4}{7875} + \cdots \right) \,, \\
\nonumber  Y_{\mathbf{3},2}^{(-2)} &=& \frac{7125 q^{1/5}}{2048 \sqrt{2} \pi^3}\left( \frac{\Gamma(3, \frac{16 \pi y}{5})}{ q}+ \frac{2368 \Gamma(3, \frac{36 \pi y}{5})}{2187 q^2}+ \frac{48 \Gamma(3, \frac{56 \pi y}{5})}{49 q^3}+ \frac{439040 \Gamma(3, \frac{76 \pi y}{5})}{390963 q^4} + \cdots  \right) \\
\nonumber&& + \frac{125 q^{1/5}}{16 \sqrt{2} \pi^3}\left( 1 + \frac{91 q}{108}+ \frac{1332 q^2}{1331}+ \frac{3641 q^3}{4096}+ \frac{988 q^4}{1029}+ \frac{3843 q^5}{4394} + \cdots \right) \,, \\
\nonumber  Y_{\mathbf{3},3}^{(-2)} &=& \frac{125 q^{4/5}}{32 \sqrt{2} \pi^3}\left( \frac{\Gamma(3, \frac{4 \pi y}{5})}{ q}+ \frac{91 \Gamma(3, \frac{24 \pi y}{5})}{108 q^2}+ \frac{1332 \Gamma(3, \frac{44 \pi y}{5})}{1331 q^3}+ \frac{3641 \Gamma(3, \frac{64 \pi y}{5})}{4096 q^4} + \cdots  \right) \\
\nonumber&& + \frac{7125 q^{4/5}}{1024 \sqrt{2} \pi^3}\left( 1 + \frac{2368 q}{2187}+ \frac{48 q^2}{49}+ \frac{439040 q^3}{390963}+ \frac{5915 q^4}{6156}+ \frac{520320 q^5}{463391} + \cdots \right) \,, \\
\nonumber  Y_{\mathbf{3}',1}^{(-2)} &=& \frac{y^3}{3} + \frac{31 }{16 \pi^3}\left( \frac{\Gamma(3, 4 \pi y)}{ q}+ \frac{441 \Gamma(3, 8 \pi y)}{496 q^2}+ \frac{91 \Gamma(3, 12 \pi y)}{93 q^3}+ \frac{57 \Gamma(3, 16 \pi y)}{64 q^4} + \cdots  \right) \\
\nonumber &&+\frac{\pi L(3,\chi_3)}{12\sqrt{5}L(4,\chi_3)} + \frac{31q }{8 \pi^3}\left( 1 + \frac{441 q}{496}+ \frac{91 q^2}{93}+ \frac{57 q^3}{64}+ \frac{126 q^4}{125} + \cdots \right)  \,, \\
\nonumber  Y_{\mathbf{3}',2}^{(-2)} &=& - \frac{1625 q^{2/5}}{432 \sqrt{2} \pi^3}\left( \frac{\Gamma(3, \frac{12 \pi y}{5})}{ q}+ \frac{945 \Gamma(3, \frac{32 \pi y}{5})}{1024 q^2}+ \frac{29646 \Gamma(3, \frac{52 \pi y}{5})}{28561 q^3}+ \frac{4921 \Gamma(3, \frac{72 \pi y}{5})}{5616 q^4} + \cdots  \right) \\
\nonumber && - \frac{875 q^{2/5}}{128 \sqrt{2} \pi^3}\left( 1 + \frac{2736 q}{2401}+ \frac{247 q^2}{252}+ \frac{39296 q^3}{34391}+ \frac{1332 q^4}{1331}+ \frac{151840 q^5}{137781} + \cdots \right) \,, \\
\nonumber  Y_{\mathbf{3}',3}^{(-2)} &=& - \frac{875 q^{3/5}}{256 \sqrt{2} \pi^3}\left( \frac{\Gamma(3, \frac{8 \pi y}{5})}{ q}+ \frac{2736 \Gamma(3, \frac{28 \pi y}{5})}{2401 q^2}+ \frac{247 \Gamma(3, \frac{48 \pi y}{5})}{252 q^3}+ \frac{39296 \Gamma(3, \frac{68 \pi y}{5})}{34391 q^4} + \cdots  \right) \\
\nonumber && - \frac{1625 q^{3/5}}{216 \sqrt{2} \pi^3}\left( 1 + \frac{945 q}{1024}+ \frac{29646 q^2}{28561}+ \frac{4921 q^3}{5616}+ \frac{164241 q^4}{158171}+ \frac{263169 q^5}{285376} + \cdots \right) \,, \\
\nonumber  Y_{\mathbf{5},1}^{(-2)} &=& \frac{y^3}{3} + \frac{315 }{416 \pi^3}\left( \frac{\Gamma(3, 4 \pi y)}{ q}+ \frac{9 \Gamma(3, 8 \pi y)}{8 q^2}+ \frac{28 \Gamma(3, 12 \pi y)}{27 q^3}+ \frac{73 \Gamma(3, 16 \pi y)}{64 q^4} + \cdots  \right) \\
\nonumber&&+\frac{5\pi}{312} \frac{\zeta(3)}{\zeta(4)} + \frac{315q }{208 \pi^3}\left( 1 + \frac{9 q}{8}+ \frac{28 q^2}{27}+ \frac{73 q^3}{64}+ \frac{2521 q^4}{2625} + \cdots \right) \,, \\
\nonumber  Y_{\mathbf{5},2}^{(-2)} &=& - \frac{45625 \sqrt{3} q^{1/5}}{26624 \sqrt{2} \pi^3}\left(\frac{\Gamma(3, \frac{16 \pi y}{5})}{ q}+ \frac{48448 \Gamma(3, \frac{36 \pi y}{5})}{53217 q^2}+ \frac{24768 \Gamma(3, \frac{56 \pi y}{5})}{25039 q^3}+ \frac{439040 \Gamma(3, \frac{76 \pi y}{5})}{500707 q^4} + \cdots  \right) \\
\nonumber && - \frac{625 \sqrt{3} q^{1/5}}{208 \sqrt{2} \pi^3}\left( 1 + \frac{7 q}{6}+ \frac{1332 q^2}{1331}+ \frac{4681 q^3}{4096}+ \frac{1376 q^4}{1323}+ \frac{9891 q^5}{8788} + \cdots \right) \,,\\
\nonumber  Y_{\mathbf{5},3}^{(-2)} &=& - \frac{4375 q^{2/5}}{936 \sqrt{6} \pi^3}\left( \frac{\Gamma(3, \frac{12 \pi y}{5})}{ q}+ \frac{15795 \Gamma(3, \frac{32 \pi y}{5})}{14336 q^2}+ \frac{4239 \Gamma(3, \frac{52 \pi y}{5})}{4394 q^3}+ \frac{757 \Gamma(3, \frac{72 \pi y}{5})}{672 q^4} + \cdots  \right) \\
\nonumber&& - \frac{5625 \sqrt{3} q^{2/5}}{1664 \sqrt{2} \pi^3}\left( 1 + \frac{2752 q}{3087}+ \frac{511 q^2}{486}+ \frac{4368 q^3}{4913}+ \frac{1332 q^4}{1331}+ \frac{163520 q^5}{177147} + \cdots \right) \,, \\
\nonumber  Y_{\mathbf{5},4}^{(-2)} &=& - \frac{5625 \sqrt{3} q^{3/5}}{3328 \sqrt{2} \pi^3}\left( \frac{\Gamma(3, \frac{8 \pi y}{5})}{ q}+ \frac{2752 \Gamma(3, \frac{28 \pi y}{5})}{3087 q^2}+ \frac{511 \Gamma(3, \frac{48 \pi y}{5})}{486 q^3}+ \frac{4368 \Gamma(3, \frac{68 \pi y}{5})}{4913 q^4} + \cdots  \right) \\
\nonumber && - \frac{4375 q^{3/5}}{468 \sqrt{6} \pi^3}\left( 1 + \frac{15795 q}{14336}+ \frac{4239 q^2}{4394}+ \frac{757 q^3}{672}+ \frac{82134 q^4}{85169}+ \frac{84753 q^5}{76832} + \cdots \right) \,, \\
\nonumber  Y_{\mathbf{5},5}^{(-2)} &=& - \frac{625 \sqrt{3} q^{4/5}}{416 \sqrt{2} \pi^3}\left( \frac{\Gamma(3, \frac{4 \pi y}{5})}{ q}+ \frac{7 \Gamma(3, \frac{24 \pi y}{5})}{6 q^2}+ \frac{1332 \Gamma(3, \frac{44 \pi y}{5})}{1331 q^3}+ \frac{4681 \Gamma(3, \frac{64 \pi y}{5})}{4096 q^4} + \cdots  \right) \\
&&  - \frac{45625 \sqrt{3} q^{4/5}}{13312 \sqrt{2} \pi^3}\left( 1 + \frac{48448 q}{53217}+ \frac{24768 q^2}{25039}+ \frac{439040 q^3}{500707}+ \frac{455 q^4}{438}+ \frac{1560960 q^5}{1780397} + \cdots \right)\,,
\end{eqnarray}}
where $\zeta(s)=\sum_{k=1}^{\infty}k^{-s}$  is the Riemann zeta function of $s$ and $L(k,\chi_3) = \sum_{n=1}^{\infty} n^{-k} \chi_3(n)$ is the $L$-function with a Dirichlet character $\chi_3$ modulo $5$, and the value of the character $\chi_3(n)$ is given by
\begin{eqnarray}
\chi_3(n) =  \begin{cases}
1 \,,~~&n\equiv \pm 1 \,~({\rm mod}\, 5)  \\
-1 \,,~~&n\equiv \pm 2 \,~({\rm mod}\, 5)  \\
0 \,,~~&n\equiv 0 \,~({\rm mod}\, 5)
\end{cases} \,.
\end{eqnarray}
One can organize the weight $k_Y=-3$ polyharmonic Maa{\ss} forms of level $5$ into two sextuplets $\bm{\widehat {6}}$ of $A^{\prime}_5$ as follow,
\begin{footnotesize}
\begin{eqnarray}
\nonumber Y_{\bm{\widehat {6}}I,1}^{(-3)} &=& \frac{y^4}{4} - \frac{26949 }{75008 \pi^4}\left( \frac{\Gamma(4, 4 \pi y)}{ q}+ \frac{39751 \Gamma(4, 8 \pi y)}{35932 q^2}+ \frac{54544 \Gamma(4, 12 \pi y)}{55971 q^3}+ \frac{2524981 \Gamma(4, 16 \pi y)}{2299648 q^4} + \cdots  \right)  \\
\nonumber &&\hskip-0.4in + c_{\bm{\widehat {6}}I,1}^{(-3)} - \frac{80403 }{37504 \pi^4}\left( q + \frac{29718 q^2}{26801}+ \frac{2134034 q^3}{2170881}+ \frac{7611307 q^4}{6861056}+ \frac{16796801 q^5}{16750625}+ \frac{18884971 q^6}{17367048} + \cdots \right) \,,  \\
\nonumber Y_{\bm{\widehat {6}}I,2}^{(-3)} &=& \frac{46293 }{75008 \pi^4}\left( \frac{\Gamma(4, 4 \pi y)}{ q}+ \frac{1141 \Gamma(4, 8 \pi y)}{1187 q^2}+ \frac{3779162 \Gamma(4, 12 \pi y)}{3749733 q^3}+ \frac{3783867 \Gamma(4, 16 \pi y)}{3950336 q^4} + \cdots  \right)    \\
\nonumber &&\hskip-0.4in + c_{\bm{\widehat {6}}I,2}^{(-3)} + \frac{138621 }{37504 \pi^4}\left( q + \frac{51019 q^2}{52808}+ \frac{1255696 q^3}{1247589}+ \frac{11375899 q^4}{11828992}+ \frac{28906207 q^5}{28879375}+ \frac{29089397 q^6}{29942136} + \cdots \right) \,,  \\
\nonumber Y_{\bm{\widehat {6}}I,3}^{(-3)} &=&  \frac{99432789375 q^{1/5}}{83961254912}\left( \frac{562760251 \
\Gamma(4, \frac{16 \pi y}{5})}{3393972544 \pi^4 q}+ \frac{238242075668 \Gamma(4, \frac{36 \pi y}{5})}{1043805649743 \pi^4 q^2}+ \frac{21176407993 \Gamma(4, \frac{56 \pi y}{5})}{127327001221 \pi^4 q^3} + \cdots  \right)  \\
\nonumber &&\hskip-0.4in + \frac{58125 q^{1/5}}{37504 \pi^4}\left( 1 + \frac{16301 q}{20088}+ \frac{14642 q^2}{14641}+ \frac{1546351 q^3}{2031616}+ \frac{6300382 q^4}{6028911}+ \frac{5413261 q^5}{7083128}+ \frac{923522 q^6}{923521} + \cdots \right) \,, \\
\nonumber Y_{\bm{\widehat {6}}I,4}^{(-3)} &=& \frac{3773125 q^{2/5}}{3037824 \sqrt{2} \pi^4}\left( \frac{\Gamma(4, \frac{12 \pi y}{5})}{ q}+ \frac{23567355 \Gamma(4,\frac{32 \pi y}{5})}{24727552 q^2}+ \frac{171198117 \Gamma(4,\frac{52 \pi y}{5})}{172422757 q^3}+ \frac{3773341 \Gamma(4, \frac{72 \pi y}{5})}{3911976 q^4} + \cdots  \right)   \\
\nonumber && \hskip-0.4in + \frac{2139375 q^{2/5}}{300032 \sqrt{2} \pi^4}\left( 1 + \frac{2842096 q}{2739541}+ \frac{1484891 q^2}{1478736}+ \frac{98888176 q^3}{95297461}+ \frac{14642 q^4}{14641}+ \frac{633643520 q^5}{606374181}+ \frac{65281 q^6}{65536} + \cdots \right)\,,   \\
\nonumber Y_{\bm{\widehat {6}}I,5}^{(-3)} &=& \frac{238125 q^{3/5}}{300032 \sqrt{2} \pi^4}\left( \frac{\Gamma(4, \frac{8 \pi y}{5})}{ q}+ \frac{275512 \Gamma(4, \frac{28 \pi y}{5})}{304927 q^2}+ \frac{965887 \Gamma(4, \frac{48 \pi y}{5})}{987552 q^3}+ \frac{9577272 \Gamma(4, \frac{68 \pi y}{5})}{10607167 q^4} + \cdots  \right)   \\
\nonumber &&\hskip-0.4in + \frac{2130625 q^{3/5}}{506304 \sqrt{2} \pi^4}\left( 1 + \frac{7869555 q}{6981632}+ \frac{99471969 q^2}{97364449}+ \frac{1635433 q^3}{1472688}+ \frac{974680209 q^4}{953977969}+ \frac{2364222411 q^5}{2095362304}+ \frac{14642 q^6}{14641} + \cdots \right) \,,  \\
\nonumber Y_{\bm{\widehat {6}}I,6}^{(-3)} &=& \frac{73125 q^{4/5}}{75008 \pi^4}\left( \frac{\Gamma(4, 4 \pi y/5)}{ q}+ \frac{2957 \Gamma(4, 24 \pi y/5)}{2916 q^2}+ \frac{14642 \Gamma(4, 44 \pi y/5)}{14641 q^3}+ \frac{2588119 \Gamma(4, 64 \pi y/5)}{2555904 q^4} + \cdots  \right)   \\
\nonumber&&\hskip-0.4in + \frac{56870625 q^{4/5}}{9601024 \pi^4}\left( 1 + \frac{3108608 q}{3158757}+ \frac{73111712 q^2}{72824731}+ \frac{3903344640 q^3}{3952766251}+ \frac{3267485 q^4}{3275748}+ \frac{3026350080 q^5}{3064648573} + \cdots \right)\,, \\
\nonumber Y_{\bm{\widehat {6}}II,1}^{(-3)} &=& \frac{46207 }{75008 \pi^4}\left( \frac{\Gamma(4, 4 \pi y)}{ q}+ \frac{51019 \Gamma(4, 8 \pi y)}{52808 q^2}+ \frac{1255696 \Gamma(4, 12 \pi y)}{1247589 q^3}+ \frac{11375899 \Gamma(4, 16 \pi y)}{11828992 q^4} + \cdots  \right)   \\
\nonumber &&\hskip-0.4in + c_{\bm{\widehat {6}}I,2}^{(-3)} + \frac{138879 }{37504 \pi^4}\left( q + \frac{1141 q^2}{1187}+ \frac{3779162 q^3}{3749733}+ \frac{3783867 q^4}{3950336}+ \frac{741187 q^5}{741875}+ \frac{1121153 q^6}{1153764} + \cdots \right)\,,  \\
\nonumber Y_{\bm{\widehat {6}}II,2}^{(-3)} &=& \frac{y^4}{4} + \frac{26801 }{75008 \pi^4}\left( \frac{\Gamma(4, 4 \pi y)}{ q}+ \frac{29718 \Gamma(4, 8 \pi y)}{26801 q^2}+ \frac{2134034 \Gamma(4, 12 \pi y)}{2170881 q^3}+ \frac{7611307 \Gamma(4, 16 \pi y)}{6861056 q^4} + \cdots  \right)  \\
\nonumber &&\hskip-0.4in -c_{\bm{\widehat {6}}I,1}^{(-3)} + \frac{80847 }{37504 \pi^4}\left( q + \frac{39751 q^2}{35932}+ \frac{54544 q^3}{55971}+ \frac{2524981 q^4}{2299648}+ \frac{430691 q^5}{431875}+ \frac{729929 q^6}{671652} + \cdots \right)  \\
\nonumber Y_{\bm{\widehat {6}}II,3}^{(-3)} &=& \frac{18956875 q^{1/5}}{19202048 \pi^4}\left( \frac{\Gamma(4, \frac{16 \pi y}{5})}{ q}+ \frac{3108608 \Gamma(4, \frac{36 \pi y}{5})}{3158757 q^2}+ \frac{73111712 \Gamma(4, \frac{56 \pi y}{5})}{72824731 q^3}+ \frac{3903344640 \Gamma(4, \frac{76 \pi y}{5})}{3952766251 q^4} + \cdots  \right)   \\
\nonumber &&\hskip-0.4in + \frac{219375 q^{1/5}}{37504 \pi^4}\left( 1 + \frac{2957 q}{2916}+ \frac{14642 q^2}{14641}+ \frac{2588119 q^3}{2555904}+ \frac{22682474 q^4}{22754277}+ \frac{4529267 q^5}{4455516} + \cdots \right) \,, \\
\nonumber Y_{\bm{\widehat {6}}II,4}^{(-3)} &=& \frac{2130625 q^{2/5}}{3037824 \sqrt{2} \pi^4}\left( \frac{\Gamma(4, \frac{12 \pi y}{5})}{ q}+ \frac{7869555 \Gamma(4,\frac{32 \pi y}{5})}{6981632 q^2}+ \frac{99471969 \Gamma(4,\frac{52 \pi y}{5})}{97364449 q^3}+ \frac{1635433 \Gamma(4, \frac{72 \pi y}{5})}{1472688 q^4} + \cdots  \right)   \\
\nonumber &&\hskip-0.4in + \frac{714375 q^{2/5}}{150016 \sqrt{2} \pi^4}\left( 1 + \frac{275512 q}{304927}+ \frac{965887 q^2}{987552}+ \frac{9577272 q^3}{10607167}+ \frac{14642 q^4}{14641}+ \frac{178904320 q^5}{202479021} + \cdots \right)\,,  \\
\nonumber Y_{\bm{\widehat {6}}II,5}^{(-3)} &=& - \frac{713125 q^{3/5}}{600064 \sqrt{2} \pi^4}\left( \frac{\Gamma(4, \frac{8 \pi y}{5})}{ q}+ \frac{2842096 \Gamma(4, \frac{28 \pi y}{5})}{2739541 q^2}+ \frac{1484891 \Gamma(4, \frac{48 \pi y}{5})}{1478736 q^3}+ \frac{98888176 \Gamma(4, \frac{68 \pi y}{5})}{95297461 q^4} + \cdots  \right)   \\
\nonumber &&\hskip-0.4in - \frac{3773125 q^{3/5}}{506304 \sqrt{2} \pi^4}\left( 1 + \frac{23567355 q}{24727552}+ \frac{171198117 q^2}{172422757}+ \frac{3773341 q^3}{3911976}+ \frac{1677370437 q^4}{1689400117}+ \frac{3535069473 q^5}{3710678272} + \cdots \right) \,, \\
\nonumber Y_{\bm{\widehat {6}}II,6}^{(-3)} &=& - \frac{19375 q^{4/5}}{75008 \pi^4}\left( \frac{\Gamma(4, 4 \pi y/5)}{ q}+ \frac{16301 \Gamma(4, 24 \pi y/5)}{20088 q^2}+ \frac{14642 \Gamma(4, 44 \pi y/5)}{14641 q^3}+ \frac{1546351 \Gamma(4, 64 \pi y/5)}{2031616 q^4} + \cdots  \right)   \\
&&\hskip-0.4in - \frac{11311875 q^{4/5}}{9601024 \pi^4}\left( 1 + \frac{54486272 q}{39582513}+ \frac{4843072 q^2}{4828411}+ \frac{344739840 q^3}{262075531}+ \frac{1385585 q^4}{1303128}+ \frac{1870991360 q^5}{1422342091} + \cdots \right)\,,
\end{eqnarray}
\end{footnotesize}
where the constant terms are
\begin{equation}
 c_{\bm{\widehat {6}}I,1}^{(-3)} =  -0.0239102, \qquad
c_{\bm{\widehat {6}}I,2}^{(-3)} = 0.0367462 \,.
\end{equation}
The weight $k=-4$ polyharmonic Maa{\ss} forms of level $5$ are of the following form:
{\small \begin{eqnarray}
\nonumber Y_{\mathbf{1}}^{(-4)}&=& \frac{y^5}{5} + \frac{63\Gamma(5,4\pi y)}{128\pi^5 q} + \frac{2079\Gamma(5,8\pi y)}{4096\pi^5 q^2} + \frac{427\Gamma(5,12\pi y)}{864\pi^5 q^3} + \frac{66591\Gamma(5,16\pi y)}{131072 \pi^5 q^4} + \cdots   \\
\nonumber &&+\frac{\pi}{80}\frac{\zeta(5)}{\zeta(6)} + \frac{q}{\pi^5}\left(\frac{189 }{16} + \frac{6237 q}{512 } + \frac{427 q^2}{36 } + \frac{199773 q^3}{16384} + \cdots \right)\,, \\
\nonumber  Y_{\mathbf{3},1}^{(-4)} &=& \frac{y^5}{5} + \frac{7815 }{34304 \pi^5}\left( \frac{\Gamma(5, 4 \pi y)}{ q}+ \frac{24211 \Gamma(5, 8 \pi y)}{25008 q^2}+ \frac{378004 \Gamma(5, 12 \pi y)}{379809 q^3}+ \frac{993 \Gamma(5, 16 \pi y)}{1024 q^4} + \cdots  \right) \\
\nonumber &&+\frac{\pi L(5,\chi_3)}{80\sqrt{5}L(6,\chi_3)} + \frac{23445q }{4288 \pi^5}\left( 1 + \frac{24211 q}{25008}+ \frac{378004 q^2}{379809}+ \frac{993 q^3}{1024}+ \frac{4882813 q^4}{4884375} + \cdots \right) \,,  \\
\nonumber  Y_{\mathbf{3},2}^{(-4)} &=& - \frac{15515625 q^{1/5}}{35127296 \sqrt{2} \pi^5}\left( \frac{\Gamma(5, \frac{16 \pi y}{5})}{ q}+ \frac{60218368 \Gamma(5, \frac{36 \pi y}{5})}{58635657 q^2}+ \frac{5557184 \Gamma(5, \frac{56 \pi y}{5})}{5563117 q^3} + \cdots  \right) \\
\nonumber && - \frac{46875 q^{1/5}}{4288 \sqrt{2} \pi^5}\left( 1 + \frac{3751 q}{3888}+ \frac{161052 q^2}{161051}+ \frac{1016801 q^3}{1048576}+ \frac{1355684 q^4}{1361367} + \cdots \right) \,, \\
\nonumber  Y_{\mathbf{3},3}^{(-4)} &=& - \frac{15625 q^{4/5}}{34304 \sqrt{2} \pi^5}\left( \frac{\Gamma(5, 4 \pi y/5)}{ q}+ \frac{3751 \Gamma(5, 24 \pi y/5)}{3888 q^2}+ \frac{161052 \Gamma(5, 44 \pi y/5)}{161051 q^3} + \cdots  \right) \\
\nonumber && - \frac{46546875 q^{4/5}}{4390912 \sqrt{2} \pi^5}\left( 1 + \frac{60218368 q}{58635657}+ \frac{5557184 q^2}{5563117}+ \frac{2535526400 q^3}{2458766307}+ \frac{3844775 q^4}{3860784} + \cdots \right) \,, \\
\nonumber  Y_{\mathbf{3}',1}^{(-4)} &=& \frac{y^5}{5} - \frac{3905 }{17152 \pi^5}\left( \frac{\Gamma(5, 4 \pi y)}{ q}+ \frac{48453 \Gamma(5, 8 \pi y)}{49984 q^2}+ \frac{5731 \Gamma(5, 12 \pi y)}{5751 q^3}+ \frac{993 \Gamma(5, 16 \pi y)}{1024 q^4} + \cdots  \right) \\
\nonumber &&-\frac{\pi L(5,\chi_3)}{80\sqrt{5}L(6,\chi_3)} - \frac{11715q }{2144 \pi^5}\left( 1 + \frac{48453 q}{49984}+ \frac{5731 q^2}{5751}+ \frac{993 q^3}{1024}+ \frac{3126 q^4}{3125} + \cdots \right) \,, \\
\nonumber  Y_{\mathbf{3}',2}^{(-4)} &=& \frac{1890625 q^{2/5}}{4167936 \sqrt{2} \pi^5}\left( \frac{\Gamma(5, \frac{12 \pi y}{5})}{ q}+ \frac{7721325 \Gamma(5,\frac{32 \pi y}{5})}{7929856 q^2}+ \frac{45111978 \Gamma(5,\frac{52 \pi y}{5})}{44926453 q^3} + \cdots  \right) \\
\nonumber && + \frac{1453125 q^{2/5}}{137216 \sqrt{2} \pi^5}\left( 1 + \frac{537792 q}{521017}+ \frac{40051 q^2}{40176}+ \frac{45435392 q^3}{44015567}+ \frac{161052 q^4}{161051} + \cdots \right)\,, \\
\nonumber  Y_{\mathbf{3}',3}^{(-4)} &=& \frac{484375 q^{3/5}}{1097728 \sqrt{2} \pi^5}\left( \frac{\Gamma(5, \frac{8 \pi y}{5})}{ q}+ \frac{537792 \Gamma(5, \frac{28 \pi y}{5})}{521017 q^2}+ \frac{40051 \Gamma(5, \frac{48 \pi y}{5})}{40176 q^3} + \cdots  \right) \\
\nonumber && + \frac{1890625 q^{3/5}}{173664 \sqrt{2} \pi^5}\left( 1 + \frac{7721325 q}{7929856}+ \frac{45111978 q^2}{44926453}+ \frac{1823017 q^3}{1881792}+ \frac{71092323 q^4}{70799773} + \cdots \right) \,, \\
\nonumber  Y_{\mathbf{5},1}^{(-4)} &=& \frac{y^5}{5} - \frac{1563 }{15872 \pi^5}\left( \frac{\Gamma(5, 4 \pi y)}{ q}+ \frac{33 \Gamma(5, 8 \pi y)}{32 q^2}+ \frac{244 \Gamma(5, 12 \pi y)}{243 q^3}+ \frac{1057 \Gamma(5, 16 \pi y)}{1024 q^4} + \cdots  \right) \\
\nonumber &&-\frac{13\pi}{5208} \frac{\zeta(5)}{\zeta(6)} - \frac{4689q }{1984 \pi^5}\left( 1 + \frac{33 q}{32}+ \frac{244 q^2}{243}+ \frac{1057 q^3}{1024}+ \frac{1625521 q^4}{1628125} + \cdots \right)\,, \\
\nonumber  Y_{\mathbf{5},2}^{(-4)} &=& \frac{3303125 \sqrt{3} q^{1/5}}{16252928 \sqrt{2} \pi^5}\left( \frac{\Gamma(5, \frac{16 \pi y}{5})}{ q}+ \frac{60716032 \Gamma(5, \frac{36 \pi y}{5})}{62414793 q^2}+ \frac{17749248 \Gamma(5, \frac{56 \pi y}{5})}{17764999 q^3} + \cdots  \right) \\
\nonumber && + \frac{9375 \sqrt{3} q^{1/5}}{1984 \sqrt{2} \pi^5}\left( 1 + \frac{671 q}{648}+ \frac{161052 q^2}{161051}+ \frac{1082401 q^3}{1048576}+ \frac{4101152 q^4}{4084101} + \cdots \right) \,, \\
\nonumber  Y_{\mathbf{5},3}^{(-4)} &=& \frac{190625 q^{2/5}}{321408 \sqrt{6} \pi^5}\left( \frac{\Gamma(5, \frac{12 \pi y}{5})}{ q}+ \frac{8219475 \Gamma(5,\frac{32 \pi y}{5})}{7995392 q^2}+ \frac{45112221 \Gamma(5,\frac{52 \pi y}{5})}{45297746 q^3} + \cdots  \right) \\
\nonumber && + \frac{309375 \sqrt{3} q^{2/5}}{63488 \sqrt{2} \pi^5}\left( 1 + \frac{48896 q}{50421}+ \frac{64477 q^2}{64152}+ \frac{1376832 q^3}{1419857}+ \frac{161052 q^4}{161051} + \cdots \right)\,, \\
\nonumber  Y_{\mathbf{5},4}^{(-4)} &=& \frac{103125 \sqrt{3} q^{3/5}}{507904 \sqrt{2} \pi^5}\left( \frac{\Gamma(5, \frac{8 \pi y}{5})}{ q}+ \frac{48896 \Gamma(5, \frac{28 \pi y}{5})}{50421 q^2}+ \frac{64477 \Gamma(5, \frac{48 \pi y}{5})}{64152 q^3} + \cdots  \right) \\
\nonumber && + \frac{190625 q^{3/5}}{13392 \sqrt{6} \pi^5}\left( 1 + \frac{8219475 q}{7995392}+ \frac{45112221 q^2}{45297746}+ \frac{652223 q^3}{632448}+ \frac{391007898 q^4}{392616923} + \cdots \right) \,, \\
\nonumber  Y_{\mathbf{5},5}^{(-4)} &=& \frac{3125 \sqrt{3} q^{4/5}}{15872 \sqrt{2} \pi^5}\left( \frac{\Gamma(5, 4 \pi y/5)}{ q}+ \frac{671 \Gamma(5, 24 \pi y/5)}{648 q^2}+ \frac{161052 \Gamma(5, 44 \pi y/5)}{161051 q^3} + \cdots  \right) \\
&&\hskip-0.2in + \frac{9909375 \sqrt{3} q^{4/5}}{2031616 \sqrt{2} \pi^5}\left( 1 + \frac{60716032 q}{62414793}+ \frac{17749248 q^2}{17764999}+ \frac{2535526400 q^3}{2617236643}+ \frac{687775 q^4}{684936} + \cdots \right)\,.
\end{eqnarray}}
We can organize the weight $k=-5$ polyharmonic Maa{\ss} forms of level $5$ into two independent sextets $Y^{(-5)}_{\bm{\widehat {6}}I}$ and  $Y^{(-5)}_{\bm{\widehat {6}}II}$ of $A^{\prime}_{5}$
\begin{footnotesize}
\begin{eqnarray}
\nonumber Y_{\bm{\widehat {6}}I,1}^{(-5)} &=& \frac{y^{6}}{6}+\frac{2640903}{86708224 \pi ^6 }\left[\frac{\Gamma (6,4 \pi  y)}{q}+\frac{14446237 \Gamma (6,8 \pi  y)}{14084816 q^2}+\frac{213394328 \Gamma (6,12 \pi  y)}{213913143 q^3}+ \frac{3696743927 \Gamma (6,16 \pi  y)}{3605712896 q^4}+\cdots\right]  \\
\nonumber &&\hskip-0.4in + c_{\bm{\widehat {6}}I,1}^{(-5)} +\frac{39605205q}{10838528 \pi ^6}\left(1+\frac{10833291 q}{10561388}+\frac{1920794798 q^2}{1924812963}+\frac{11092486969 q^3}{10814861312}+\frac{41259765347 q^4}{41255421875}+\cdots\right)\,,  \\
\nonumber Y_{\bm{\widehat {6}}I,2}^{(-5)} &=& -\frac{4343919}{86708224 \pi ^6}\left[\frac{\Gamma (6,4 \pi  y)}{q}+\frac{11472573 \Gamma (6,8 \pi  y)}{11583784 q^2}+\frac{3169437206 \Gamma (6,12 \pi  y)}{3166716951 q^3}+\frac{5872648801 \Gamma (6,16 \pi  y)}{5930897408 q^4}+\cdots\right]   \\
\nonumber &&\hskip-0.4in + c_{\bm{\widehat {6}}I,2}^{(-5)}-\frac{65153715 q}{10838528 \pi ^6}\left(1+\frac{137688499 q}{138994592}+\frac{352114616 q^2}{351830061}+\frac{17619366097 q^3}{17791307776}+\frac{67871093581 q^4}{67868453125}+\cdots\right)  \,,  \\
\nonumber Y_{\bm{\widehat {6}}I,3}^{(-5)} &=& -\frac{6527296875 q^{1/5}}{355156885504 \pi ^6}\left[\frac{\Gamma \left(6,\frac{16 \pi  y}{5}\right)}{q}+\frac{238603563008 \Gamma \left(6,\frac{36 \pi  y}{5}\right)}{222007883427 q^2}+\frac{16386167552 \Gamma \left(6,\frac{56 \pi  y}{5}\right)}{16382505601 q^3}+\cdots\right]   \\
\nonumber &&\hskip-0.4in -\frac{25546875 q^{1/5}}{10838528 \pi ^6}\left(1+\frac{2394179 q}{2542752}+\frac{1771562 q^2}{1771561}+\frac{1711120429 q^3}{1828716544}+\frac{9400770538 q^4}{9348507189}+\cdots\right)\,, \\
\nonumber Y_{\bm{\widehat {6}}I,4}^{(-5)} &=& -\frac{3169234375 q^{2/5}}{31605147648 \sqrt{2} \pi ^6}\left[\frac{\Gamma \left(6,\frac{12 \pi  y}{5}\right)}{q}+\frac{52609148865 \Gamma \left(6,\frac{32 \pi  y}{5}\right)}{53170929664 q^2}+\frac{5788230111 \Gamma \left(6,\frac{52 \pi  y}{5}\right)}{5793056191 q^3}+\cdots\right] \\
\nonumber && \hskip-0.4in -\frac{4130390625 q^{2/5}}{346832896 \sqrt{2} \pi ^6}\left(1+\frac{2093200192 q}{2073328327}+\frac{822725873 q^2}{822218688}+\frac{429455616832 q^3}{425376378487}+\frac{1771562 q^4}{1771561}+\cdots\right) \,,   \\
\nonumber Y_{\bm{\widehat {6}}I,5}^{(-5)} &=& -\frac{86671875 q^{3/5}}{1387331584 \sqrt{2} \pi ^6}\left[\frac{\Gamma \left(6,\frac{8 \pi  y}{5}\right)}{q}+\frac{212084896 \Gamma \left(6,\frac{28 \pi  y}{5}\right)}{217533001 q^2}+\frac{516350869 \Gamma \left(6,\frac{48 \pi  y}{5}\right)}{517601664 q^3}+\frac{43511994016 \Gamma \left(6,\frac{68 \pi  y}{5}\right)}{44630365081 q^4}+\cdots\right]   \\
\nonumber &&\hskip-0.4in -\frac{9603359375 q^{3/5}}{1316881152 \sqrt{2} \pi ^6}\left(1+\frac{16559209485 q}{16111763456}+\frac{594667492947 q^2}{593325842707}+\frac{1960982731 q^3}{1911698496}+\frac{18238169358027 q^4}{18197015583547}+\cdots\right) \,,  \\
\nonumber Y_{\bm{\widehat {6}}I,6}^{(-5)} &=& -\frac{6984375 q^{4/5}}{86708224 \pi ^6}\left[\frac{\Gamma \left(6,\frac{4 \pi  y}{5}\right)}{q}+\frac{5232041 \Gamma \left(6,\frac{24 \pi  y}{5}\right)}{5213808 q^2}+\frac{1771562 \Gamma \left(6,\frac{44 \pi  y}{5}\right)}{1771561 q^3}+\frac{2508717269 \Gamma \left(6,\frac{64 \pi  y}{5}\right)}{2499805184 q^4}+\cdots\right]   \\
\nonumber &&\hskip-0.4in -\frac{430650234375 q^{4/5}}{44394610688 \pi ^6}\left(1+\frac{324231467008 q}{325497160827}+\frac{216226103168 q^2}{216173096209}+\frac{86136866242560 q^3}{86444030630521}+\frac{2380578655 q^4}{2381323536}+\cdots\right)\,, \\
\nonumber Y_{\bm{\widehat {6}}II,1}^{(-5)} &=&  -\frac{4343581}{86708224 \pi ^6}\left[\frac{\Gamma (6,4 \pi  y)}{q}+\frac{137688499 \Gamma (6,8 \pi  y)}{138994592 q^2}+\frac{352114616 \Gamma (6,12 \pi  y)}{351830061 q^3}+\frac{17619366097 \Gamma (6,16 \pi  y)}{17791307776 q^4}+\cdots\right]  \\
\nonumber &&\hskip-0.4in +c_{\bm{\widehat {6}}I,2}^{(-5)}-\frac{65158785 q}{10838528 \pi ^6}\left(1+\frac{11472573 q}{11583784}+\frac{3169437206 q^2}{3166716951}+\frac{5872648801 q^3}{5930897408}+\frac{22623697973 q^4}{22624578125}+\cdots\right) \,,  \\
\nonumber Y_{\bm{\widehat {6}}II,2}^{(-5)} &=&  -\frac{2640347}{86708224 \pi ^6}\left[\frac{\Gamma (6,4 \pi  y)}{q}+\frac{10833291 \Gamma (6,8 \pi  y)}{10561388 q^2}+\frac{1920794798 \Gamma (6,12 \pi  y)}{1924812963 q^3}+\frac{11092486969 \Gamma (6,16 \pi  y)}{10814861312 q^4}+\cdots\right] \\
\nonumber &&\hskip-0.4in +\frac{y^6}{6}- c_{\bm{\widehat {6}}I,1}^{(-5)} -\frac{39613545 q}{10838528 \pi ^6}\left(1+\frac{14446237 q}{14084816}+\frac{213394328 q^2}{213913143}+\frac{3696743927 q^3}{3605712896}+\frac{13753255301 q^4}{13754703125}+\cdots\right)  \\
\nonumber Y_{\bm{\widehat {6}}II,3}^{(-5)} &=& -\frac{28710015625 q^{1/5}}{355156885504 \pi ^6}\left[\frac{\Gamma \left(6,\frac{16 \pi  y}{5}\right)}{q}+\frac{324231467008 \Gamma \left(6,\frac{36 \pi  y}{5}\right)}{325497160827 q^2}+\frac{216226103168 \Gamma \left(6,\frac{56 \pi  y}{5}\right)}{216173096209 q^3}+\cdots\right]   \\
\nonumber &&\hskip-0.4in -\frac{104765625 q^{1/5}}{10838528 \pi ^6}\left(1+\frac{5232041 q}{5213808}+\frac{1771562 q^2}{1771561}+\frac{2508717269 q^3}{2499805184}+\frac{38324712254 q^4}{38337456087}+\cdots\right) \,, \\
\nonumber Y_{\bm{\widehat {6}}II,4}^{(-5)} &=& -\frac{1920671875 q^{2/5}}{31605147648 \sqrt{2} \pi ^6}\left[\frac{\Gamma \left(6,\frac{12 \pi  y}{5}\right)}{q}+\frac{16559209485 \Gamma \left(6,\frac{32 \pi  y}{5}\right)}{16111763456 q^2}+\frac{594667492947 \Gamma \left(6,\frac{52 \pi  y}{5}\right)}{593325842707 q^3}+\cdots\right]  \\
\nonumber &&\hskip-0.4in -\frac{1300078125 q^{2/5}}{173416448 \sqrt{2} \pi ^6}\left(1+\frac{212084896 q}{217533001}+\frac{516350869 q^2}{517601664}+\frac{43511994016 q^3}{44630365081}+\frac{1771562 q^4}{1771561}+\cdots\right)\,,  \\
\nonumber Y_{\bm{\widehat {6}}II,5}^{(-5)} &=& \frac{275359375 q^{3/5}}{2774663168 \sqrt{2} \pi ^6}\left[\frac{\Gamma \left(6,\frac{8 \pi  y}{5}\right)}{q}+\frac{2093200192 \Gamma \left(6,\frac{28 \pi  y}{5}\right)}{2073328327 q^2}+\frac{822725873 \Gamma \left(6,\frac{48 \pi  y}{5}\right)}{822218688 q^3}+\frac{429455616832 \Gamma \left(6,\frac{68 \pi  y}{5}\right)}{425376378487 q^4}+\cdots\right]  \\
\nonumber &&\hskip-0.4in \frac{15846171875 q^{3/5}}{1316881152 \sqrt{2} \pi ^6}\left(1+\frac{52609148865 q}{53170929664}+\frac{5788230111 q^2}{5793056191}+\frac{4686827323 q^3}{4731641568}+\frac{30001249459719 q^4}{30026267401759}+\cdots\right) \,, \\
\nonumber Y_{\bm{\widehat {6}}II,6}^{(-5)} &=& \frac{1703125 q^{4/5}}{86708224 \pi ^6}\left[\frac{\Gamma \left(6,\frac{4 \pi  y}{5}\right)}{q}+\frac{2394179 \Gamma \left(6,\frac{24 \pi  y}{5}\right)}{2542752 q^2}+\frac{1771562 \Gamma \left(6,\frac{44 \pi  y}{5}\right)}{1771561 q^3}+\frac{1711120429 \Gamma \left(6,\frac{64 \pi  y}{5}\right)}{1828716544 q^4}+\cdots\right]   \\
&&\hskip-0.4in \frac{97909453125 q^{4/5}}{44394610688 \pi ^6}\left(1+\frac{238603563008 q}{222007883427}+\frac{16386167552 q^2}{16382505601}+\frac{7001430589440 q^3}{6551091883369}+\frac{1089351445 q^4}{1082800224}+\cdots\right) \,,
\end{eqnarray}
\end{footnotesize}
where
\begin{equation}
 c_{\bm{\widehat {6}}I,1}^{(-5)}=0.00388957\,, \qquad
 c_{\bm{\widehat {6}}I,2}^{(-5)} =-0.00619737 \,.
\end{equation}
We summarize the polyharmonic Maa{\ss} form multiplets of level $N=5$ and weights $k=-5$ to $k=6$ in talbe~\ref{tab:MF_summary}.

%\newpage
\begin{table}[t!]
	\centering
\renewcommand{\tabcolsep}{1.5mm}
	\begin{tabular}{|c|c||c|c|}
		\hline  \hline
		Weight 	&  $Y^{(k)}_{\bm{r}}$ & Weight 	&  $Y^{(k)}_{\bm{r}}$  \\  \hline
		& & &  \\[-0.18in]

     $k=-5$ & $Y^{(-5)}_{\bm{\widehat {6}}I}$,  $Y^{(-5)}_{\bm{\widehat {6}}II}$  &   $k=1$ & $Y^{(1)}_{\bm{\widehat {6}}I}$, $Y^{(1)}_{\bm{\widehat {6}}II}$  \\
	& & &     \\[-0.18in] \hline
		& & &  \\[-0.18in]

		$k=-4$ &  $Y_{\bm{1}}^{(-4)}$,  $Y^{(-4)}_{\bm{3}}$,$Y^{(-4)}_{\bm{3^{\prime}}}$,  $Y^{(-4)}_{\bm{5}}$ & 	$k=2$ &  $Y^{(2)}_{\bm{1}}$, $Y^{(2)}_{\bm{3}}$, $Y^{(2)}_{\bm{3^{\prime}}}$, $Y^{(2)}_{\bm{5}}$ \\
		& & &     \\[-0.18in] \hline
		& & &  \\[-0.18in]

      $k=-3$ & $Y^{(-3)}_{\bm{\widehat {6}}I}$,  $Y^{(-3)}_{\bm{\widehat {6}}II}$ &  $k=3$ & $Y^{(3)}_{\bm{\widehat {4}}},~ Y^{(3)}_{\bm{\widehat {6}}I},~Y^{(3)}_{\bm{\widehat {6}}II}$\\
    	& & &    \\[-0.18in] \hline
		& & &   \\[-0.18in]
		
		$k=-2$ &  $Y^{(-2)}_{\bm{1}}$, $Y^{(-2)}_{\bm{3}}$, $Y^{(-2)}_{\bm{3^{\prime}}}$, $Y^{(-2)}_{\bm{5}}$  & 	$k=4$ &  $Y_{\bm{1}}^{(4)}$,  $Y^{(4)}_{\bm{3}}$, $Y^{(4)}_{\bm{3^{\prime}}}$,  $Y^{(4)}_{\bm{4}}$,  $Y^{(4)}_{\bm{5}I}$,  $Y^{(4)}_{\bm{5}II}$ \\
		& & &    \\[-0.18in] \hline
		& & &  \\[-0.18in]

   $k=-1$ & $Y^{(-1)}_{\bm{\widehat {6}}I}$,  $Y^{(-1)}_{\bm{\widehat {6}}II}$  &  $k=5$ & $Y^{(5)}_{\bm{\widehat {2}}},~ Y^{(5)}_{\bm{\widehat {2}^{\prime}}}, ~Y^{(5)}_{\bm{\widehat {4}}},~Y^{(5)}_{\bm{\widehat {6}}I},~Y^{(5)}_{\bm{\widehat {6}}II},~Y^{(5)}_{\bm{\widehat {6}}III}$\\
	& & &    \\[-0.18in] \hline
		& & & \\[-0.18in]
		
		$k=0$ &   $Y^{(0)}_{\bm{1}}$,  $Y^{(0)}_{\bm{3}}$, $Y^{(0)}_{\bm{3^{\prime}}}$, $Y^{(0)}_{\bm{5}}$  & $k=6$ & $Y_{\bm{1}}^{(6)}$,  $Y^{(6)}_{\bm{3}I}$, $Y^{(6)}_{\bm{3}II}$, $Y^{(6)}_{\bm{3^{\prime}}I}$,  $Y^{(6)}_{\bm{3^{\prime}}II}$,  $Y^{(6)}_{\bm{4}I}$,  $Y^{(6)}_{\bm{4}II}$, $Y^{(6)}_{\bm{5}I}$,  $Y^{(6)}_{\bm{5}II}$ \\[0.05in] \hline \hline

	\end{tabular}
\caption{\label{tab:MF_summary}Polyharmonic Maa{\ss} form multiplets of level $5$ of weight $k=-5,\,-4,\cdots\,5,\,6$, the subscript $\bm{r}$ denotes the transformation property under finite group $A^{\prime}_{5}$.  }
\end{table}

\end{appendix}

%\vskip 2cm
%
%\bibliographystyle{utphys}
%\bibliography{references}

\providecommand{\href}[2]{#2}\begingroup\raggedright\endgroup

\end{document}